\documentclass[a4paper]{article}
\usepackage[english]{babel}
\usepackage[utf8x]{inputenc}
\usepackage[T1]{fontenc}
\usepackage{authblk}
\usepackage[a4paper,top=1.0in,bottom=1.0in,left=1.25 in,right=1.25in,marginparwidth=3cm]{geometry}

\usepackage{amsmath, amssymb}
\usepackage{graphicx}
\usepackage{hyperref}
\hypersetup{colorlinks=true, allcolors=blue}
\usepackage[round,sort]{natbib}
\usepackage{enumerate}
\usepackage{algorithm}
\usepackage{algpseudocode}
\usepackage{pdfpages}
\usepackage{chngpage}

\usepackage{appendix}
\usepackage{nomencl}
\makenomenclature
\newcommand{\R}{\mathbb{R}}			% Real numbers
\newcommand{\dif}{\mathop{}\!\mathrm{d}}
\newcommand{\D}{\mathcal{D}}	
\DeclareMathOperator{\ind}{\mathbb{I}}
\newcommand{\e}{\mathrm{e}}         % Exponential
\newcommand{\x}{\boldsymbol{x}}

\newcommand{\thetas}{\boldsymbol{\theta}}

\newcommand{\mus}{\boldsymbol{\mu}}
\newcommand{\nus}{\boldsymbol{\nu}}
\newcommand{\fs}{\boldsymbol{f}}
\newcommand{\ps}{\boldsymbol{p}}

\newcommand{\omegas}{\boldsymbol{\omega}}
\newcommand{\Omegas}{\boldsymbol{\Omega}}
\newcommand{\Ks}{\boldsymbol{K}}
\newcommand{\us}{\boldsymbol{u}}

\newcommand{\diff}{\mathrm{d}}

\newcommand{\GP}{\mathcal{GP}}
\newcommand{\bx}{\boldsymbol{x}}
\newcommand{\bu}{\boldsymbol{u}}
\newcommand{\bnu}{\boldsymbol{\nu}}

\newcommand{\barlambda}{\bar{\lambda}}
\newcommand{\calX}{\mathcal{X}}
\newcommand{\calT}{\mathcal{T}}

\newcommand{\EE}[2]{\mathbb{E}_{#1}\left[#2\right]}
\newcommand{\PG}{p_{\scriptscriptstyle \mathrm{PG}}}

\title{GP-ETAS: Semiparametric Bayesian inference for the spatio-temporal Epidemic Type Aftershock Sequence model}
\author[1]{Christian Molkenthin\thanks{Contributed equally, e-mail: {\tt molkenth@uni-potsdam.de}, {\tt christian.research@mailbox.org}}}
\author[2]{Christian Donner$^\ast$}
\author[1]{Sebastian Reich} 
\author[1]{Gert Z\"oller}
\author[3]{Sebastian Hainzl}
\author[1]{Matthias Holschneider} 
\author[4]{Manfred Opper}
\affil[1]{University of Potsdam, Institute of Mathematics}
\affil[2]{ETH Z\"urich, Swiss Data Science Center}
\affil[3]{Helmholtz Centre Potsdam, GFZ German Research Centre for Geosciences}
\affil[4]{TU Berlin, Institute of Software Engineering and Theoretical Computer Science}

\begin{document}
	\maketitle
	
	\begin{abstract}
		The spatio-temporal Epidemic Type Aftershock Sequence (ETAS) model is widely used to describe the self-exciting nature of earthquake occurrences. While traditional inference methods provide only point estimates of the model parameters, we aim at a full Bayesian treatment of model inference, allowing naturally to incorporate prior knowledge and uncertainty quantification of the resulting estimates.
		Therefore, we introduce a highly flexible, non-parametric representation for the spatially varying ETAS background intensity through a Gaussian process (GP) prior. Combined with classical triggering functions this results in a new model formulation, namely the GP-ETAS model. We enable tractable and efficient Gibbs sampling by deriving an augmented form of the GP-ETAS inference problem.  This novel sampling approach allows us to assess the posterior model variables conditioned on observed earthquake catalogues, 
		i.e., the spatial background intensity and the parameters of the triggering function. 
		Empirical results on two synthetic data sets indicate that GP-ETAS outperforms standard models and thus demonstrate the predictive power for observed earthquake catalogues  including uncertainty quantification for the estimated parameters. Finally, a case study for the l'Aquila region, Italy, with the devastating event on 6 April 2009, is presented.  
	\end{abstract}
	
	%#####################################################################################################
	\section{Introduction}
	\label{sec:introduction}
	%#####################################################################################################
	Point process models are
	often used in statistical seismology for describing 
	the occurrence of earthquakes (point data) in a spatio-temporal setting. 
	The most widely used one is the Epidemic Type Aftershock Sequence (ETAS) model, first introduced as 
	a temporal point process model \citep{Ogata1988},  and later enhanced to the currently predominantly employed spatio-temporal version \citep{Ogata1998}. 
	Main applications are seismic forecasting or the characterisation
	of earthquake clustering in a particular 
	geographical region and topics alike \citep[e.g.,][]{jordan2011operational}.
	The ETAS model is an example of a self-exciting, spatio-temporal, marked point process, which is a particular Hawkes process model, extending the temporal Hawkes process proposed by \cite{hawkes1971}. 
	Self-excitation means that one event can trigger a series of subsequent follow-up events (offspring), as in the case of earthquakes, 
	main shocks and aftershocks. The ETAS model assigns the earthquake magnitude as an additional mark to each event. 
	Besides its primary application in seismology, the Hawkes process is utilised in several other domains , e.g.~finance \citep{Bacry2015,Filimonov2015}, crime \citep{Mohler2011,Porter2010}, neuronal activities \citep{Gerhard2017}, social networks \citep{Zhao2015,zhou2013learning}, genomes \citep{Reynaud-Bouret2010}, transportation \citep{Hu2017}.

	\medskip
	
	The ETAS model is characterised by its conditional intensity function, that is, the rate of arriving events 
	conditioned on the history of previous events. This time-dependent conditional intensity function itself consists of two parts,
	(i) a stationary background intensity $\mu$ of a Poisson process, which models 
	the arrival of spontaneous (exogenous) events, and 
	(ii) a time-dependent triggering function $\varphi$ which encodes the form of self-excitation 
	by adding a positive impulse response for each event, that is, a spontaneous jump which decays gradually at time progresses.
	An alternative approach interprets the stationary Hawkes process (i.e.,~the ETAS model) as 
	a Poisson cluster process or branching process \citep{HawkesAndOakes1974}, 
	which leads to the concept of a non-observable, underlying branching structure (latent variable).
	Each event has either a direct parent from which it was generated 
	or is background; this yields an ordered branching structure useful for designing
	simulation and inference algorithms, e.g. \citep{Zhuang2002,Veen2008}.
	\medskip
	
	The fitting of an ETAS model to data entails learning the conditional intensity function.
	Most currently used ETAS models employ a \emph{parametric} form for the background $\mu$ and the 
	triggering function $\varphi$.
	The parameters are then calibrated via maximum-likelihood estimation (MLE), 
	maximising the classical likelihood function for point processes. 
	Unfortunately, MLE has no simple analytical form. 
	Alternatively, different numerical optimisation methods are employed 
	involving, e.g., an Expectation-Maximisation (EM, \cite{Dempster1977}) algorithm using the latent branching structure  \citep{Ogata1998,Veen2008,Lippiello2014,Lombardi2015}.
	
	\medskip
	
	\emph{Non-parametric} methods have also been suggested previously
	to fit the conditional intensity function (or parts of it). For example,
	\cite{Zhuang2002}  and \cite{Adelfio2014} fit simultaneously 
	a non-parametric background intensity via kernel density estimation 
	and a classical parametric triggering kernel;
	\cite{Marsan2008} consider a constant background intensity combined with a non-parametric histogram estimator
	of the triggering kernel;
	\cite{Mohler2011} suggest non-parametric kernel density estimators 
	for both the components, background $\mu$\nomenclature{$\mu$}{Background intensity} and offspring $\varphi$\nomenclature{$\varphi$}{Triggering function}; and
	\cite{Fox2016} propose a non-parametric kernel density estimator for the background and 
	non-parametric histogram estimation for the triggering kernel. Furthermore,
	\cite{Bacry2014} suggest a non-Bayesian, non-parametric way 
	of estimating the triggering function of a Hawkes process 
	based on Wiener Hopf integral equation;
	\cite{Kirchner2017} presents a non-Bayesian non-parametric estimation procedure for a 
	multivariate Hawkes process based on an integer-valued autoregressive model.
	\medskip
	
	Uncertainty quantification of the ETAS model remains challenging.
	Most estimation techniques  deliver a point estimate for its 
	conditional intensity function and
	uncertainty quantification is usually
	achieved by relying on standard errors of estimated ETAS parameters, 
	based on the Hessian \citep{RATHBUN1996,Wang2010,Ogata1978}.
	This approach requires that the observational window is long enough 
	(sufficiently large sample size), otherwise it may lead to an 
	underestimation of parameter uncertainties.
	Moreover, standard errors based on Hessians cannot be obtained in the non-parametric case.
	Another approach to uncertainty quantification relies on various 
	bootstrap techniques based on many forward simulations, e.g.,
	\cite{Fox2016}.
	Ad hoc variants for quantifying uncertainty have also been devised, e.g., 
	by the solutions of multiple optimisation runs of the MLE, e.g.~\cite{Lombardi2015}.
	\medskip
	
	None of the aforementioned uncertainty quantification methods are fully satisfactory 
	and we believe that a fully semi-parametric Bayesian framework
	is worthwhile pursuing, which allows one to incorporate prior knowledge.
	The posterior distribution effectively encodes the uncertainty of 
	the quantities arising from data and a prior distribution. 
	However, this poses a challenge for a spatio-temporal ETAS model, 
	as there is no known conjugate structure, that is, the posterior can not be obtained in closed-form.
	One way to deal with this problem is to employ Monte Carlo sampling techniques, e.g.~via Markov chain Monte Carlo 
	(MCMC). However implementing MCMC remains challenging for \emph{non- or semi-parametric} conditional intensity functions.
	Several studies have suggested Bayesian methods for the temporal, or multivariate Hawkes process,
	either based on parametric forms of the conditional intensity function \citep{Rasmussen2013,Ross2018} 
	or for non-parameteric versions \citep{Donnet2018,Zhang2019,zhang2019variational,linderman2015scalable,Zhou2019a}. But
	these studies rarely consider the spatio-temporal ETAS model and only with strong simplifications, e.g., a constant background intensity 
	$\mu$ \citep{Rasmussen2013}. Recently, however, \citep{Kolev2020} considered an inhomogeneous  background intensity modelled via a Dirichlet process.
	\medskip 
	
	It is desirable to estimate the spatially dependent background $\mu$ of an ETAS model fully \emph{non-parametrically} 
	as it is often difficult to specify an appropriate functional form a priori. The background intensity (also called long-term component)
	is of particular importance for seismic hazard assessment and seismic forecasting.
	It is often preferred to maintain a specific \emph{parametric} 
	triggering function $\varphi$ (e.g., modified Omori law \cite{utsu1961statistical,omori1894after})
	as there is a long tradition for interpreting and comparing this particular parametric form in different settings, 
	regions, etc. Thus, one faces two main issues for the development of a suitable Bayesian inference approach:
	\begin{itemize}
		\item[(i)] providing a Bayesian non-parametric way of modelling the background intensity $\mu$, and
		\item[(ii)] creating a fully Bayesian inference algorithm for the resulting ETAS models including its parametric triggering component 
		$\varphi$.
	\end{itemize}
	\medskip
	
	We address these two issues in this paper by first formulating a Bayesian non-parametric %scheme
	approach to the estimation of the background intensity $\mu$ via a Gaussian process (GP) prior.
	Secondly, we propose and implement a computationally tractable approach for the implied Bayesian inference problem 
	by introducing auxiliary variables: a latent branching structure, a latent Poisson process, and latent P\'olya--gamma random variables.
	More specifically, we suggest to model the background intensity $\mu$ \emph{non-parametrically} by sigmoid 
	transformed realisations of a GP prior, i.e., as a Sigmoid-Gauss-Cox-Process (SGCP, \cite{Adams2009}), 
	which is a doubly stochastic Poisson process. No specific functional form has to be chosen for the intensity function, and the prior
	fully specifies the chosen GP. \cite{Adams2009} proposed a Bayesian
	inference scheme via MCMC for SGCPs. However, the suggested scheme is computationally demanding and convergence is slow. 
	Our paper relies instead on the work of \cite{Donner2018} who recently enhanced Bayesian inference for SGCPs substantially 
	by data augmentation with P\'olya--gamma random variables \citep{Polson2013}. 
	The triggering function $\varphi$ is modelled in a classical \emph{parametric} way, 
	which together with the SGCP model for $\mu$ leads to a novel semi-parametric 
	ETAS model formulation, which we denote as GP-ETAS. In order to implement such an approach, 
	we need to address a number of computational challenges:
	\begin{itemize}
		\item[(i)] the background intensity $\mu$ and the triggering function $\varphi$ 
		are not directly separable in the likelihood; 
		\item[(ii)] intractable integrals for the posterior computation when $\mu$ is modelled as SGCP, and
		\item[(iii)] handling a non-Gaussian point process likelihood
		while using a Gaussian process prior. 
	\end{itemize}
	
	\medskip
	
	We show how these challenges can be resolved by data augmentation
	(introducing auxiliary variables), which strongly simplifies the Bayesian inference problem.
	It effectively allows us to construct an efficient MCMC sampling scheme for 
	the posterior involving an overall Gibbs sampler \citep{Geman1984} 
	consisting of three main steps, each conditioned on the previous: 
	\begin{itemize}
		\item[(a)] conditionally sampling the latent branching structure which 
		factorises the likelihood function into background and triggering component;
		\item[(b)] conditionally sampling the posterior of the background intensity $\mu$
		from explicit conditional densities easy to sample from; and 
		\item[(c)] conditionally sampling the parameters of the triggering function $\varphi$
		by employing Metropolis-Hastings (MH) \citep{Hastings1970} steps.
	\end{itemize}
	
	\medskip
	
	The remainder of this paper is structured as follows: First we describe the classical spatio-temporal ETAS model; 
	secondly we introduce our GP-ETAS model including a simulation algorithm;
	thirdly the Bayesian inference approach is presented;
	fourthly empirical results based on synthetic and real data illustrate practical aspects of the framework.
	The paper concludes with a discussion and some final remarks.
	
	%####################################################################################
	\section{Background}
	\label{sec:model}
	%####################################################################################
	We start with a review of the classical spatio-temporal ETAS model, which we will use as a benchmark for comparison.
	
	%------------------------------------------------------------------------------------------------------------------------
	\subsection{Classical ETAS model}
	\label{sec:model_classical_ETAS}
	%------------------------------------------------------------------------------------------------------------------------
	The ETAS model  \citep{Ogata1998}, 
	describes a stochastic process, which generates point pattern over some domain 
	$\calX \times \calT \times \mathcal{M}$\nomenclature{$\calX$}{Spatial domain}\nomenclature{$\calT$}{Time domain}\nomenclature{$\mathcal{M}$}{Magnitude domain},
	where $\calT \times \calX$ is the time-space window and 
	$\mathcal{M}$ the mark space of the process.
	Realisations of this point process are denoted by 
	$\D=\{(t_i,\boldsymbol{x}_i,m_i)\}_{i=1}^{N_{\D}}$\nomenclature{$\D$}{Data set}, 
	which in seismology can be interpreted as an 
	earthquake catalog consisting of $N_{\D}$\nomenclature{$N_{\D}$}{Number of elements/events in set ${\D}$} observed events.
	$\D$ is usually ordered in time (time series),
	$t_i \in \calT \subseteq \R_{> 0}$ 
	is the time of the $i$th event (time of the earthquake),
	$\x_i \in \calX \subseteq \R^2$ 
	is the corresponding location (longitude and latitude of the epicenter), and
	$m_i \in \mathcal{M} \subseteq \R$ 
	the corresponding mark (the magnitude of the earthquake).
	\nomenclature{$m_i$}{Earthquake magnitude of event $i$}
	\nomenclature{$t_i$}{Event time of event $i$}
	\nomenclature{$\x_i$}{Position of event $i$}
	
	\subsubsection{Interpretations}
	There are two equivalent interpretations of the ETAS model (Hawkes process). We briefly discuss both.
	
	\paragraph{Conditional intensity function}
	One way to define the ETAS model is by a conditional intensity function, which models 
	the infinitesimal rate of expected arrivals around $(t,\x)$ given the history 
	$H_t=\{(t_i,\x_i,m_i): t_i<t\}$\nomenclature{$H_t$}{History until time $t$} of the process until time $t$. 
	The earthquake magnitudes $m_i\ge m_0$ are not influenced by $H_t$ and 
	are modeled as independent %of all other model components 
	each following an exponential distribution $p_M(m|\beta)=\beta\e^{-\beta(m-m_0)}, \ \beta>0$,
	and $m_0$ is the magnitude of completeness,
	(cut-off magnitude) a threshold above which all events are observed
	(complete data)
	\nomenclature{$\beta$}{Rate parameter of magnitudes}
	\nomenclature{$m_0$}{Magnitude of completeness}.
	The conditional ETAS intensity function can be written as
	\citep{Ogata1998}
	\nomenclature{$\lambda$}{ETAS intensity} with $\thetas=(\thetas_\mu,\thetas_\varphi)$\nomenclature{$\thetas_\varphi$}{Parameters of triggering function $\varphi$}
	\begin{align}\label{eq:lambda_tx}
	\lambda(t,\x|H_t,\thetas_\mu,\thetas_\varphi) = \mu(\x|\thetas_\mu)+ \sum_{i: t_i < t}\varphi(t-t_i,\x-\x_i|m_i,\thetas_\varphi),
	\end{align}
	a set of parameters. Here the background intensity
	$\mu(\x|\thetas_\mu): \R^2 \rightarrow [0,\infty)$ defines a non-homogeneous Poisson process in space 
	but stationary in time with $\thetas_\mu$ as the required parameters, while
	$ \varphi(t-t_i,\x-\x_i|m_i,\thetas_\varphi):\R^4\rightarrow[0,\infty)$ 
	is the triggering function,
	modeling the rate of aftershocks (self-exciting process) 
	following an event at $(t_i,\x_i)$ with magnitude $m_i$,
	controlled by the parameters $\thetas_\varphi$.
	Specific parametric representations of $\mu(\cdot)$ and $\varphi(\cdot)$ for the ETAS model 
	will be discussed  in Section \ref{sec:ingredientsEtas}.
	
	\paragraph{Latent branching structure}
	\label{para:branching}
	
	Another interpretation of a Hawkes process (with the ETAS model being a particular example) 
	is as Poisson cluster- or branching process  \citep{HawkesAndOakes1974},
	leading to the concept of an underlying branching structure, that is,
	a non observable latent random variable $z_i$ 
	for each event $i$.
	Events are structured in an ensemble of trees, 
	either having a parent, which is one of the previous events 
	or being spontaneous, called background. The latent variable is typically modelled 
	as taking integer values in a discrete set $z_i\in \{0,1,...,i-1\}$, where
	\nomenclature{$z_i$}{Branching structure index}
	\begin{align*}
	z_i = 
	\begin{cases}
	0 &\text{event $i$ is background}\\
	j>0 &\text{event $i$ is direct offspring (aftershock) of event $j$ at $t_j<t_i$}.
	\end{cases}
	\end{align*}
	Background events $z_i=0$ occur according to a Poisson process with intensity $\mu(\x)$ 
	and form cluster centres, i.e., initial points for branching trees.
	Within each branching tree, an existing event at $t_j$ 
	can produce direct offspring at $t>t_j$ 
	according to an inhomogeneous Poisson process with rate  
	$\lambda_j(t|t_j,\x_j,m_j)=\varphi(t-t_j,\x-\x_j|m_j,\thetas_\varphi)$.
	The overall intensity $\lambda(t,\x|H_t)$ is the sum of all the offspring Poisson processes 
	$\sum_j \lambda_j$ with $t_j<t$ and the background Poisson process $\mu(\x)$ (Poisson superposition), as given in (\ref{eq:lambda_tx}).
	
	\medskip
	
	The latent branching structure cannot be observed. However, by its construction 
	(superposition of $i$ Poisson processes at $t_i$) the probability $p_{i0}=p(z_i=0)$ (background event) is \citep[see, e.g.,][]{Zhuang2002},
	\begin{align}\label{eq:p_i_BG}
	p_{i0} = \frac{\mu(\x_i|\thetas_\mu)}{\lambda(t_i,\x_i|H_{t_i},\thetas_\mu,\thetas_\varphi)},
	\end{align}
	while the probability  $p_{ij}=p(z_i=j)$ (event $j$ triggered event $i$, $j>0$) is,
	\begin{align}\label{eq:p_ij_offspring}
	p_{ij} = \frac{\varphi(t_i-t_j,\x_i-\x_j|m_j,\thetas_\varphi)}{\lambda(t_i,\x_i|H_{t_i},\thetas_\mu,\thetas_\varphi)},
	\end{align}
	with $p_{i0}+\sum_j p_{ij}=1$.
	\nomenclature{$p_{i0}$}{Probability being background event}
	\nomenclature{$p_{ij}$}{Probability that event $j$ triggered event $i$}
	\medskip
	
	%------------------------------------------------------------------------------------------------------------------------
	\subsubsection{Components of the ETAS model}
	\label{sec:ingredientsEtas}
	%------------------------------------------------------------------------------------------------------------------------
	
	This section sketches the components (background and triggering function)  as 
	given in \eqref{eq:lambda_tx}.
	
	\paragraph{Background intensity}
	The background intensity $\mu(\x)$ is usually modelled either as piecewise constant function 
	over a rectangular grid (or specific polygones, seismo-tectonic units) 
	with $L$ cells (e.g., in \cite{Veen2008,Lombardi2015}),
	\begin{align}\label{eq:mu_k}
	\mu(\x \vert \thetas_\mu)=\mu_l
	\end{align}
	if $\x$ is in grid cell $l$, $l=1,...,L$; 
	or via a weighted kernel density estimator with variable bandwidth, as
	suggested by \cite{Zhuang2002},
	\begin{align}\label{eq:mu_kde}
	\mu_{\mathrm{kde}}(\x) = \frac{1}{|\calT|}\sum_{i=1}^{N_{\D}}p_{i0}k_{d_i}(\x-\x_i).
	\end{align}
	Here, $|\calT|$ is the length of the observational time window,
	$p_{i0}$
	is the probability that event $i$ is background as defined in \eqref{eq:p_i_BG}, 
	$d_i=\max\{d_{min},r_{i,n_p}\}$ is the variable bandwidth determined for event $i$ 
	corresponding to the distance $r_{i,n_p}$ of its number of nearest neighbours  $n_p$, 
	where $d_{min}$ is some minimal bandwidth, and
	$k_d(\cdot)$
	is an isotropic, bivariate Gaussian kernel function. 
	There are different suggestions to select $n_p$; 
	\cite{Zhuang2002} propose to choose $n_p$  between 10 and 100,
	and state that estimated parameters only change slightly if $n_p$ is changed in the range of 15–100; 
	\cite{Zhuang2011} suggests based on cross-validation experiments, 
	that an optimal $n_p$ is in the range $3∼6$ for Japan.
	The minimal bandwidth is commonly chosen as 
	$d_{min}\in [0.02,0.05]$ degrees, 
	which is in the range of the localisation error \citep{Zhuang2002}. 
	\nomenclature{$n_p$}{Number of nearest neighbours used in adaptive kernel density estimation}
	
	\medskip
	
	Background parameters to be estimated are 
	$\thetas_\mu=(\mu_1,\mu_2,...,\mu_L)$  in the first case
	and the scaled kernel density estimator  
	$\mu_{\mathrm{kde}}(\x)$ given through estimated background probabilities 
	$\{p_{i0}\}_{i=1}^{N_\D}$ in the second case, respectively. For non-parametric models of $\mu$ as in~\eqref{eq:mu_kde} we neglect the explicit dependency on $\thetas_\mu$ in our notation, but the reader should keep in mind, that in such cases $\mu$ depends on a varying (potentially infinite) number of parameters.  
	
	\paragraph{Parametric triggering function}
	The triggering function $\varphi(t-t_i,\x-\x_i|m_i,\thetas_\varphi)$ of the ETAS model is usually a non-negative 
	parametric function, which is separable in space and time,
	and depends on $m_i$ and $\thetas_\varphi$.
	There are numerous suggested parameterisations. See, for example, \cite{Ogata1998,Console2003,Zhuang2002,Ogata2006}.
	One of the most common parametrisations is provided by
	\begin{align}\label{eq:var_phi}
	\varphi(t-t_i,\x-\x_i|m_i,\thetas_\varphi) = 
	\kappa(m_i|K_0,\alpha)
	g(t-t_i|c,p)
	s(\x-\x_i|m_i,\thetas_s).
	\end{align}
	The first term $\kappa(\cdot)$ 
	is proportional to the aftershock productivity 
	(or Utsu law, \cite{utsu1970aftershocks}) of event i with $m_i$,
	\nomenclature{$K_0$}{Parameter of aftershock productivity law}
	\nomenclature{$\alpha$}{Parameter of aftershock productivity law}
	\begin{align}\label{eq:kappa_m}
	\kappa(m_i|K_0,\alpha)&=K_0e^{\alpha(m_i-m_0)},
	\end{align}
	and $K_0$ is called productivity coefficient.
	The second term 
	$g(\cdot)$\nomenclature{$g$}{Temporal distribution of aftershocks} describes the temporal distribution of aftershocks (offspring); a power law decay 
	proportional to the modified Omori Utsu law \citep{omori1894after,utsu1961statistical}, and $t-t_i>0$ is the elapsed time since 
	the parent event (main shock), that is,
	\nomenclature{$c$}{Parameter of temporal aftershock distribution}\nomenclature{$p$}{Parameter of temporal aftershock distribution}
	\begin{align}\label{eq:gt}
	g(t-t_i|c,p)=(t-t_i+c)^{-p}.
	\end{align}
	Finally, the third term $s(\cdot)$ is a probability density function for the spatial distribution
	of the direct aftershocks (offspring) around the triggering event at $\x_i$.
	Often, one of the following probability density functions are employed.
	One distinguishes between 
	a short range decay, which uses an 
	isotropic Gaussian distribution 
	with covariance 
	$d_1^2 e^{\alpha(m_i-m_0)}\boldsymbol{I}$
	\citep{Ogata1998,Zhuang2002};
	and a long range decay following a Pareto distribution
	\citep{Ogata2006,Kagan2002},
	\nomenclature{$s$}{Spatial aftershock distribution}
	\nomenclature{$q$}{Parameter of spatial aftershock distribution}
	\nomenclature{$d$}{Parameter of spatial aftershock distribution}
	\nomenclature{$\gamma$}{Parameter of spatial aftershock distribution}
	\nomenclature{$\sigma_m$}{Magnitude dependent scale function of spatial aftershock distribution}
	\begin{align}\label{eq:s_x_pwl}
	s(\x-\x_i|m_i,d,\gamma,q)=
	\frac{q-1}{\pi \sigma_m(m_i)}
	\left[
	1+\frac{(\x-\x_i)^\top(\x-\x_i)}{\sigma_m(m_i)}
	\right]^{-q},
	\end{align}
	where 
	$\sigma_m(m_i)=d^2 10^{2\gamma(m_i)}$. 
	
	\medskip
	
	The unknown parameters to be estimated are 
	$\thetas_\varphi=(K_0,\alpha,c,p,d_1)$, or
	$\thetas_\varphi=(K_0,\alpha,c,p,d,\gamma,q)$ 
	depending on which version of $s(\cdot)$ is used. Note that
	$q>1$ and the rest of the parameters are strictly positive.

	%------------------------------------------------------------------------------------------------------------------------
	\subsubsection{Parameter estimation via MLE}\label{app:MLE_ETAS_classical}
	%------------------------------------------------------------------------------------------------------------------------
	The likelihood function observing $\D$ under the spatio-temporal 
	ETAS model is given in \eqref{eq:L}; it is usually analytically intractable for simple direct optimisation.
	Numerical optimisation methods (e.g., quasi-Newton methods as in \citep{Ogata1988,Ogata1998},
	using an EM algorithm \citep{Veen2008} or simulated annealing 
	\citep{Lombardi2015,Lippiello2014}) are usually employed.
	Often the integral term related to the triggering function in \eqref{eq:L} 
	using \eqref{eq:lambda_tx} is approximated as
	$\int_{\calT_i}\int_{\calX}\sum_{i: t_i < t}\varphi(t-t_i,\x-\x_i|m_i,\thetas_\varphi)\dif \x \dif t \leq \int_{\calT_i}\int_{\R^2}\sum_{i: t_i < t}\varphi(t-t_i,\x-\x_i|m_i,\thetas_\varphi)\dif \x \dif t$, 
	by integrating over $\R^2$ in space instead of an arbitrary $\calX$ \citep{Schoenberg2013}.
	The introduced bias is small and 
	often negligible \citep{Schoenberg2013,Lippiello2014}
	while the computations are greatly simplified as 
	$\int_{\R^2}s(\x-\x_i|m_i)\dif \x =1$. We also use this approximation.
	Computational and numerical details of MLE using (\ref{eq:L}) are given in \cite{Ogata1998}.
	Instead of directly maximising (\ref{eq:L}),
	one can augment the likelihood function by a the
	latent branching structure $Z$  and apply an EM algorithm
	for MLE \citep{Veen2008,Mohler2011}, which is supposed 
	to be advantageous, e.g. 
	regarding stability and convergence \citep{Veen2008}.
	\medskip
	
	%------------------------------------------------------------------------------------------------------------------------
	\section{Bayesian GP-ETAS model}
	\label{sec:GP-ETAS}
	%------------------------------------------------------------------------------------------------------------------------
	Our goal is to improve the inference of  
	the spatio-temporal ETAS model in order to allow
	for comprehensive uncertainty quantification. 
	Despite the availability of powerful MLE based inference methods 
	\citep[see, e.g.,][]{Ogata1998,Veen2008,Lippiello2014,Lombardi2015}, 
	we believe that a Bayesian framework can complement existing methods
	and will provide a more reliable quantification of uncertainties.
	
	%%%%%%%%%%%%%%%%%%%%
	%
	\subsection{GP-ETAS model specification}
	%
	%%%%%%%%%%%%%%%%%%%%%
	
	We introduce a novel formulation of the spatio-temporal ETAS model, which models the background rate $\mu(\x)$ 
	in a Bayesian \textit{non-parametric} way
	via a GP \citep{Williams2006gaussian}, while the triggering function $\varphi(\cdot)$ assumes still a classical parametric form (modified Omori law \eqref{eq:var_phi}). As we will see subsequently, we are able to perform Bayesian inference for this model via Monte Carlo sampling despite its complex form.
	
	\medskip
	
	While the conditional intensity function of the GP-ETAS model is still given by~\eqref{eq:lambda_tx}, the background intensity is a priori defined by
	\begin{align}\label{eq:GP-ETAS}
	\mu(\x) = \bar{\lambda} \sigma(f(\x)) = \frac{\bar{\lambda} }{1+e^{-f(\x)}},
	\end{align}
	
	where $\sigma(\cdot)$\nomenclature{$\sigma$}{Logistic sigmoid} is the logistic sigmoid function, $\bar{\lambda}$\nomenclature{$\bar{\lambda}$}{Upper bound for GP-ETAS background intensity} a positive scalar, and $f(\bx)$\nomenclature{$f$}{Gaussian process function} an arbitrary scalar function mapping $\x \in\calX$ to the real line $\R$. Since $\sigma:\R\rightarrow [0,1]$ the background intensity of the GP-ETAS model is bounded from above by $\bar{\lambda}$, i.e., $\mu(\bx)\in[0,\bar{\lambda}]$ for any $\bx\in\calX$. 
	
	\medskip
	
	For the function $f(\bx)$ the GP-ETAS model assumes a Gaussian process prior, which implies that 
	the prior over any discrete set of $J$ function values
	$\fs=\{f(\x_i)\}_{i=1}^J$\nomenclature{$\fs$}{Gaussian process at a finite set of points.} at positions 
	$\{\x_1,\x_2,...,\x_J\}$ 
	is a $J$ dimensional Gaussian 
	distribution
	$\mathcal{N}(\fs|\mus_f,\Ks_{\fs,\fs}),$\nomenclature{$\mus_f$}{Mean of GP prior}
	where $\mus_f$\nomenclature{$\mus_f$}{Mean of GP prior} is the prior mean and
	$\Ks_{\fs,\fs}\in\R^{J\times J}$\nomenclature{$\Ks_{\fs,\fs}$}{Covariance matrix of GP prior between $\fs$} is the covariance matrix
	between function values at positions $\x_i$. The matrix $\Ks_{\fs,\fs}$ is
	built from the covariance function (kernel) $k(\x,\x'|\nus)$ 
	such that
	$\Ks_{i,j}=k(\x_i,\x_j|\nus)$, where $\nus$ are hyperparameters.
	We set $\mus_f =0$ and employ a  Gaussian covariance function
	\nomenclature{$k$}{Covariance function of GP prior}
	\begin{align}\label{eq:cov_fun}
	k(\x,\x'|\nus)=\nu_0 \prod_{i=1}^2 e^{-\frac{(\x-\x')^2}{2\nu_{i}^2}},
	\end{align}
	where
	$ \nu_0$ is the so called amplitude and $(\nu_1,\nu_2)$ are the length scales,
	representing a distance in input space over which the function values become weakly correlated.
	Note that the parameter $\barlambda$ and the hyperparameters $\nus$\nomenclature{$\nus$}{Parameters of covariance function $k$} are also to be inferred from the data. For an in--depth treatment of GPs we refer to \cite{Williams2006gaussian}.
	
	\medskip
	
	The complete specification of the prior model of GP-ETAS
	including the hyperparameters is now as follows:
	\begin{subequations}
		\begin{align}
		\nus & \sim p_{\nus}  \text{, a prior on $\nus$  (exponential distribution)}\\
		f &\sim \GP \text{ prior with zero mean and a covariance function}\\
		\barlambda & \sim p_{\barlambda} \text{, a prior on $\barlambda$ (gamma distribution)}\\
		\label{eq:generative_model_GP-ETAS}
		\mu | \barlambda,f,\nus & \sim \text{prior model on $\mu$ as defined in \eqref{eq:GP-ETAS}}\\ 
		\thetas_\varphi &\sim p_{\thetas_\varphi} \text{, a prior on $\thetas_\varphi$ of a triggering function $\varphi(\cdot)$ (uniform distribution)}
		\end{align}
	\end{subequations}
	The corresponding observational model is
	\begin{align}
	\D | \mu, \thetas_\varphi \sim \text{ Hawkes process with ETAS intensity function given by GP-ETAS in  \eqref{eq:lambda_tx},\eqref{eq:GP-ETAS}},
	\end{align}
	where $\D$ is the data. Note that some quantities are independent by construction, e.g., $\nus$ and $\barlambda$, 
	$f$ and $\barlambda$.
	
	\medskip
	
	Without the triggering function in the intensity function~\eqref{eq:lambda_tx} the GP-ETAS model would be equivalent to the 
	SGCP model which is used to describe an inhomogeneous Poisson process~\citep{Adams2009} 
	because of its favourable statistical properties \citep{kirichenko2015optimality}.
	\medskip
	
	In the following we 
	sketch how to generate data from the GP-ETAS model. A full description of the Bayesian inference problem is provided 
	in Section \ref{sec:Inference}.

	%------------------------------------------------------------------------------------------------------------------------
	\subsection{Simulating the GP-ETAS model}
	\label{sec:data_simulation}
	%------------------------------------------------------------------------------------------------------------------------
	Data $\D=\{(t_i,\boldsymbol{x}_i,m_i)\}_{i=1}^{N_{\D}}$
	can be easily simulated from the GP-ETAS model 
	using the latent branching structure of the 
	point process. 
	We propose a procedure which consists of two parts:
	\begin{enumerate}
		\item Generate all \emph{background} events
		$\D_0=\{(t_i,\boldsymbol{x}_i,m_i,z_i=0)\}_{i=1}^{N_{\D_0}}$\nomenclature{$\D_0$}{Subset of data $\D$ containing only background events}
		from a SGCP in equation (\ref{eq:GP-ETAS}) as explained in \cite{Adams2009}.
		\item Sample all \emph{aftershock} events (offspring) 
		given $\D_0$ in possibly several generations
		denoted as
		$\D_\varphi=\{(t_i,\boldsymbol{x}_i,m_i,z_i \neq 0)\}_{i=1}^{N_{\D_\varphi}}$
		and add them to obtain $\D=\D_0 \cup \D_\varphi$.
	\end{enumerate}
	The above procedure 
	can be implemented based on the
	\emph{thinning} algorithm  \citep{Lewis1976};
	a variant of 
	rejection sampling for point processes.
	
	\medskip
	
\begin{figure}
		\centering
		\includegraphics[width=0.9\textwidth]{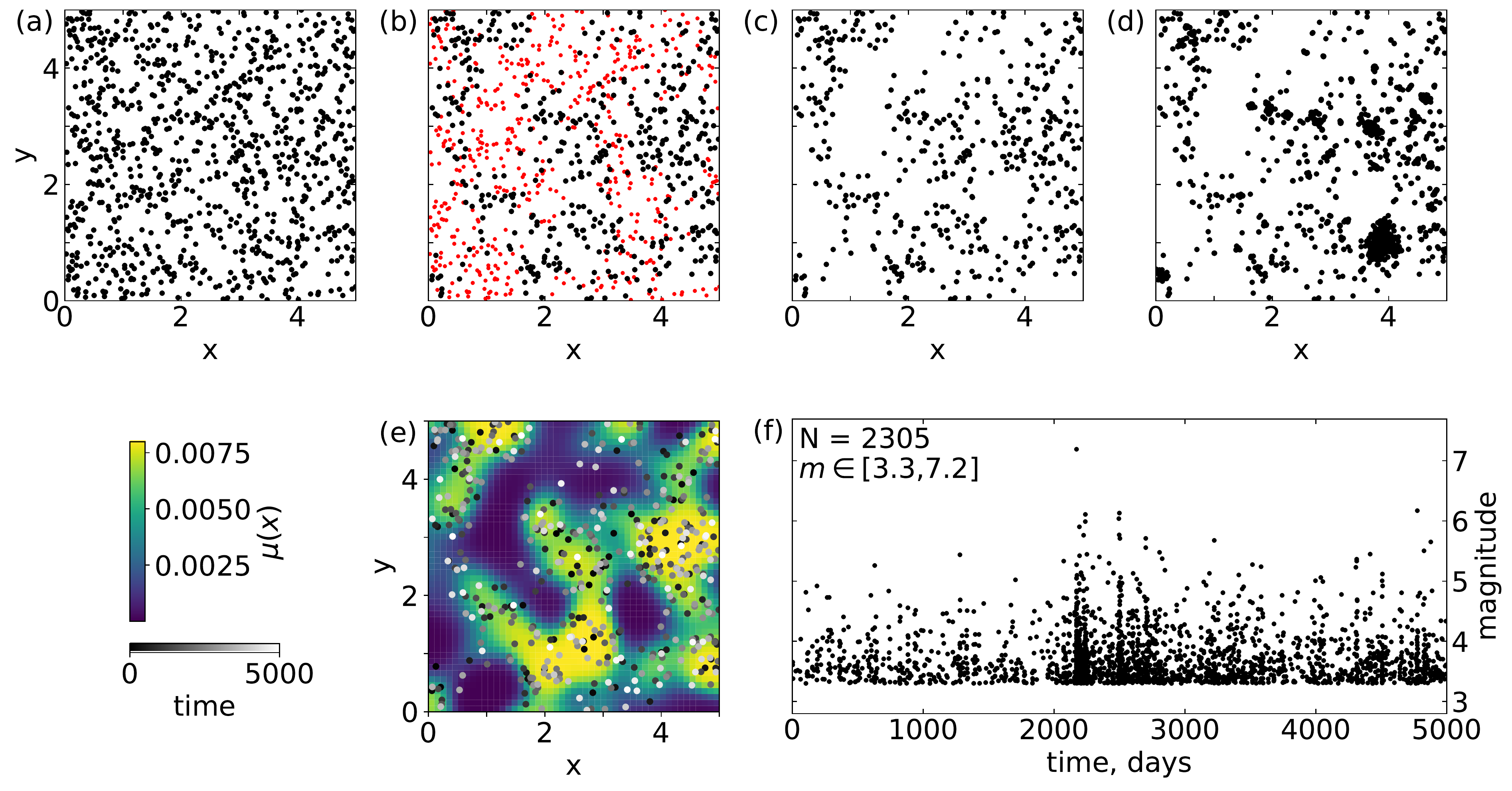}
		\caption{The Figure depicts the different steps of a forward simulation of the generative GP-ETAS model.
			(a) Events of a homogeneous Poisson process with intensity $\barlambda$ are generated
			($\barlambda=0.008, N=988$).
			(b) One retains events according to an inhomogeneous Poisson process with the desired 
			intensity $\mu(\x)=\barlambda\sigma(f(\x))$ by randomly
			deleting events (red dots) via \emph{thinning}.
			(c) The background events (black dots from (b)) are denoted by $\D_0$ ($N_{\D_0}=481$),
			(d) After adding aftershocks (offspring events) $\D_\varphi$ to $\D_0$ in accordance with the triggering function $\varphi(\cdot)$
			one obtains finally the simulated data $\D$ ($N_{\D}=2305$) of the spatio-temporal GP-ETAS. 
			(e) Shows the background intensity $\mu(\x)=\barlambda\sigma(f(\x))$ together with the generated background events. Gray scaling of the dots refers to the event times.
			(f) Depicts the simulated data as a synthetic earthquake catalogue in time.
		}
		\label{fig:sim_GP-ETAS}
	\end{figure}

	After choosing $\barlambda$, $\nus$, $\thetas_\varphi$ and a mark distribution $p(m)$, the simulation procedure of $\D \in \calX \times \calT \times \mathcal{M}$ can be summarised as follows: 
	\emph{First part:} 
	One uses the upper bound $\barlambda$ to generate 
	positions $\{\x_j\}_{j=1}^J$ of
	events from a homogeneous Poisson process with mean $|\calX||\calT|\barlambda$ which provide candidate background events 
	(Figure \ref{fig:sim_GP-ETAS} a). Subsequently 
	a Gaussian process $\fs$ is sampled from the prior $\mathcal{N}(\fs|\boldsymbol{0},\Ks_{\fs,\fs})$ 
	based on $\{\x_j\}_{j=1}^J$ using $\eqref{eq:cov_fun}$. The values $\mu(\x_j)$ can be computed using \eqref{eq:GP-ETAS}.
	Afterwards events, which do not follow an inhomogeneous Poisson process with intensity $\mu(\x)$ as given by \eqref{eq:GP-ETAS},
	are randomly deleted via \textit{thinning} (Figure \ref{fig:sim_GP-ETAS} b). 
	The remaining $N_{\D_0}$ events are background events (Figure \ref{fig:sim_GP-ETAS} c). The event times $\{t_i\}_{i=1}^{N_{\D_0}}$ are sampled from a uniform distribution $\mathcal{U}(|\calT|)$ and 
	the marks $\{m_i\}_{i=1}^{N_{\D_0}}$ from an exponential distribution, e.g., Gutenberg-Richter relation. Finally one obtains $\D_0$.
	\emph{Second part:} Given the background events $\D_0$, the aftershock events (offsprings) 
	of all generations are added to $\D_0$ in accordance with the triggering function $\varphi(\cdot)$ and using the mark distribution which yields 
	$\D$ (Figure \ref{fig:sim_GP-ETAS} d).
	
	\medskip

	The overall simulation algorithm is described in detail in the Appendix \ref{app:simGP-etas}, 
	and is visualised in Figure \ref{fig:sim_GP-ETAS}. 
	
	\nomenclature{$\D_\varphi$}{Subset of data $\D$ containing all offspring events, $\D \backslash \D_0$}

	%####################################################################################
	\section{Bayesian inference}
	\label{sec:Inference}
	%####################################################################################
	In this section, we address the Bayesian inference problem of our spatio-temporal GP-ETAS model.
	The objective is to estimate the random conditional intensity function \eqref{eq:lambda_tx} 
	in a Bayesian way including uncertainties, i.e.~the joint posterior $p(\mu,\thetas_\varphi| \D)$, where $\mu$ denotes the entire random field of the background intensity as in \eqref{eq:generative_model_GP-ETAS}.
	
	\medskip
	
	The likelihood of observing a point pattern 
	$\D=\{(t_i,\x_i,m_i)\}_{i=1}^{N_{\D}}$ 
	under the GP-ETAS model \eqref{eq:GP-ETAS} is given by the point process likelihood
	\begin{align}\label{eq:L}
	p(\D |\mu,\thetas_\varphi)
	&=
	\prod_{i=1}^{N_{\D}}\lambda(t_i,\x_i|\mu(\x_i),\thetas_\varphi)\exp
	\left(
	-\int_{\calT}\int_{\calX}
	\lambda(t,\x|\mu(\x),\thetas_\varphi)\diff \x  \diff t
	\right),
	\end{align}
	where the intensity $\lambda(\cdot)$ is given by~\eqref{eq:GP-ETAS}, 
	and the dependencies on $H_t$, $H_{t_i}$ are omitted for notational convenience.
	
	\medskip
	
	Assuming a joint prior distribution denoted here by 
	$p(\mu,\thetas_\varphi)$ for simplicity, the posterior distribution 
	becomes
	\begin{align}\label{eq:post}
	p(\mu,\thetas_\varphi | \D) \propto  
	p(\D | \mu,\thetas_\varphi)
	p(\mu,\thetas_\varphi).
	\end{align}
	This posterior is intractable in practice and hence standard inference techniques are not directly applicable.
	More precisely, the following three main challenges arise:
	\begin{enumerate}[(i)]
		\item The background intensity $\mu$ and triggering function $\varphi(\cdot|\thetas_\varphi)$ cannot be treated separately in the likelihood function (\ref{eq:L}).
		\item The likelihood (\ref{eq:L}) includes an intractable integral inside the exponential term.  Furthermore, 
		normalisation of \eqref{eq:post} requires an intractable marginalisation over $\mu$ and $\thetas_\varphi$.
		Thus, the posterior distribution is {\it doubly intractable} \citep{Murray2006}.
		\item We assume a Gaussian process prior for modelling the background rate. However, the point process likelihood (\ref{eq:L}) 
		is non-Gaussian, which makes the functional form of the posterior nontrivial to treat in practice.
	\end{enumerate}
	We approach these challenges by data augmentation based on the work of 
	\citet{Donner2018,Adams2009,Polson2013}. We will find that this augmentation simplifies the inference problem substantially.
	The following three auxiliary random variables are introduced:
	\begin{enumerate}[(1.)]
		\item A \emph{latent branching structure} $Z$\nomenclature{$Z$}{Branching structure}, as described in 
		Section \ref{para:branching}, decouples $\mu$
		and $\thetas_\varphi$ in the likelihood function \citep[e.g.,][]{Veen2008}.
		\item A \emph{latent Poisson process} $\Pi$\nomenclature{$\Pi$}{Latent Poisson process}
		enables an unbiased estimation of the  integral term in 
		the likelihood function that depends on $\mu$.
		\item We make use of the fact, that the logistic sigmoid function can be written as
		an infinite scale mixture of Gaussians 
		using latent \emph{P\'olya--gamma random variables} 
		$\omegas \sim \PG(\omega)$~\citep{Polson2013}\nomenclature{$\PG$}{P\'olya--gamma density}\nomenclature{$\omega$}{P\'olya--gamma variable}, defined in Appendix \ref{app:PG_def}. This leads to a likelihood 
		representation, which is conditional conjugate to all the priors 
		including the Gaussian process prior for the background 
		component of the likelihood function \citep{Donner2018}.
	\end{enumerate}
	\medskip
	These three augmentations allow one to implement a Gibbs sampling procedure \citep{Geman1984}  that produces 
	samples from the posterior distribution in~\eqref{eq:post}. More precisely, random samples are generated in a Gibbs sampler 
	by drawing one variable (or a block of variables) from the conditional posterior given all the other variables.
	Hence, we need to derive the required conditional posterior distributions as outlined next.
	\medskip
	
	The suggested sampler consists of three modules
	using the solutions (data augmentations) sketched above:
	\textit{sampling the latent branching structure}, \textit{inference of the background} $\mu$, and
	\textit{inference of the triggering} $\thetas_\varphi$.
	Our overall  \textit{Gibbs sampling algorithm}  of the posterior distribution 
	is summarised in Algorithm~\ref{algo:Gibbs_sampler}.
	After an initial burn-in
	(a sufficiently long run of the three modules (Section \ref{sec:Decoupling of the likelihood: latent branching structure} -- \ref{sec:Inference for the parameters of the triggering function}),
	the generated samples converge to 
	the desired joint posterior distribution 
	$p(\mu,\thetas_\varphi|\D)$.
	\medskip
	
	In the following, we discuss some important
	aspects of the three modules of the Gibbs sampler
	which the sampler runs repeatedly trough. 
	
	%------------------------------------------------------------------------------------------------------------------------
	\subsection{Sampling the latent branching structure} 
	\label{sec:Decoupling of the likelihood: latent branching structure}
	%------------------------------------------------------------------------------------------------------------------------
	\paragraph{Augmentation by the latent branching structure.}~
	We consider an auxiliary variable $z_i$ for each data point $i$, which represents 
	the latent branching structure as defined in Section \ref{para:branching}.
	Recall that it gives the time index of the parent event. If $z_i=0$ then the event is a spontaneous background event. 
	The likelihood 
	$p(\D,Z| \mu,\thetas_\varphi)$
	of the augmented model 
	can be written as
	\begin{equation}\label{eq:L_aug_1_Z}
	\begin{split}
	p(\D,Z| \mu,\thetas_\varphi) 
	=&
	\underbrace{\prod_{i=1}^{N_{\D}}
		\mu(\x_i)^{\ind(z_i=0)}
		\exp\left(-|\calT|\int_{\calX}\mu(\x)	\diff \x  \right)}_
	{\text{(a)=$p(\D_0 \vert Z,\mu)$}}
	\\ 
	& \times
	\underbrace{\prod_{i=1}^{N_{\D}}
		\prod_{j=1}^{i-1}
		\varphi_{ij}(\thetas_\varphi )
		^{\ind(z_i=j)}
		\exp\left(-\int_{\calT_i}\int_{\calX}
		\varphi_{i}(\thetas_\varphi)
		\diff \x  \diff t\right)}_
	{\text{(b)=$p(\D\vert  Z, \thetas_\varphi)$}}
	p(Z),
	\end{split}
	\end{equation}
	where
	$\ind(\cdot)$ denotes the indicator function, 
	i.e., $\ind(z_i=j)$ takes the value 1 for all $z_i=j$
	and 0 otherwise, 
	$\varphi_{ij}(\thetas_\varphi )=\varphi(t_i-t_j,\x_i-\x_j | m_j,\thetas_\varphi)$, 
	$\varphi_{i}(\thetas_\varphi )=\varphi(t-t_i,\x-\x_i | m_i,\thetas_\varphi)$,
	$\calT_i =[t_i,|\calT|] \subset \calT$, and
	all possible branching structures are equally likely, i.e.~$p(Z) = \mbox{const}$. Furthermore, $\D_0=\{\x_i\}_{i:z_i=0}$ denotes 
	the set of $N_{\D_0}$ background events. Note, that marginalizing over $Z$ in \eqref{eq:L_aug_1_Z} recovers
	\eqref{eq:L}, because $\sum_{z_i=0}^{i-1}\mu(\x_i)^{\ind(z_i=0)}\prod_{j=1}^{i-1}
	\varphi_{ij}(\thetas_\varphi )
	^{\ind(z_i=j)}=\lambda(t_i,\x_i\vert\mu(\x_i),\thetas_\phi)$. The augmented likelihood 
	factorises into two independent  components, 
	(a) a likelihood component for the 
	background intensity 
	which depends on 
	$\mu$
	(first two terms on the rhs of \eqref{eq:L_aug_1_Z})
	and 
	(b) a likelihood component of the triggering function
	which depends on 
	$\thetas_\varphi$
	(last two terms on the rhs of \eqref{eq:L_aug_1_Z}).
	\medskip
	
	From \eqref{eq:L_aug_1_Z} one can derive the conditional distribution 
	of $z_i$  given all the other variables. Note that all $z_i$'s are independent.
	The conditional distribution is proportional to a categorical distribution,
	\begin{equation}\label{eq:Zcond}
	p(z_i| \D,\mu(\x_i),\thetas_\varphi) 
	\propto
	\left[
	\mu(\x_i)
	\right]^{\ind(z_i=0)}
	\prod_{j=1}^{i-1}
	\left[
	\varphi_{ij}(\thetas_\varphi) 
	\right]^{\ind(z_i=j)}\\ 
	=
	\prod_{j=0}^{i-1}
	p_{ij}^{\ind(z_i=j)},
	\end{equation}
	with the probabilities $p_{ij}$ given by \eqref{eq:p_i_BG} and \eqref{eq:p_ij_offspring} and which
	we collect in a vector $\ps_i\in \mathbb{R}^i$. 
	
	\medskip
	
	From \eqref{eq:Zcond} one can see that the latent branching structure 
	at the $k$th iteration of the Gibbs sampler 
	is sampled from a categorical distribution,
	\begin{align}\label{eq:cond_post_Z}
	&\forall i=1,...,N_{\D} &z_i^{(k)}|\D,(\mu(\x_i),\thetas_\varphi) ^{(k-1)}
	\sim 
	\mathrm{Categorical}(\ps_i).
	\end{align}
	Here $(\mu(\x_i),\thetas_\varphi)^{(k-1)}$ denotes the values of $\mu(\x_i)$ and $\thetas_\varphi$ from
	the previous iteration.

	%------------------------------------------------------------------------------------------------------------------------
	\subsection{Inference for the background intensity} 
	\label{sec:Inference for the background intensity}
	%------------------------------------------------------------------------------------------------------------------------
	Given an instance of a branching structure $Z$, 
	the background intensity  in~\eqref{eq:L_aug_1_Z} depends on events $i$ for which $z_i=0$ only. 
	One finds that the resulting term is a Poisson likelihood of the form
	\begin{align}\label{eq:bg likelihood}
	p(\D_0\vert f,\barlambda, Z) 
	=
	\prod_{i=1:z_i=0}^{N_\D}\barlambda\sigma(f_i)
	\exp\left(-|\calT|\int_{\calX}\barlambda\sigma(f(\bx))d\bx\right),
	\end{align}
	where $\mu(\x)$ has been replaced by \eqref{eq:GP-ETAS}  and $f_i=f(\x_i)$ has been used for notational
	convenience.
	\medskip
	
	Because of the aforementioned problems in Section \ref{sec:Inference}, sampling the conditional posterior 
	$p(f,\bar\lambda|\D_0, Z)$ is still non trivial and require further augmentations which we describe next.
	%\medskip
	
	%%%%%%%%%%%%%%%%%%%%%%%%%%%%%%%%%%%%%%%%%%%%%

	\paragraph{Augmentation by a latent Poisson process.}~
	We can resolve issue (ii) from Section \ref{sec:Inference} by introducing an independent 
	latent Poisson process $\Pi=\{\x_l\}_{l=N_\D + 1}^{N_{\D\cup\Pi}}$
	on the data space with rate 
	$\hat\lambda(\x)=\barlambda(1-\sigma(f(\x)))=\barlambda(\sigma(-f(\x)))$ 
	using $1-\sigma(z)=\sigma(-z)$. The points in $\D$, $\Pi$ form the joint set $\D\cup\Pi$ 
	with cardinality $N_{\D\cup\Pi}$. Note, that the number of elements in $\Pi$, .i.e. $N_{\Pi}$, is also a random variable. 
	The joint likelihood of $\D_0$ and the new random variable $\Pi$ is,
	\begin{equation}\label{eq:L_aug2_latentPoisson}
	p(\D_0,\Pi\vert f, \barlambda, Z) 
	=
	\prod_{i=1:z_i=0}^{N_\D}
	\barlambda\sigma(f_i)
	\prod_{l=N_\D + 1}^{N_{\D\cup\Pi}}
	\bar \lambda\sigma(-f_l)\exp\left(-|\calX||\calT|\bar\lambda\right),
	\end{equation}
	where $f_l=f(\x_l)$.
	Thus, by introducing the latent Poisson process $\Pi$, we obtain 
	a likelihood representation of the augmented system, where the former intractable integral 
	inside the exponential term disappears, i.e.~reduces to a constant.
	\nomenclature{$N_{\D\cup\Pi}$}{Number of elements/events in set $\D\cup\Pi$}
	\nomenclature{$N_{\D_0}$}{Number of elements/events in set $\D_0$, background events}
	
	\medskip
	
	Let us provide further intuition for this augmentation. Due to the fact that $\barlambda\sigma(f(\x))$ is bounded, 
	one can simply use the superposition property of Poisson processes
	to construct the latent $\Pi$. The basic idea is to introduce an independent $\Pi$ 
	with bounded intensity $\hat\lambda(\x)$ such that its superposition
	with the inhomogeneous Poisson process of the background 
	results in a homogeneous Poisson process with intensity $\bar\lambda$.
	Hence, 
	$\hat\lambda(\x)+\bar\lambda\sigma(f(\x))=\bar\lambda$, 
	and one obtains
	$\hat\lambda(\x)=\barlambda(1-\sigma(f(\x)))=\barlambda\sigma(-f(\x))$
	for the intensity of the latent Poisson process. 
	Writing the joint likelihood for events 
	in $i$ with $z_i=0$ and events from the latent $\Pi$ we get \eqref{eq:L_aug2_latentPoisson}.
	\medskip
	
	More rigorously one can derive the latent Poisson process $\Pi$ following \cite{Donner2018}. 
	Note that \eqref{eq:bg likelihood} implies
	\begin{equation}
	\exp\left(-|\calT|\int_{\calX}\barlambda\sigma(f(\x))\diff\bx\right)
	=  \exp\left(\int_{\calT}\int_{\calX}\barlambda(\sigma(-f(\x))-1)\diff\bx\diff t\right) = \EE{\barlambda}{
		\prod_{\x_l\in \Pi}
		\sigma(-f(\bx_l))},
	\end{equation} 
	where the expectation is over random sets $\Pi$ with respect to a Poisson process measure with rate $\barlambda$ 
	on the space-time window of the data $\calT \times \calX$.
	Here, one uses Campbell's theorem~\citep{Kingman2005}.
	Writing the likelihood parts depending on $f$ and $\barlambda$  in~\eqref{eq:L_aug_1_Z} in terms  of the new random variable $\Pi$ we get \eqref{eq:L_aug2_latentPoisson}. Note that marginalisation over the augmented variable $\Pi$ leads  back to 
	the background likelihood in \eqref{eq:bg likelihood} conditioned on the branching structure $Z$.
	
	%%%%%%%%%%%%%%%%%%%%%%%%%%%%%%%%%%%%%%%%%%%%%%%
	
	\paragraph{Augmentation by P\'olya--gamma random variables.}~
	In order to resolve issue (iii) from Section \ref{sec:Inference} we substitute the sigmoid function by 
	an infinite scale mixture of Gaussians 
	using latent \emph{P\'olya--gamma random variables} ~\citep{Polson2013}, that is,
	\begin{align}
	\label{eq:PG sigmoid}
	\sigma(z) 
	&= \frac{e^\frac{z}{2}}{2\cosh(\frac{z}{2})}
	= \frac{1}{2}e^\frac{z}{2}\int_{0}^\infty e^{-\frac{z^2}{2}\omega}\PG(\omega\vert 1,0) \diff\omega 
	,
	\end{align}
	where the new random variable $\omega$ is distributed according to the P\'olya--gamma density $\PG(\omega\vert 1,0)$, see Appendix~\ref{app:PG_def}.
	Inserting the P\'olya--gamma representation of the sigmoid function
	\eqref{eq:PG sigmoid} into~\eqref{eq:L_aug2_latentPoisson} yields
	\begin{equation}\label{eq:aug2}
	\begin{split}
	p(\D_0,\omegas_{\D},\Pi,\omegas_{\Pi} \vert \fs, \barlambda, Z)
	= & 
	\prod_{\begin{subarray}{l}
		i:z_i=0
		\end{subarray}}^{N_\D}
	\frac{\barlambda}{2} e^{\frac{f_i}{2}-\frac{f_i^2}{2}\omega_i}\PG(\omega_i\vert 1,0)\\
	& \times 
	\prod_{l=N_\D+1}^{N_{\D\cup\Pi}}
	\frac{\barlambda}{2} e^{-\frac{f_l}{2}-\frac{f_l^2}{2}\omega_l}
	\PG(\omega_l\vert 1,0)
	\exp\left(-\barlambda|\calX|T\right),
	\end{split}
	\end{equation}
	where we set the P\'olya-gamma variables of all events 
	$\omegas_{\D}=(\omega_1,\ldots,\omega_{N_\D})$\nomenclature{$\omegas_{\D}$}{P\'olya--gamma variables at data points $\D$} to 
	$\omega_i=0$ if $z_i\neq0$. For the latent Poisson process the P\'olya-gamma variables are
	denoted by $\omegas_{\Pi}=(\omega_{N_\D+1},\ldots,\omega_{N_{\D\cup\Pi}})$\nomenclature{$\omegas_{\Pi}$}{P\'olya--gamma variables at points of latent process $\Pi$}.
	The likelihood representation of the augmented system 
	~\eqref{eq:aug2} has a Gaussian form
	with respect to $\fs$ (that is, only linear or quadratic terms of $\fs$ appear in the exponential function)
	and is therefore conditionally 
	conjugate to the GP prior denoted by $p(\fs)$.
	Hence, we can implement an efficient Gibbs sampler for the background intensity function. 
	\medskip
	
	Employing a Gaussian process prior over $\fs$ and 
	a Gamma distributed prior over $\barlambda$, one gets 
	from \eqref{eq:aug2} the following conditional posteriors for the $k$th Gibbs iteration:
	\begin{subequations}\label{eq:cond_post_BG}
		\begin{align}
		\label{eq:cond_post_Pi_latent}
		\Pi^{(k)} \ | \ (\barlambda,\fs)^{(k-1)}
		&\sim
		\mathrm{PP}(\barlambda(\sigma(-f(\x)))
		\\
		\label{eq:cond_post_Omega_Pi}
		\forall \ l 
		:N_ \D+1,\ldots,N_{\D\cup\Pi}
		\ \ \ \ \ \ \
		\omega_l^{(k)}\  |  \ f_l^{(k-1)},\Pi^{(k)}
		&\sim 
		\PG(1, |f_l|)
		\\
		\label{eq:cond_post_Omega_S0}
		\forall \ i :z_i=0 \ \ \ \ \ \ \
		\omega_i^{(k)}\  |  \ f_i^{(k-1)},\D,Z^{(k)}
		&\sim 
		\PG(1, |f_i|)
		\\
		\label{eq:cond_post_upper_bound}
		\barlambda^{(k)} \  |  \ Z^{(k)},\Pi^{(k)}
		&\sim 
		\mathrm{Gamma}(N_{\D_0\cup\Pi}+\alpha_0,|\calX||\calT|+\beta_0)
		\\
		\label{eq:cond_post_GP}
		\fs^{(k)} \  |  \ \D, (\omegas_{\D},\Pi,\omegas_{\Pi},Z)^{(k)}
		&\sim 
		\mathcal{N}((\Omegas+\Ks^{-1})^{-1}\us,(\Omegas+\Ks^{-1})^{-1})
		\end{align}
	\end{subequations}
	where 
	$\fs=(\fs_{\D},\fs_\Pi) \in \R^{N_{\D\cup\Pi}}$ 
	is the Gaussian process 
	at the data locations $\D$ and $\Pi$;
	and PP($\cdot$) denotes an inhomogeneous Poisson process with intensity
	$\barlambda(\sigma(-f(\x)))$;
	$\Omegas$ 
	is a diagonal matrix with 
	$(\omegas_{\D},\omegas_{\Pi})$ 
	as diagonal entries. 
	$\Ks \in \R^{N_{\D\cup\Pi} \times N_{\D\cup\Pi}}$ 
	is the covariance matrix of the Gaussian process prior
	at positions $\D$ and $\Pi^{(k)}$. 
	It can be shown that, the vector $\bu$ is $1/2$ for all entries in $\D_0$, zero for all entries of the remaining data $\D\backslash\D_0$, and $-1/2$ for the corresponding entries of
	$\Pi$. Gamma$(\cdot)$ is a Gamma distribution, 
	where the Gamma prior has shape and rate  parameters $\alpha_0,\beta_0$.
	We used $e^{-\frac{c^2}{2}\omega} \PG(\omega\vert 1,0)\propto \PG(\omega\vert 1,c)$ 
	due to the definition of a tilted P\'olya--gamma density 
	\eqref{eq:tilted_PG}
	as given in \citep{Polson2013}, see Appendix \ref{app:PG_def}.
	Note that one does not need an explicit form of the  P\'olya--gamma density for our inference approach since it is sampling based.
	In other words, we only need an efficient way to sample from the tilted $\PG$ density 
	\eqref{eq:tilted_PG} which was provided by \citet{windle2014sampling,Polson2013}. 
	Several $\PG$ samplers are freely available for different computer languages. 
	\nomenclature{$\Omegas$}{Diagonal matrix with $(\omegas_{\D},\omegas_{\Pi})$ as diagonal entries}
	%\nomenclature{PP}{Inhomogeneous Poisson process}
	\nomenclature{$\alpha_0$}{Shape parameter of the Gamma prior of $\bar{\lambda}$}
	\nomenclature{$\beta_0$}{Rate parameter of the Gamma prior of $\bar{\lambda}$}
	\nomenclature{$\mathcal{N}$}{Normal distribution}
	
	\medskip
	
	A detailed step-by-step derivation of the conditional distributions 
	is given in the Appendix \ref{appendix:BG_gibbs_sampler_in_detail}.

	\paragraph{Hyperparameters}
	The Gaussian process covariance kernel
	given in \eqref{eq:cov_fun} depends on the 
	hyperparameters $\nus$. Compare Section \ref{sec:GP-ETAS}.
	We use exponentially distributed priors on
	$p(\nu_i)=p_{\nu_i}$, and 
	we sample $\nus$ using a standard MH algorithm as there is no closed 
	form for the conditional posterior available. The only terms where $\nus$ enter are in the 
	Gaussian process prior and hence the relevant 
	terms are
	\begin{align}
	\label{eq:sample_hypers}
	\ln p(\nus|\fs,\D,\Pi,\omegas_{\D},\omegas_{\Pi})
	= 
	- \frac{1}{2}\fs^\top \Ks_{\bnu}^{-1}\fs- \frac{1}{2}\ln \det \Ks_{\bnu} + \ln p(\nus) +{\rm const.},
	\end{align}
	where $\Ks_{\nus}$ is the Gaussian process prior covariance matrix depending on $\nus$ via $\eqref{eq:cov_fun}$.
	
	%%%%%%%%%%%%%%%%%%%%%%%%%%%%%%%%%%%%%%%%%%%%%%%%%%%%%%%%
	
	\subsubsection{Conditional predictive posterior distribution of the background intensity}\label{sec:cond_pred_post_mux}
	Given the $k$th posterior sample $(\barlambda^{(k)},\fs^{(k)},\nus^{(k)})$,
	the background intensity $\mu(\x^*)^{(k)}$ at any set of positions 
	$\{\x_i^*\} \in \calX$ (predictive conditional posterior) can be obtained 
	in the following way, see \eqref{eq:generative_model_GP-ETAS}. Conditioned on $\fs^{(k)}$ and hyperparameters $\nus^{(k)}$ 
	the latent function values $\fs^*$  can be sampled via the conditional prior $p(\fs^*|\fs^{(k)},\nus^{(k)})$
	using \eqref{eq:predictive GP} with covariance function given in \eqref{eq:cov_fun} \citep{Williams2006gaussian}.
	Using \eqref{eq:GP-ETAS} 
	one gets $\mu(\x^*)^{(k)}=\barlambda^{(k)} \sigma(\fs^*)$.
	
	%------------------------------------------------------------------------------------------------------------------------
	\subsection{Inference for the parameters of the triggering function} 
	\label{sec:Inference for the parameters of the triggering function}
	%------------------------------------------------------------------------------------------------------------------------
	Given an instance of a branching structure $Z$,
	the likelihood function in \eqref{eq:L_aug_1_Z}
	factorises in terms involving $\mu$ and terms involving $\thetas_\varphi$. The relevant terms related to $\thetas_\varphi$ are
	\begin{equation}\label{eq:DgivenZthetas}
	\begin{split}
	p(  \D \vert Z ,\thetas_\varphi)
	= &
	\prod_{i=1:z_i\neq0}^{N_\D}
	\varphi(t_i-t_{z_i},\x_i-\x_{z_i}|m_{z_i},\thetas_\varphi)\\
	& \times \prod_{i=1}^{N_{\D}}
	\exp\left(-\int_{\calT_i}\int_{\calX}
	\varphi(t-t_i,\x-\x_i | m_i,\thetas_\varphi) 
	\diff \x  \diff t\right).
	\end{split}
	\end{equation}
	The conditional posterior 
	$p(\thetas_\varphi \vert \D,Z)\propto  p(  \D \vert  Z, \thetas_\varphi)p(\thetas_\varphi)$ 
	with prior $p(\thetas_\varphi)$ has no closed form. The dimension of $\thetas_\varphi$  is usually small ($\le 7$).
	We employ MH sampling \citep{Hastings1970},
	which can be considered a nested step within the overall Gibbs sampler.
	We use a random walk MH where proposals are generated by a Gaussian
	in log space. The acceptance probability of $\thetas_\varphi^{(k)}$ based on \eqref{eq:DgivenZthetas} is given by
	\begin{equation}\label{eq:acc_triggering}
	p_{\rm accept} = 
	\min \left\lbrace 1, 
	\frac
	{p(\D \vert Z^{(k)},\thetas_\varphi^{\rm proposed})p(\thetas_\varphi^{\rm proposed})}
	{p(\D \vert Z^{(k)},\thetas_\varphi^{(k-1)})p(\thetas_\varphi^{(k-1)})}\right\rbrace.
	\end{equation}
	We take 10 proposals before we return to the overall Gibbs sampler, that is, to step in Section~\ref{sec:Decoupling of the likelihood: latent branching structure}.
	\medskip
	
	\begin{algorithm}
		\caption{Gibbs Sampler for the posterior distribution of  spatio-temporal GP-ETAS}
		\label{algo:Gibbs_sampler}
		\small
		\begin{algorithmic}[1]
			\State Initialise randomly $\bar\lambda^{(0)},\fs^{(0)},\thetas_\varphi^{(0)}$ from the priors
			\For{$k=1$ to $K$}
			\State \textbf{Factorisation of the likelihood (Section \ref{sec:Decoupling of the likelihood: latent branching structure})} 
			\State \hspace{\algorithmicindent} $\forall i=1,...,N_{\D}$ Sample latent branching structure 
			$z_i^{(k)}|\D,(\mu(\x_i),\thetas_\varphi)^{(k-1)}$\hspace{1em}\eqref{eq:cond_post_Z}
			\State \textbf{Inference of the background intensity 
				$p(\barlambda,\fs|\D_0,\omegas_{\D},\Pi,\omegas_{\Pi},Z)$ (Section \ref{sec:Inference for the background intensity})
			}
			\State \hspace{\algorithmicindent} Sample latent Poisson process 
			$\Pi^{(k)}|(\barlambda,\fs)^{(k-1)}$\hspace{1em}\eqref{eq:cond_post_Pi_latent}
			\State \hspace{\algorithmicindent} 
			$\forall l=N_\D+1,...,N_{\D\cup\Pi}$ Sample P\'olya--gamma variables
			$\omega_{l}^{(k)}|f_l^{(k-1)},\Pi^{(k)}$\hspace{1em}\eqref{eq:cond_post_Omega_Pi}
			\State \hspace{\algorithmicindent} 
			$\forall i: z_i=0 $ Sample P\'olya--gamma variables
			$\omega_{i}^{(k)}|\D,f_i^{(k-1)},Z^{(k)}$ \hspace{1em} \eqref{eq:cond_post_Omega_S0}
			\State \hspace{\algorithmicindent} 
			Set $\omega_i^{(k)} = 0$ otherwise. 		
			\State \hspace{\algorithmicindent} Sample upper bound 
			$\barlambda^{(k)}|(Z,\Pi)^{(k)}$\hspace{1em}\eqref{eq:cond_post_upper_bound}
			\State \hspace{\algorithmicindent} 
			Sample Gaussian process
			$\fs^{(k)}|\D,(\omegas_{\D},\Pi,\omegas_{\Pi},Z)^{(k)}$\hspace{1em}\eqref{eq:cond_post_GP}
			\State  \hspace{\algorithmicindent} Sample hyperparameters $\nus^{(k)}$ 
			using MH step\hspace{1em}\eqref{eq:sample_hypers}
			\State  \hspace{\algorithmicindent} $\forall i=1,...,N_{\D}$ compute $(\mu(\x_i))^{(k)}$\hspace{1em}\eqref{eq:GP-ETAS} 
			\State \textbf{Inference of the triggering function $p(\thetas_\varphi|\D,Z^{(k)})$ 
				(Section \ref{sec:Inference for the parameters of the triggering function})}
			\State  \hspace{\algorithmicindent} Sample $\thetas_\varphi^{(k)}$ using 
			MH steps\hspace{1em}\eqref{eq:acc_triggering}
			\EndFor
		\end{algorithmic}
	\end{algorithm}

	%####################################################################################
	\section{Experiments and results}
	\label{sec:Results}
	%####################################################################################
	%------------------------------------------------------------------------------------------------------------------------
	We consider two kinds of experiments where
	we evaluate the performance of our proposed Bayesian approach GP-ETAS 
	(see Section \ref{sec:GP-ETAS} and \ref{sec:Inference}).
	First we look at synthetic data, with known conditional 
	intensity $\lambda(t,\x)$, i.e. with known background intensity 
	$\mu(\x)$ and known parameters $\thetas_\varphi$ of the triggering function.
	Here, we investigate if GP-ETAS
	can recover the model underlying the data well.
	Secondly, we apply our method to observational earthquake data.
	
	\medskip
	
	\textbf{Comparison:} We compare our approach
	with the current standard spatio-temporal ETAS model which uses MLE.
	This classical ETAS model is based on kernel density estimation 
	with variable bandwidths for the background intensity $\mu(\x)$ as 
	described in Section \ref{sec:ingredientsEtas}.
	Two variations are considered:
	(1) ETAS model with standard choice of the minimal bandwidth (0.05 degrees) and $n_p=15$ the
	number of nearest neighbors used for obtaining the individual bandwidths 
	(ETAS--classical; \cite{Zhuang2002}), 
	and 
	(2) ETAS model with a minimal bandwidth given by Silverman's rule \citep{silverman1986density} and $n_p=15$ (ETAS--Silverman).
	
	\medskip
	
	\textbf{Evaluation metrics:} 
	Two metrics are used to evaluate the performances.
	The first metric is the test likelihood, which evaluates the likelihood \eqref{eq:L} 
	for a data test set  $\D^*$\nomenclature{$\D^*$}{Test set} (unseen data during the inference)
	given the inferred model on training data $\D$, 
	which is $p(\D^*|\D)=\EE{p(\mu,\thetas_\varphi|\D)}{p(\D^*|\mu,\thetas_\varphi)}$, where the expectation is over the inferred model posterior.
	The test likelihood reflects the predictive power of the different modelling approaches. 
	In the case of GP-ETAS we obtain $K$ posterior samples
	$\{(\mu^k,\thetas^k_{\varphi})\}_{k=1}^K$
	and we evaluate the 
	log expected test likelihood, 
	$\ell_{\rm test} = \ln p(\D^*|\D) \approx \ln \frac{1}{K}\sum_{k=1}^K  
	p(\D^*|\mu^{(k)},\thetas^{(k)})$.
	In the case of ETAS--classical and ETAS--Silverman we use the MLE point estimate for evaluating $\ell_{\rm test}$.
	The involved spatial integral in \eqref{eq:L} is approximated by Riemann sums
	on a $50\times50$ point grid.
	The second metric is the $\ell_2$ norm between true background intensity $\mu$ and the predicted $\hat\mu$,
	$\ell_{2} = \sqrt{\int_{\calX}(\mu(\x)-\hat\mu(\x))^2\diff \x}$.
	This is only possible for the experiments with synthetic data.
	\nomenclature{$K$}{Number of simulated posterior samples}
	\nomenclature{$\ell_{\rm test}$}{Log expected test likelihood}
	\nomenclature{$\ell_{2}$}{$\ell_2$ norm}
	%------------------------------------------------------------------------------------------------------------------------
	\subsection{Synthetic data}
\begin{figure}
		\centering
		\small
		\includegraphics[width=0.44\linewidth]{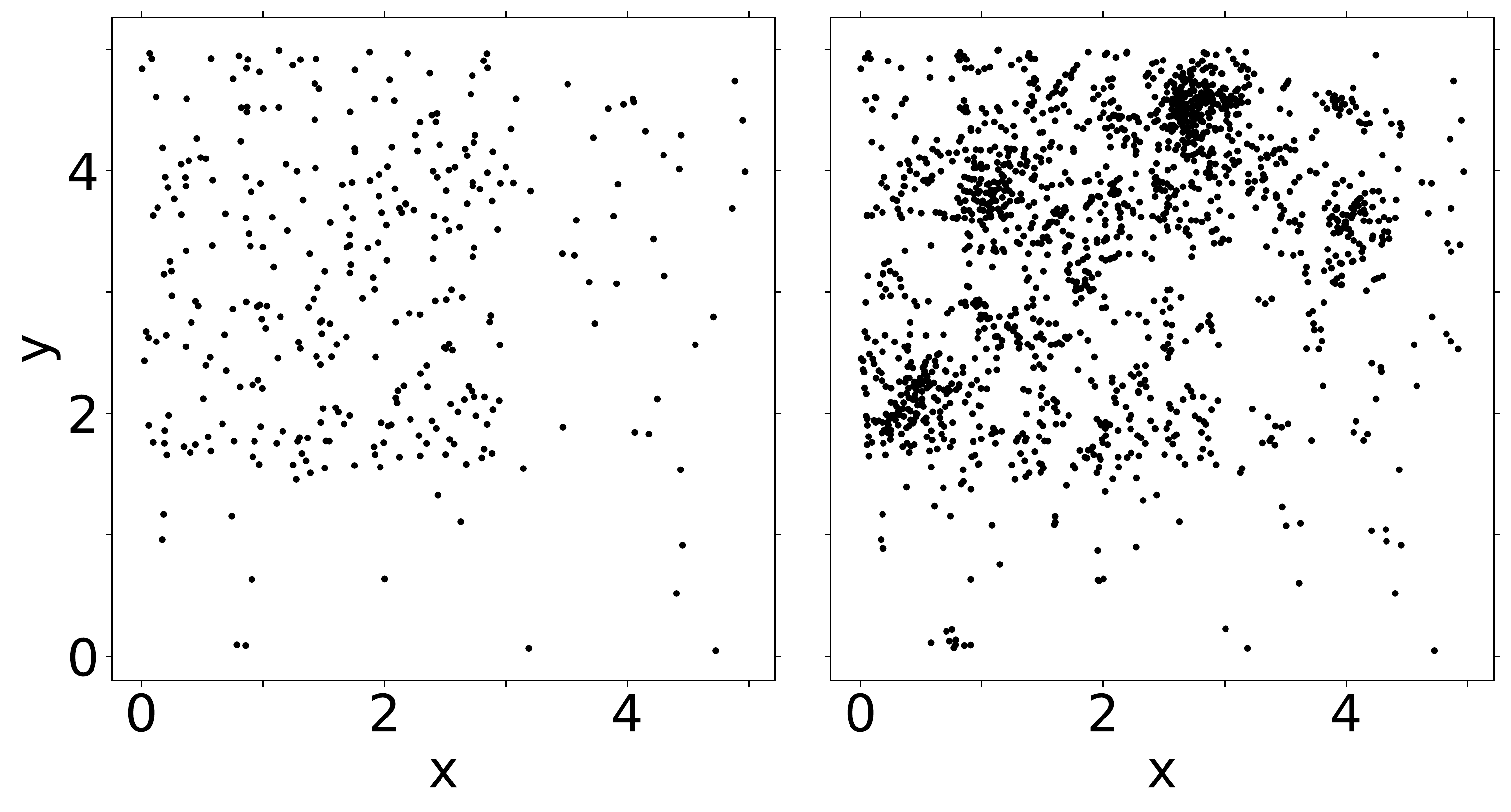}
		\includegraphics[width=0.485\linewidth]{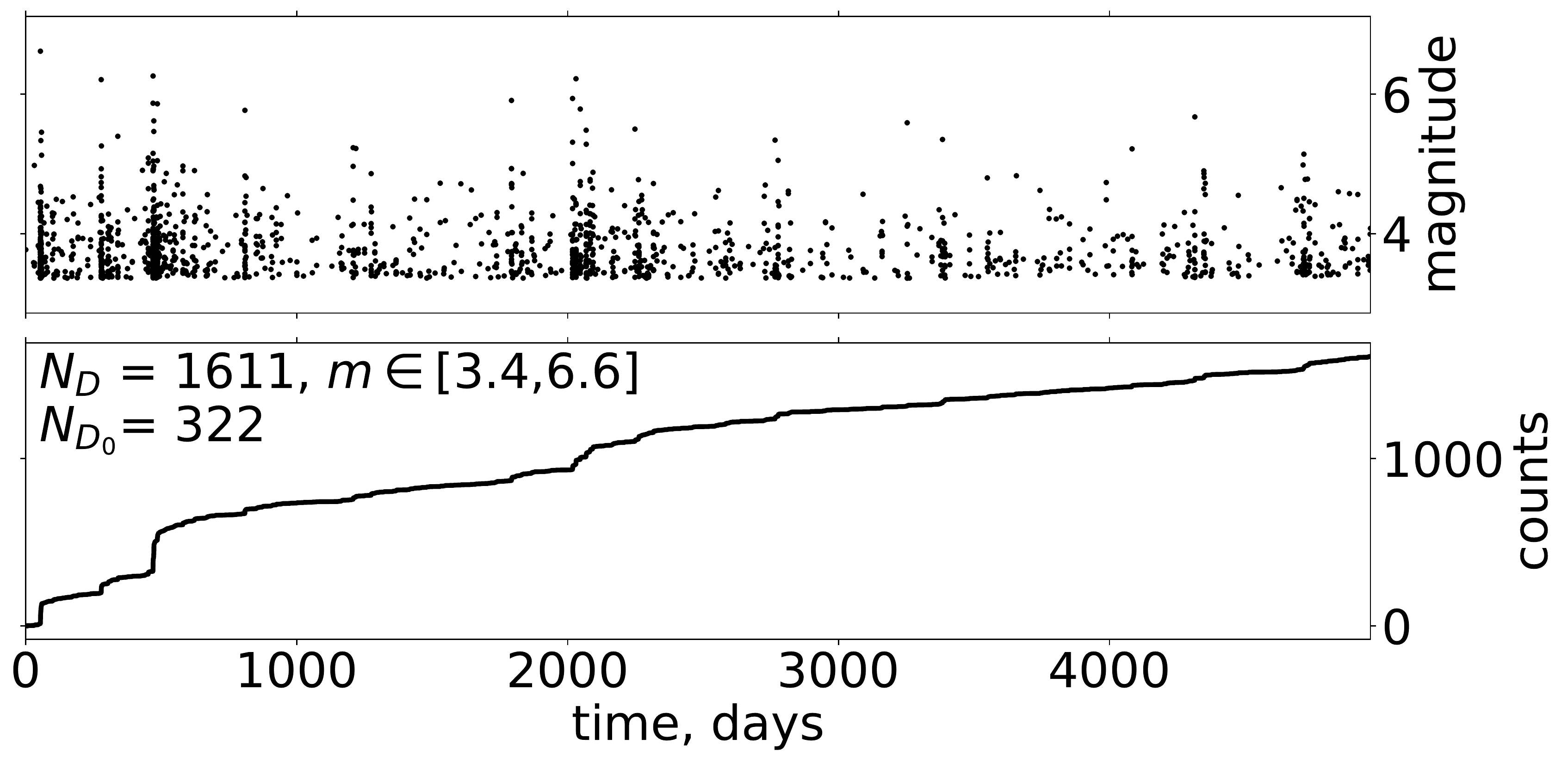}\\
		\includegraphics[width=0.44\linewidth]{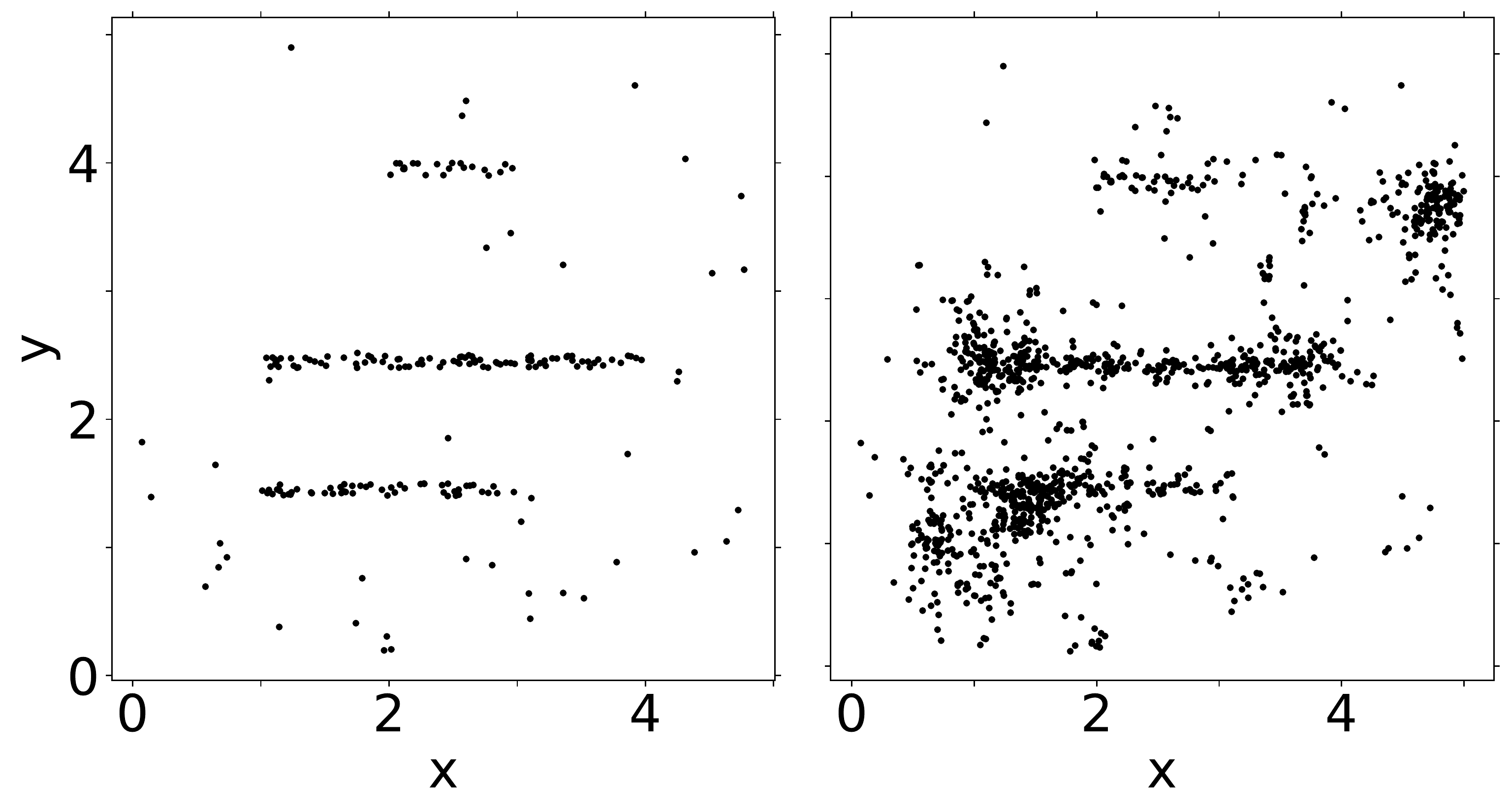}
		\includegraphics[width=0.485\linewidth]{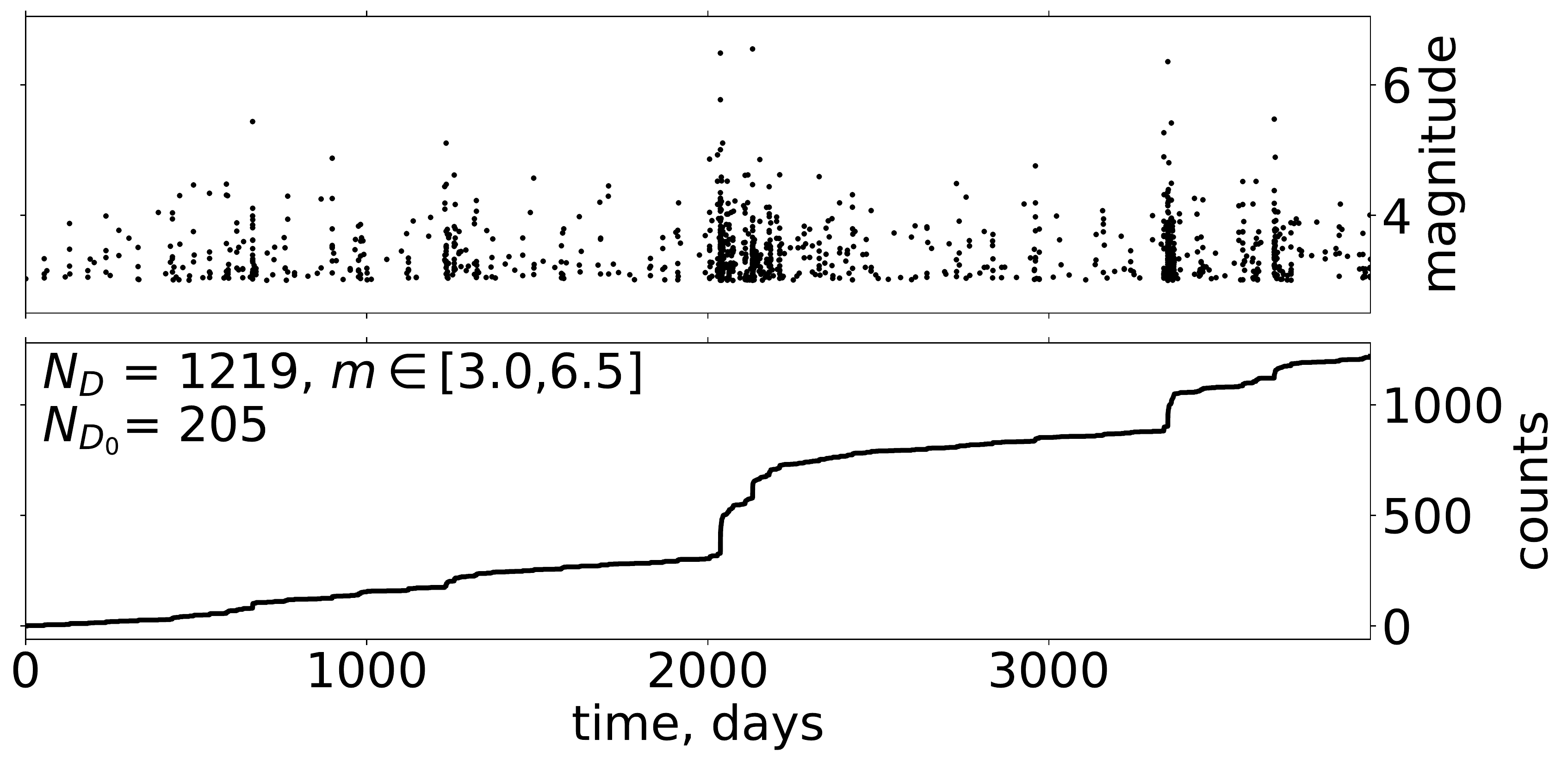}\\
		\caption{Setup of synthetic data experiments, Case 1 (first row) and Case 2 (second row): From left to right:
			background events, data set including background and offspring events, visualisation of the data as earthquake sequence over time.}
		\label{fig:fig002setting_A1}
	\end{figure}

	%------------------------------------------------------------------------------------------------------------------------
	\paragraph{General experimental setups:}
	We simulate synthetic data
	from two different conditional intensity functions which differ in $\mu(\x)$ on a spatio-temporal domain 
	$\calX\times\calT=[0,5]\times[0,5]\times[0,5000]$.
	In the first case we consider $\mu_1(\x)$ to be constant 
	over large spatial regions, e.g. large area sources (\textit{Case 1}, Figure \ref{fig:fig002setting_A1} 
	first row). In a second experiment
	we consider another particular setting 
	where $\mu_2(\x)$ is concentrated mainly on small fault-type areas
	(\textit{Case 2}, Figure \ref{fig:fig002setting_A1} second row).
	These two settings (area sources and faults) are important, typical limiting cases in 
	analysing seismicity pattern, both used in seismic hazard assessment.
	The two chosen intensity functions are,
	\begin{align}
	\mu_1(\x) &= 
	\begin{cases} 
	0.005 & \x  \in [0,3]\times[1.5,5] \\
	0.001 & \x   \in [3,5]\times[1.5,5] \\
	0.0005 & \x \in [0,5]\times[0,1.5]
	\end{cases}\\
	\mu_2(\x) &= 
	\begin{cases} 
	0.07035&  \x  \in [1,3]\times[1.4,1.5]  \\
	0.07035 &  \x  \in [1,4]\times[2.4,2.5] \\
	0.03535 &  \x  \in [2,3]\times[3.9,4]  \\
	0.00035 & \mathrm{else}
	\end{cases}
	\end{align}
	The triggering function is given in (\ref{eq:var_phi}--\ref{eq:gt}) with spatial kernel \eqref{eq:s_x_pwl}
	in both cases. 
	The magnitudes are simulated following an exponential distribution 
	$p_M(m_i) = \frac{1}{\beta} e^{-\beta(m_i-m_0)}$ 
	with $\beta=\ln(10)$ which corresponds to a 
	Gutenberg-Richter relation with b-value of 1; $m_0=3.36$ in the first case and $m_0=3$ in the second case.
	The test likelihood is computed for twelve unseen data sets, 
	simulated from the generative model and averaged.
	The simulations are done on the same spatial domain $\calX_{\rm sim}=[0,5]\times[0,5]$.
	The time window is $\calT_{\rm sim} = [0,1500]$ and $\ell_{\rm test}$ is evaluated 
	using events with event times $t_i \in [500,1500]$, 
	all previous events are taken into account in the history $H_t$.
	
	\paragraph{GP-ETAS setup:}  In GP-ETAS we need to set priors, and parameters of the Gibbs sampler,
	these are given in Table \ref{tab:GP-ETAS_setup}, see Section \eqref{sec:GP-ETAS}. 
	
	\begin{table}[]
		\centering
		\tiny
		\caption{GP-ETAS setup, prior choice.}
		\begin{tabular}{llc}
			\hline
			variable, symbol                            & prior                              & values               \\
			\hline
			latent function  $f$                       &Gaussian process prior & zero mean function; cov function \eqref{eq:cov_fun} \\
			upper bound, $\barlambda$          &  ${\rm Gamma}(\alpha_0,\beta_0)$               & $\alpha_0=1/c_s^2 , \beta_0=1/c_s^2/ \mu_{\bar\lambda}$   \\
			&												& with $c_s=1$,  $\mu_{\bar\lambda} = 2N_\D|\calX|$\\
			hyper parameters of cov function,$\nus$ 	 & exponential distribution       & $\beta_{\nu_0}=1/5$, $\beta_{\nu_1}=\beta_{\nu_2}=5/2$    \\
			parameters of the triggering function, $\thetas_\varphi$ & uniform distribution            &                  								  $K_0\in(0,10)$, $c\in(0,10)$,     \\
			&                                             & $p\in(0,10)$, $\alpha\in(0,10)$,          \\
			&											  & $d\in(0,10)$, $\gamma\in(0,10)$, \\
			&                                             & $q\in(1,10)$                  \\
			\hline
			number of posterior samples, K     &                                    & 5000                \\
			burn-in  (number of discarded initial iteration)                  &                                    & 2000                \\
			MH proposal distribution for $\nus$: Gaussian             &  & $\sigma_p=0.05$ in log units \\
			MH proposal distribution for $\thetas_\varphi$: Gaussian          &  & $\sigma_p=0.01$ in log units         \\
			number of MH steps per iteration for $\thetas_\varphi$   &  & 10   \\
			\hline
		\end{tabular}\label{tab:GP-ETAS_setup}
	\end{table}
	\nomenclature{$\sigma_p$}{Standard deviation of the proposal distribution for MH steps}

	\begin{table}
		\centering
		\small
		\caption{Case1: Parameter values $\thetas_\varphi$ of the triggering function.}
		\begin{tabular}{lcccccccc}
			\hline
			model & quantiles & $K_0$ & $c$ & $p$ & $\alpha$ & $d$ & $\gamma$ & $q$  \\ 
			\hline
			&  &  &  &  &  &  &  & \\ 
			generative  &  & 0.018 & 0.006 & 1.20 & 1.690 & 0.015  & 0.20   & 2.00 \\ 
			&  &  &  &  &  &  &  &  \\ \hline
			&  &  &  &  &  &  &  &   \\ 
			ETAS--classical &  & 0.0180 & 0.0075 & 1.23 & 1.667 & 0.018  & 0.18   & 2.06 \\ 
			&  &  &  &  &  &  &  &  \\ \hline
			&  &  &  &  &  &  &  &   \\ 
			ETAS--Silverman &  & 0.0184 & 0.0068 & 1.21 & 1.666 & 0.018  & 0.18   & 2.07 \\ 
			&  &  &  &  &  &  &  &  \\ \hline
			& median & 0.0184 & 0.0068 & 1.21 & 1.662 & 0.017  & 0.19   & 2.07 \\ 
			GP-ETAS 
			& 0.05 & 0.0164 & 0.0056 & 1.19 & 1.595 & 0.014  & 0.17   & 1.93 \\ 
			& 0.95 & 0.0203 & 0.0085 & 1.24 & 1.734 & 0.022  & 0.21   & 2.23 \\ 
			\hline 
		\end{tabular}\label{tab:params_case1}
	\end{table}
	\begin{table}
		\centering
		\small
		\caption{Case 2: Parameter values $\thetas_\varphi$ of the triggering function.}
		\begin{tabular}{lcccccccc}
			\hline
			model & quantiles & $K_0$ & $c$ & $p$ & $\alpha$ & $d$ & $\gamma$ & $q$  \\ 
			\hline
			&  &  &  &  &  &  &  & \\ 
			generative  &  & 0.018 & 0.006 & 1.20 & 1.690 & 0.015  & 0.20   & 2.00 \\ 
			&  &  &  &  &  &  &  &  \\ \hline
			&  &  &  &  &  &  &  &   \\ 
			ETAS classical &  & 0.0193 & 0.0081 & 1.21 & 1.646 & 0.015  & 0.20   & 2.08 \\ 
			&  &  &  &  &  &  &  &  \\ \hline
			&  &  &  &  &  &  &  &   \\ 
			ETAS silverman &  & 0.0214 & 0.0058 & 1.15 & 1.616 & 0.015  & 0.20   & 2.10 \\ 
			&  &  &  &  &  &  &  &  \\ \hline
			& median & 0.0194 & 0.0084 & 1.21 & 1.648 & 0.014  & 0.20   & 2.00 \\ 
			GP-ETAS & 0.05 & 0.0175 & 0.0064 & 1.19 & 1.586 & 0.011  & 0.18   & 1.89 \\
			& 0.95 & 0.0216 & 0.0103 & 1.24 & 1.703 & 0.017  & 0.22   & 2.14 \\ 
			\hline 
		\end{tabular}\label{tab:params_case2}
	\end{table}
	\begin{table}
		\centering
		\small
		\caption{Averaged test likelihood $\ell_{\rm test}$ of twelve unseen data sets (higher is better).}
		\begin{tabular}{cccccc}
			\hline
			experiment &generative model & ETAS--classical &  ETAS--Silverman & GP-ETAS \\ 
			\hline 
			Case 1 &-344.9 & -417.0 & -404.2 & \textbf{-347.8} \\ 
			Case 2 &-140.1 & -201.8 & -210.2 & \textbf{-171.8} \\ 
			\hline 
		\end{tabular} \label{tab:test_likelihood_c1c2}
	\end{table}
	
	\begin{table}
		\centering
		\small
		\caption{Comparison on  $\ell_2$ norm to the true background intensity (smaller is better).}
		\begin{tabular}{llccc}
			\hline
			experiment & criterium  & ETAS--classical & ETAS--Silverman & GP-ETAS \\ \hline
			Case 1 & $\ell_2$   & 0.0482    & 0.0297      & \textbf{0.0190} \\ 
			Case 1 & normalised & 2.53    & 1.56      & 1 \\ 
			\hline
			Case 2 & $\ell_2$   & 0.1950    & 0.2410      & \textbf{0.1282} \\ 
			Case 2 & normalized & 1.52    & 1.88      & 1 \\ 
			\hline
		\end{tabular} \label{tab:l2_BG}
	\end{table}
	
\begin{figure}
		\centering
		\small
		\includegraphics[width=0.31\linewidth]{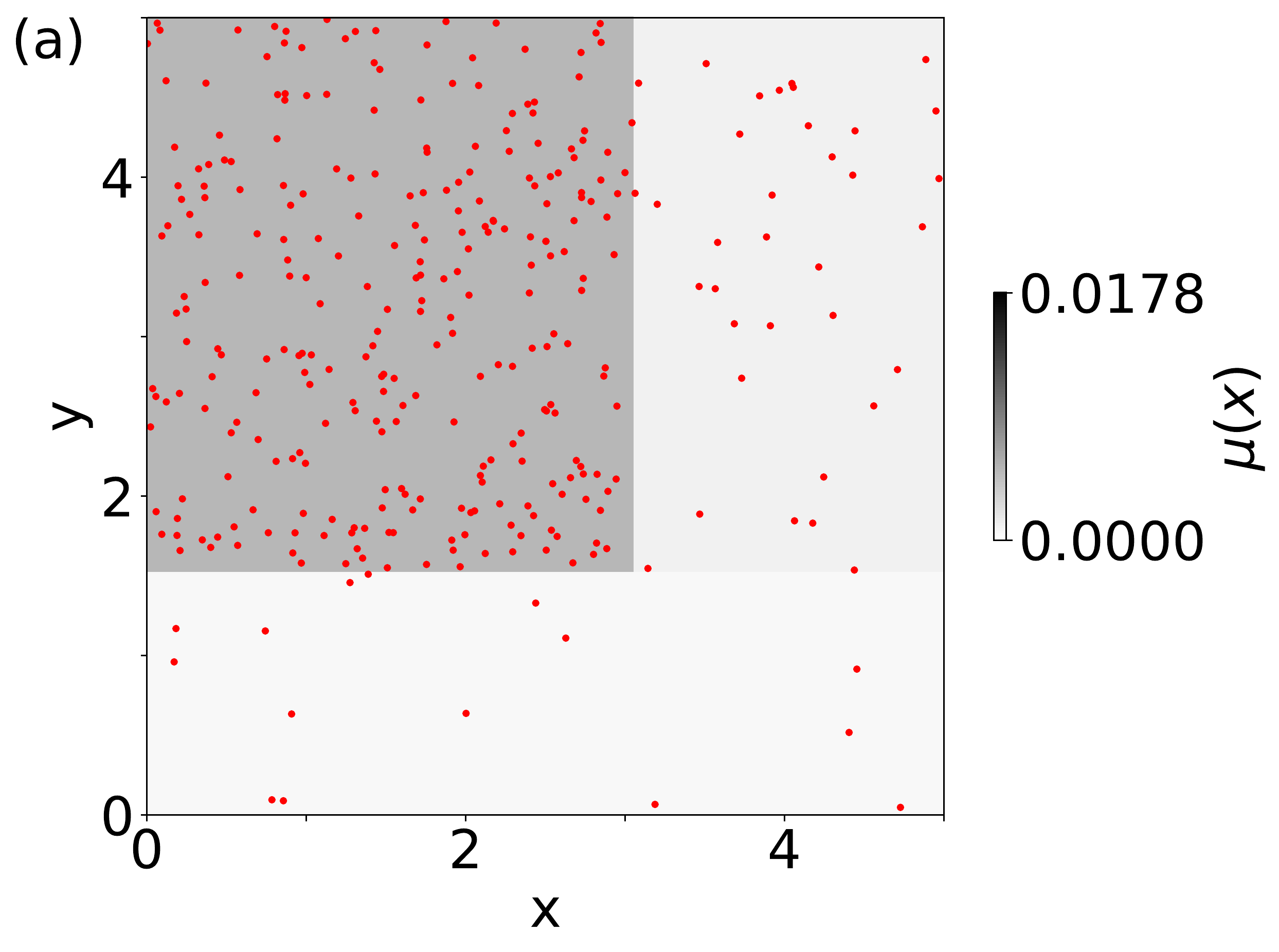}
		\includegraphics[width=0.31\linewidth]{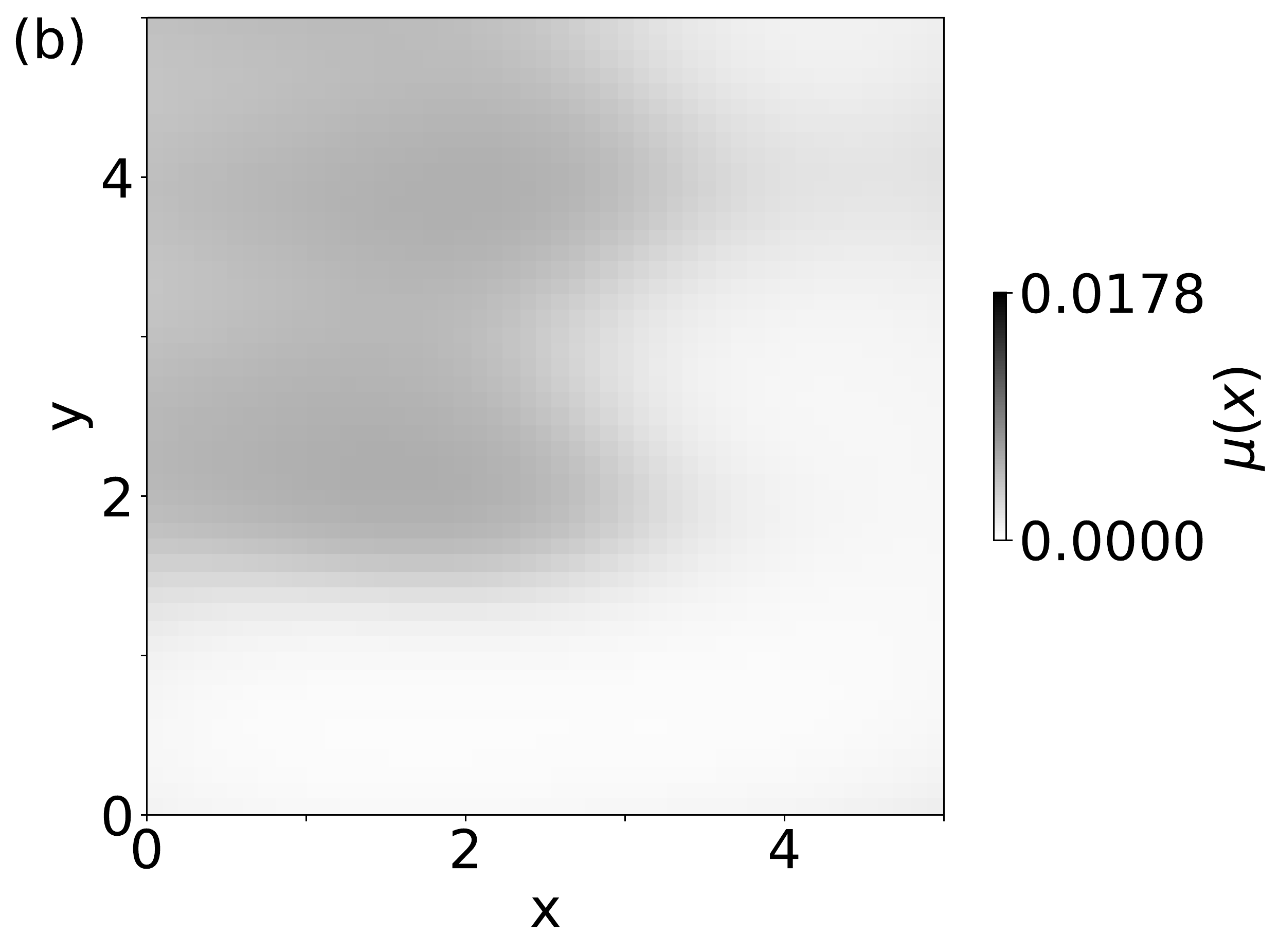}
		\includegraphics[width=0.31\linewidth]{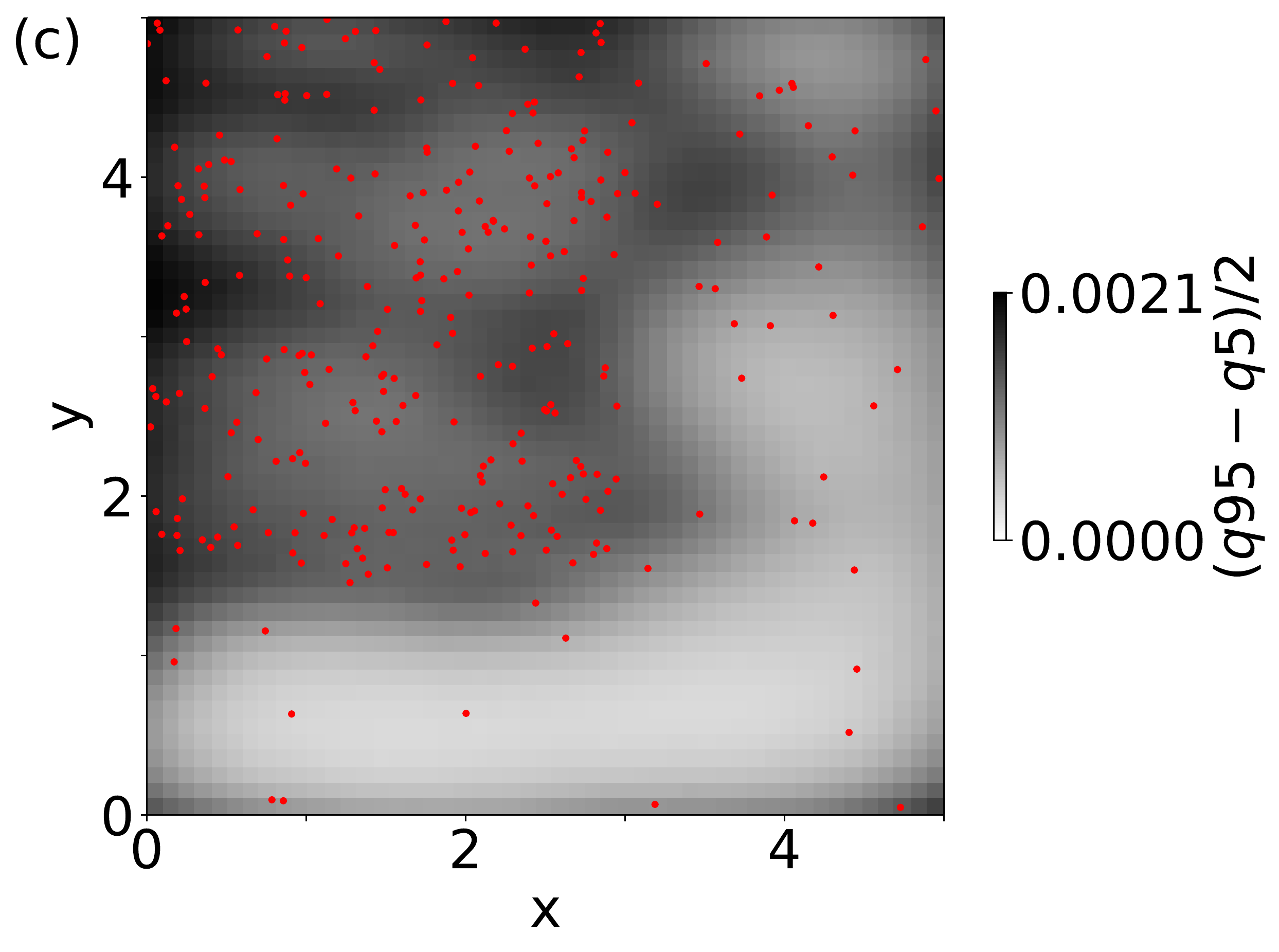}	
		\includegraphics[width=0.31\linewidth]{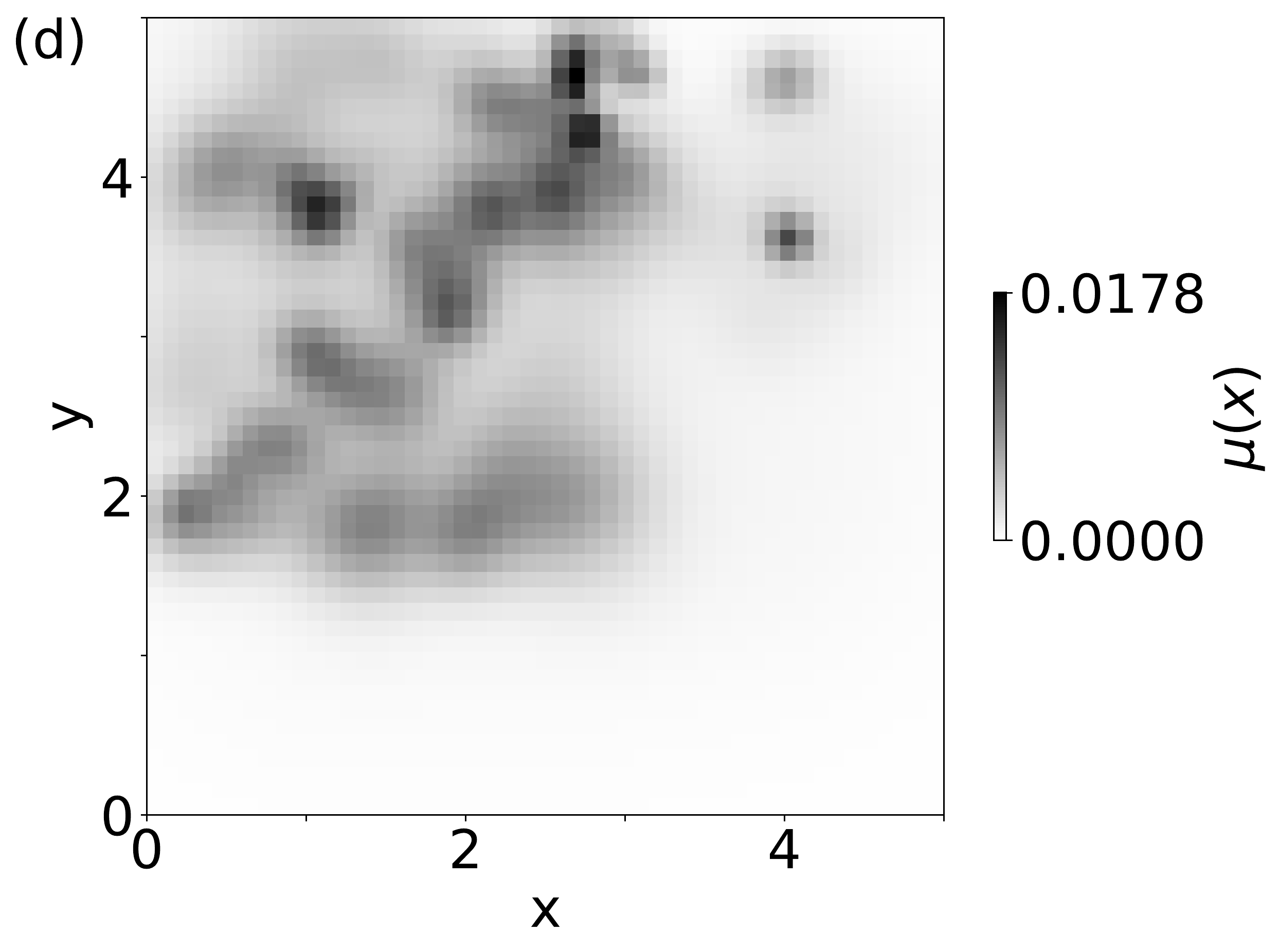}
		\includegraphics[width=0.31\linewidth]{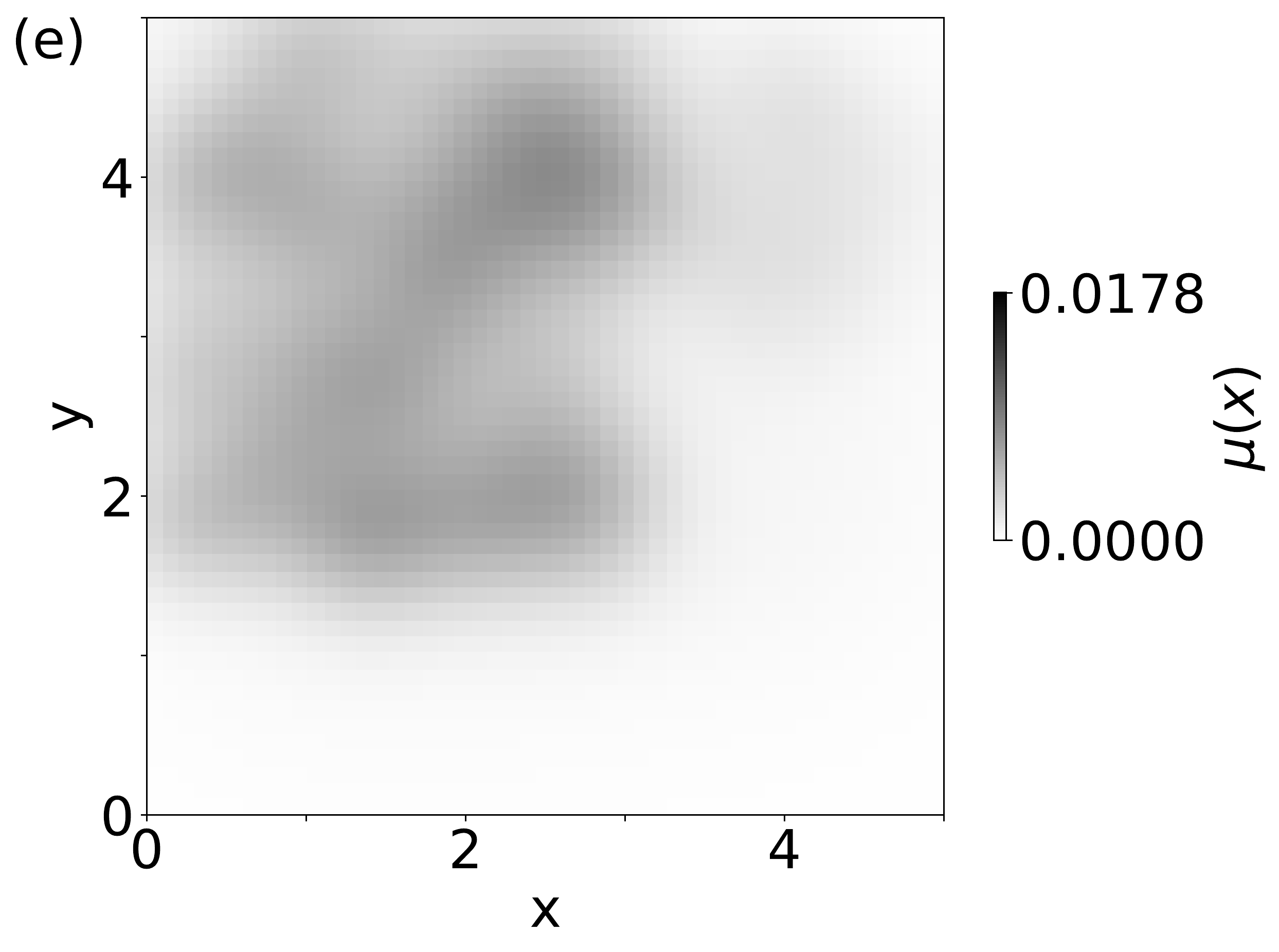}
		\includegraphics[width=0.31\linewidth]{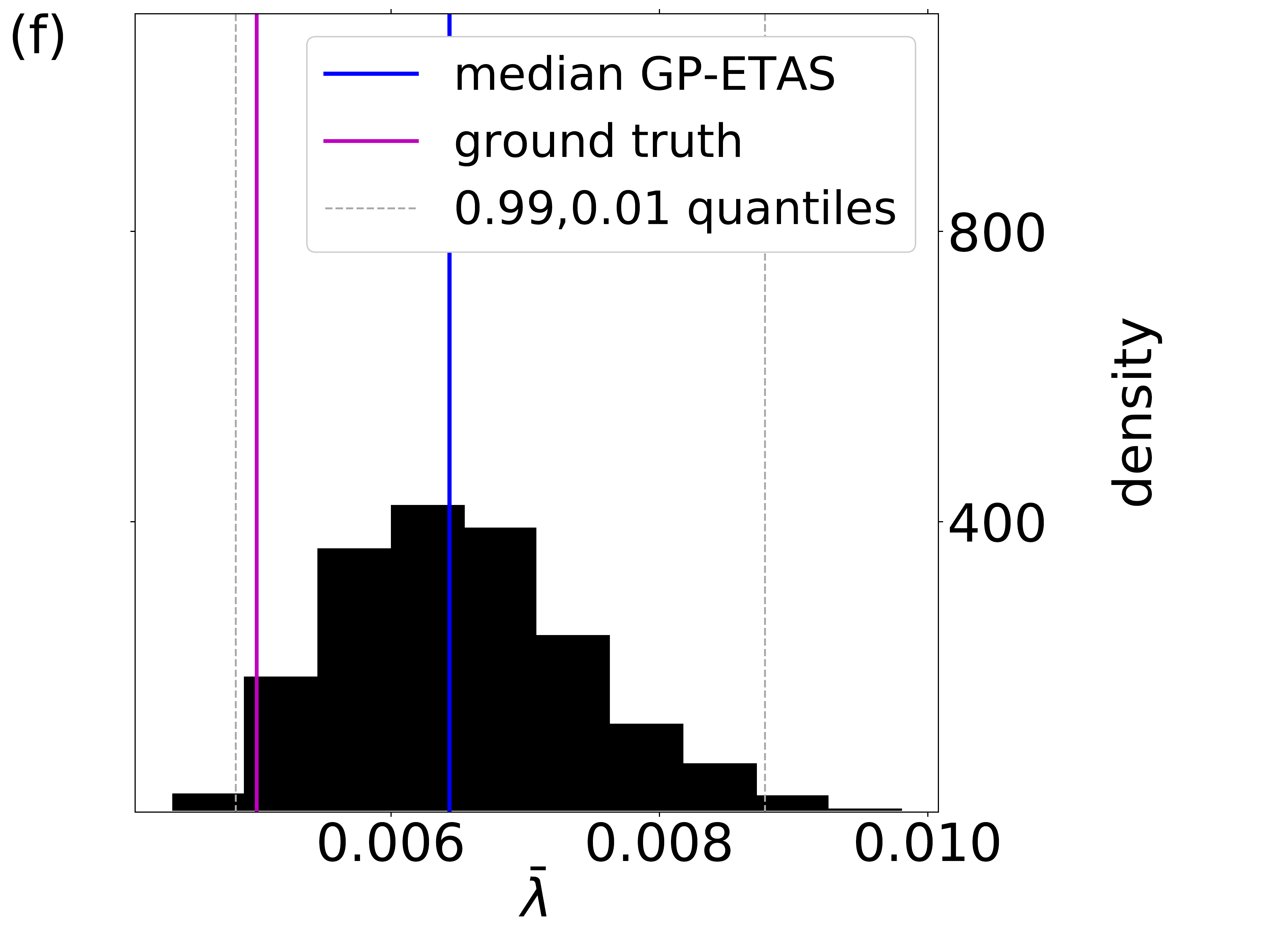}
		\includegraphics[width=\linewidth]{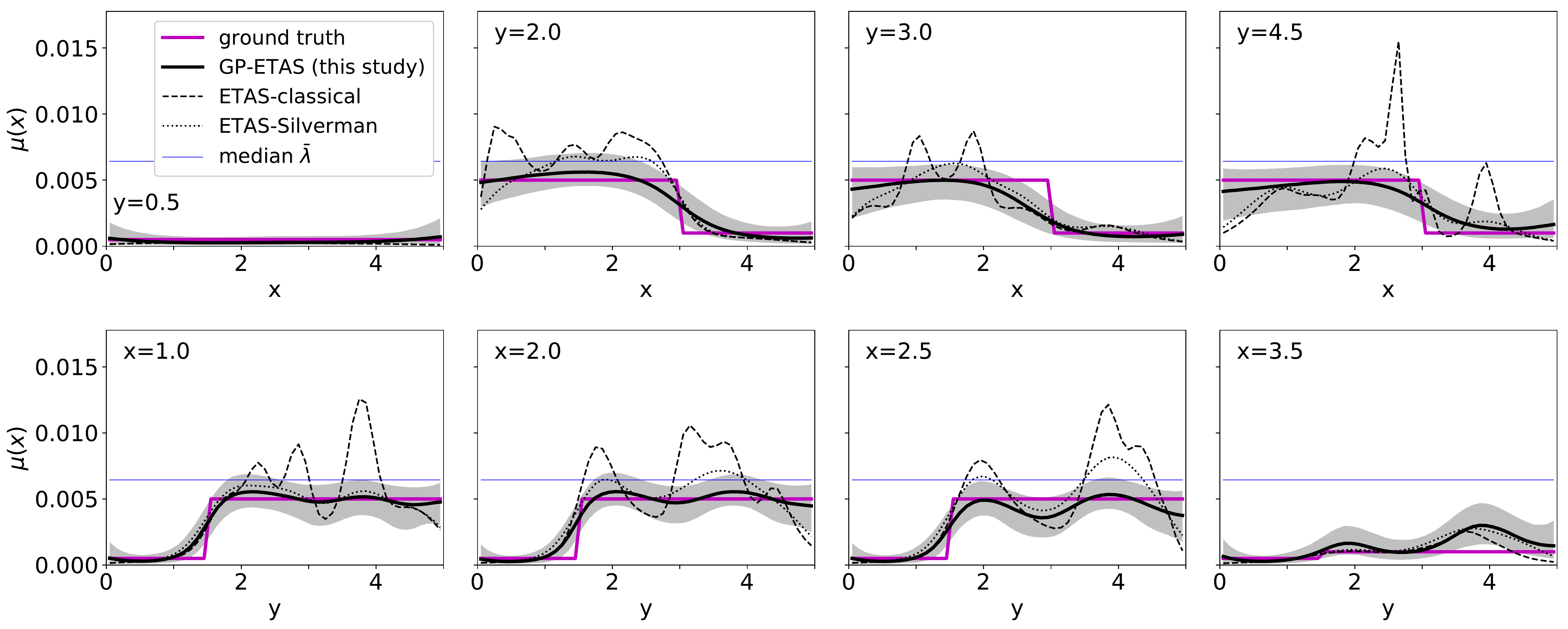}
		\caption{Experimental results of background intensity $\mu_1(\x)$ for the synthetic data of Case 1:
			\textit{First and second row}:
			(a) generative model, 
			(b) median GP-ETAS, 
			(c) uncertainty GP-ETAS  as semi inter quantile 0.05, 0.95 distance
			(d) ETAS--classical MLE, 
			(e) ETAS--Silverman MLE,
			(f) normalised histogram of the sampled upper bound $\barlambda$.
			Dots are the background events of the realisation.
			\textit{Third and fourth row}: 
			One dimensional profiles of $\mu_1(\x)$ (ground truth) and inferred results are shown.
			The profiles are at $y\in \{0.5,2,3,4.5\}$ and
			$x\in \{1,2,2.5,3.5\}$.}
		\label{fig:fig005_background_results_A1}
	\end{figure}

\begin{figure}
		\centering
		\small
		\includegraphics[width=0.31\linewidth]{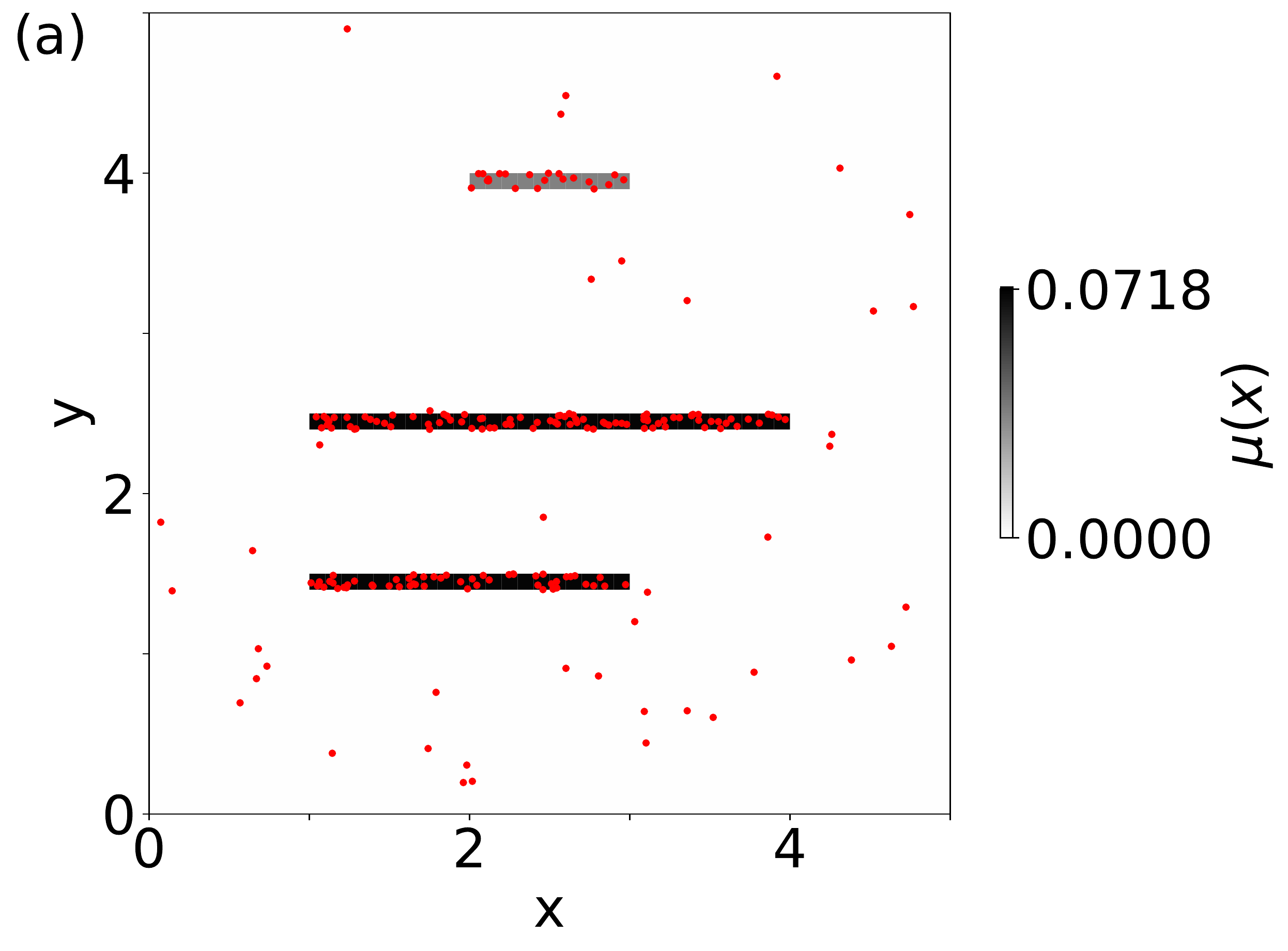}
		\includegraphics[width=0.31\linewidth]{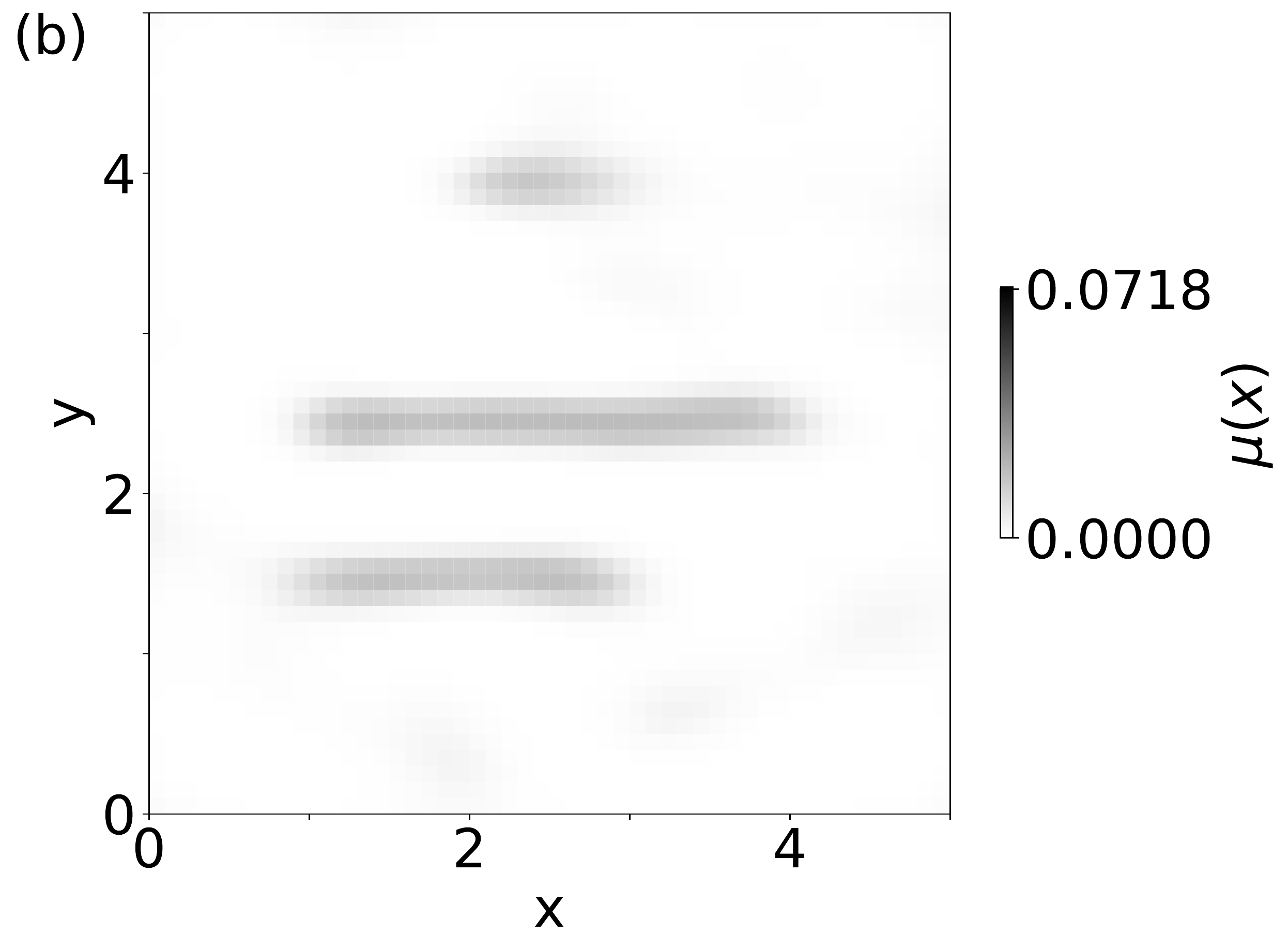}
		\includegraphics[width=0.31\linewidth]{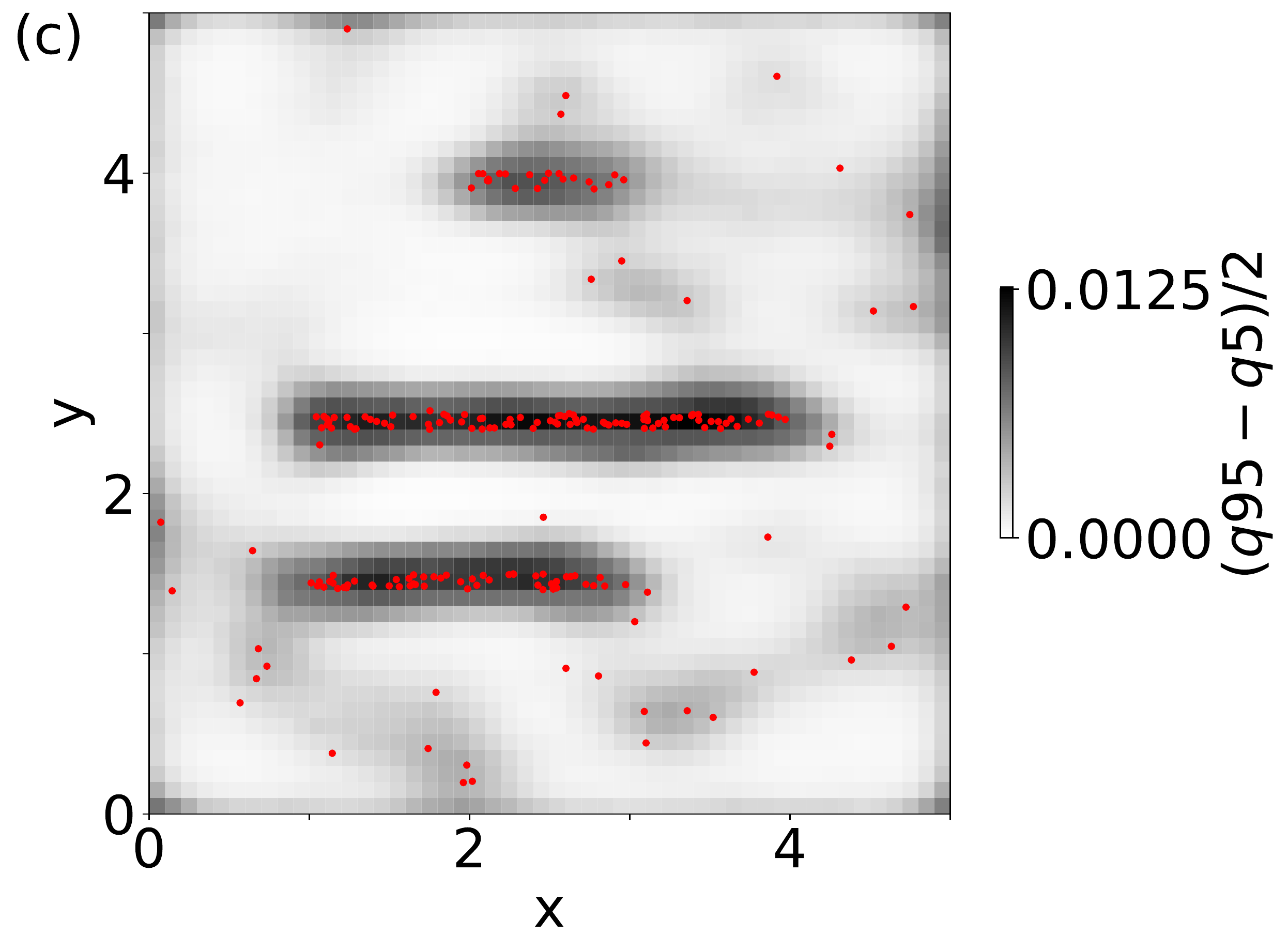}
		\includegraphics[width=0.31\linewidth]{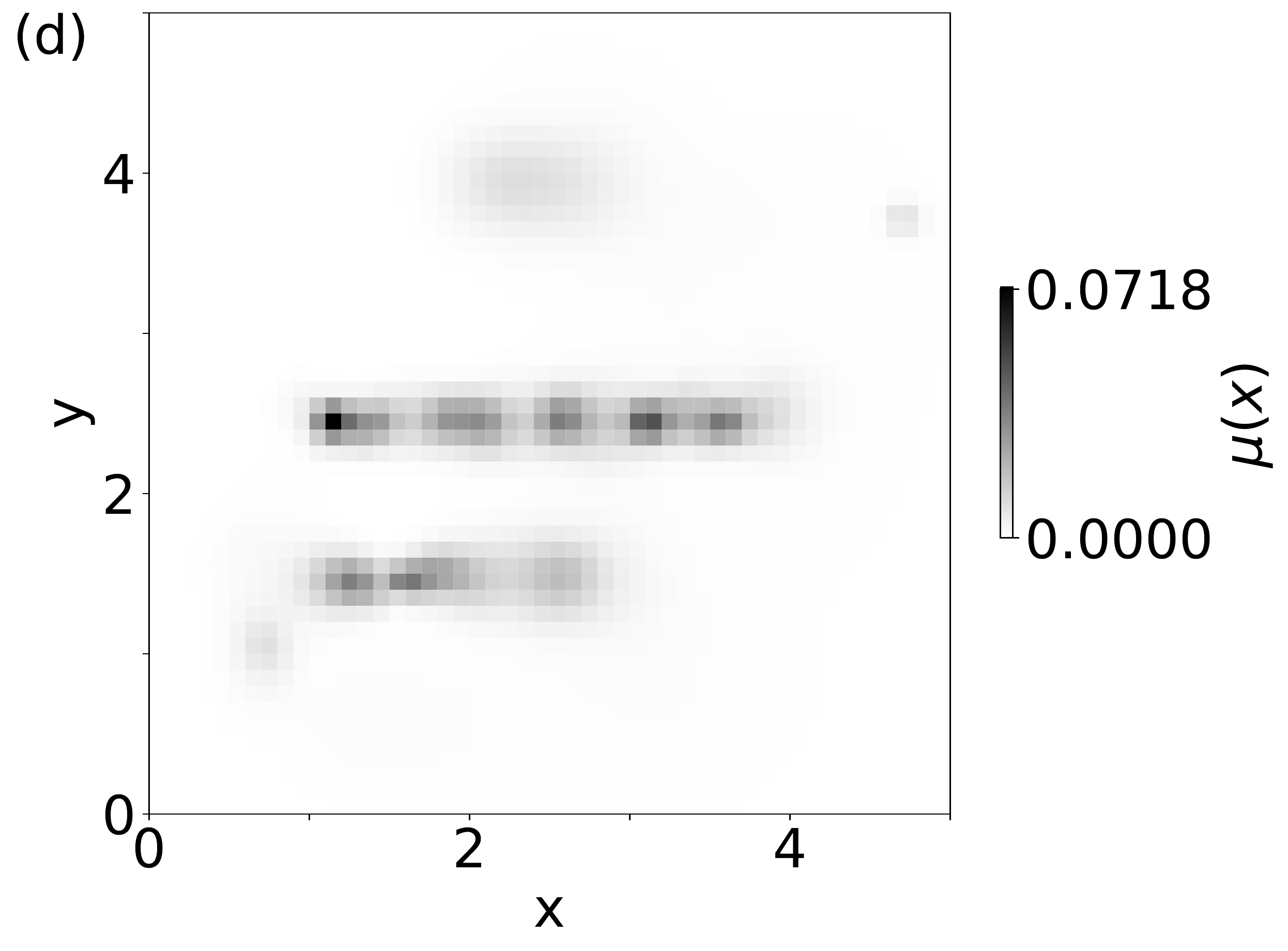}
		\includegraphics[width=0.31\linewidth]{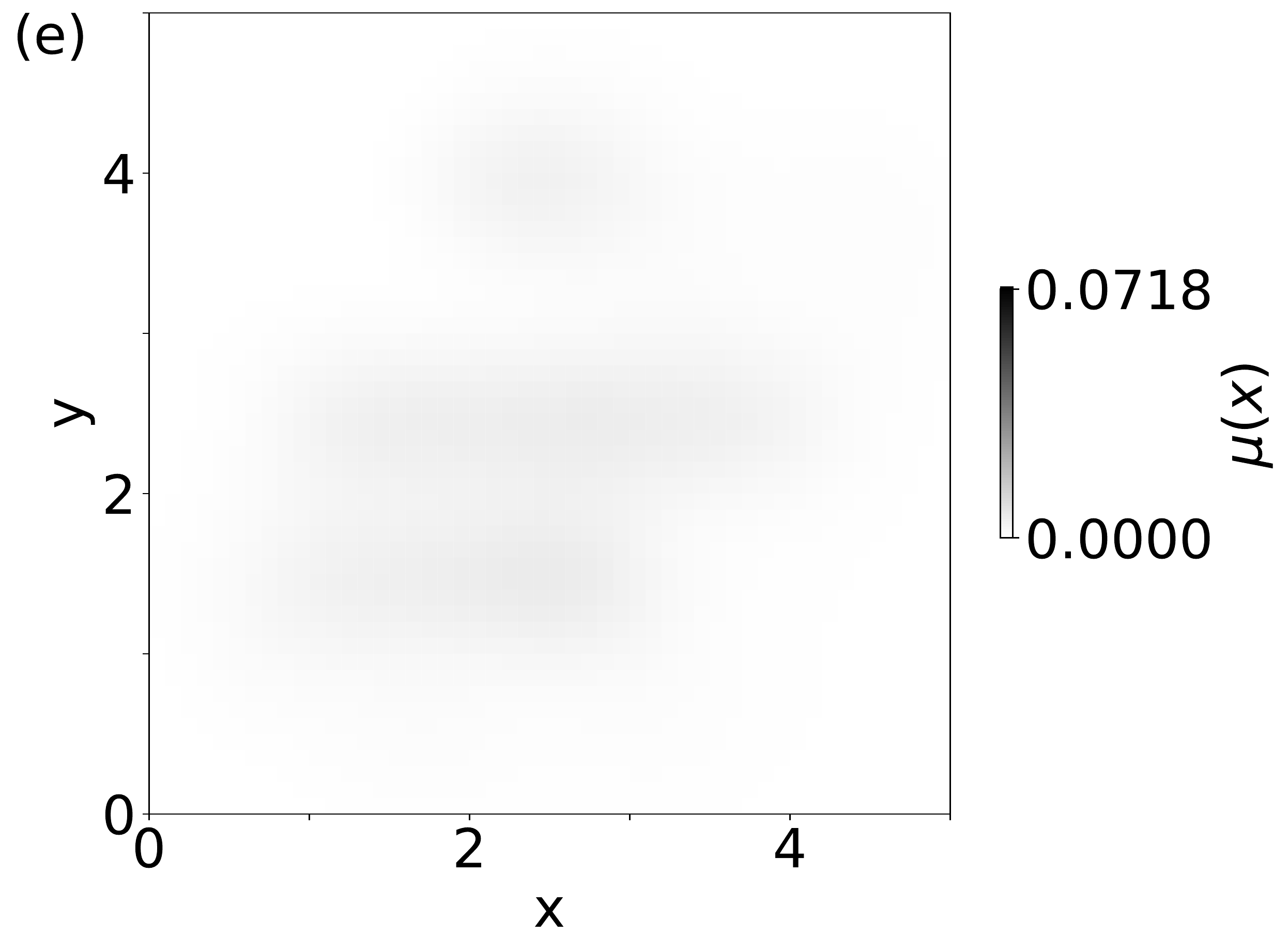}
		\includegraphics[width=0.31\linewidth]{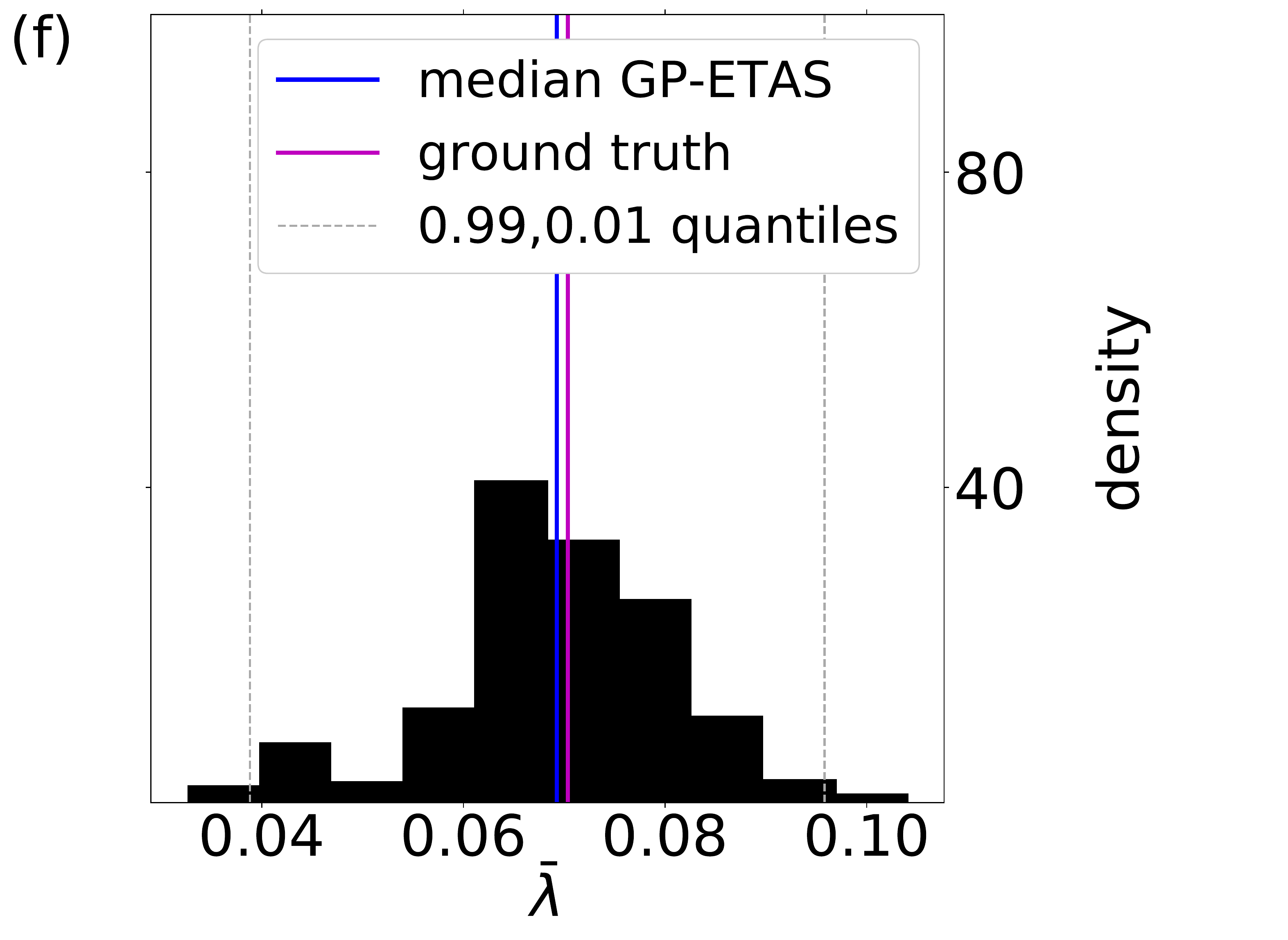}
		\includegraphics[width=\linewidth]{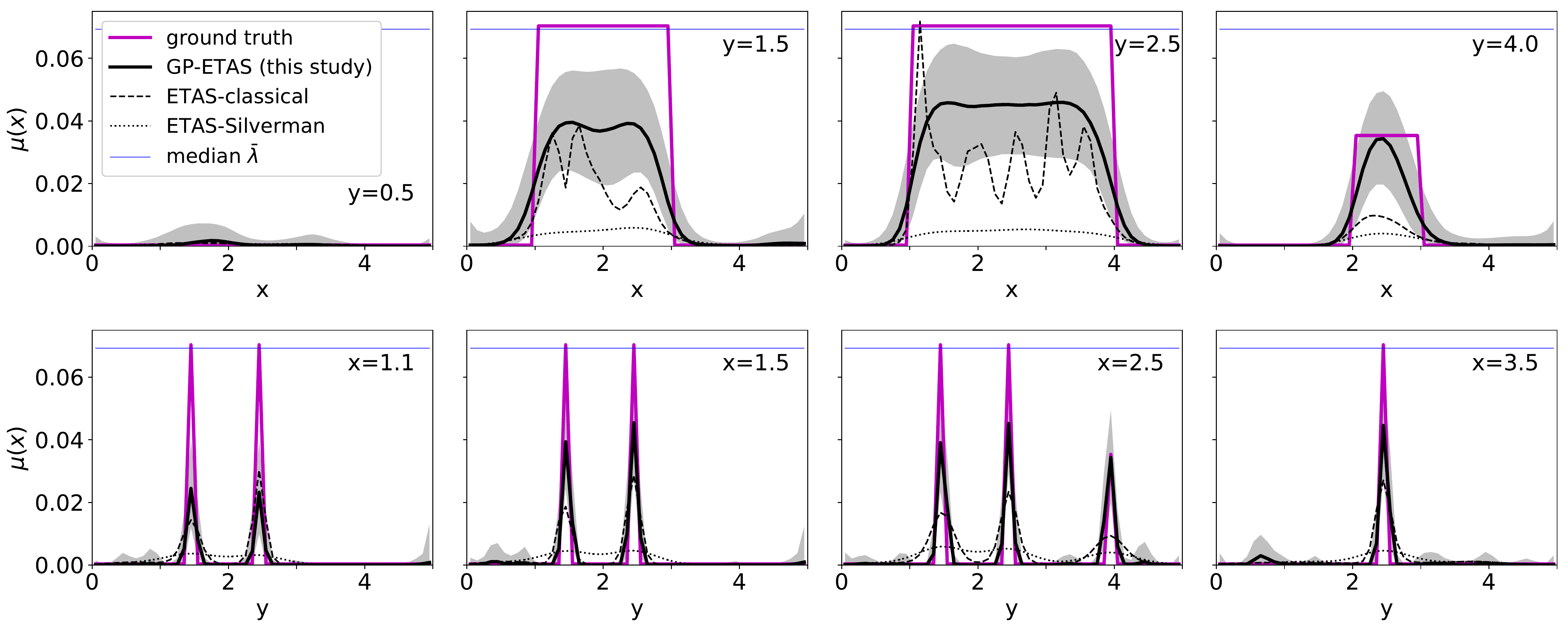}	
		\caption{Same as Figure \ref{fig:fig005_background_results_A1} but now 
			results of background intensity $\mu_2(\x)$ for the synthetic data of Case 2.
			See Figure  \ref{fig:fig005_background_results_A1} for the description of plots and lines.
					The profiles are at $y\in \{0.5,1.5,2.5,4\}$ and
					$x\in \{1.1,1.5,2.5,3.5\}$.
		}
		\label{fig:fig005_background_results_F2}
	\end{figure}

	\paragraph{Findings and interpretations:} 
	The ground truth and inferred results of $\mu(\x)$ and $\thetas_\varphi$ are 
	given in Figure \ref{fig:fig005_background_results_A1}--\ref{fig:fig005_background_results_F2} and Table \ref{tab:params_case1}--\ref{tab:params_case2}.
	Performance metrics: the averaged $\ell_{\rm test}$ of twelve unseen data sets
	and the numerically approximated error of the estimated background intensity $\ell_2$ are 
	shown in Table \ref{tab:test_likelihood_c1c2}--\ref{tab:l2_BG}. 
	Here we describe a few noteworthy aspects.
	First of all, 
	GP-ETAS recovers well the assumptions 
	both the background intensity $\mu(\x)$ and the parameters of the triggering function $\thetas_\varphi$.
	GP-ETAS outperforms the standards models for 
	both metrics  $\ell_{\rm test}$ and $\ell_2$.
	The latter fact is of particular importance, as
	it is common practice to use the declustered background intensity $\mu(\x)$ 
	for seismic hazard assessment.
	One may appreciate that 
	ETAS--classical occasionally tends to strongly overshoot the true $\mu(\x)$ (see Figure \ref{fig:fig005_background_results_A1});
	in regions with many aftershocks, e.g. near (2.8,4.5), (1.2,3.8). 
	This effect is less pronounced for ETAS--Silverman, 
	where the minimum bandwidth is broader.
	In our approach no bandwidth selection has to be made in advance, it is obtained via sampling the hyperparameters.
	One also observes, that ETAS--classical and ETAS--Silverman 
	suffer more strongly from edge effects than GP-ETAS, 
	which seems to be fairly unaffected.
	
	\medskip
	
	The parameters of the triggering function $\thetas_\varphi$ are roughly correctly identified 
	in all cases and methods.
	All the values are close to those of the generative model. 
	All the methods overestimate $c$ and $K_0$ (in Case 2), however the true values
	are still included in the uncertainty band (credible band) of GP-ETAS.
	GP-ETAS has the advantage that it provides the whole distribution of the parameters instead of 
	only a point estimate.
	The median of the obtained upper bound $\barlambda$ on $\mu(\x)$ 
	using GP-ETAS overestimates in Case 1 (underestimates in Case 2)
	the true upper bound, however, it is fairly close to the true value, 
	which is contained in the uncertainty band around $\barlambda$. 
	Therefore, $\barlambda$ can be considered in seismic hazard assessment as conservative choice.
	
	\paragraph{Computational costs:} GP-ETAS encounters approximately 
	a complexity of $\mathcal{O}((N_{\D\cup\Pi})^3)$ for the inference of $\mu$. 
	This is due to the matrix inversions involved in Gaussian process modelling.
	The estimation of $\thetas_\varphi$ is less expensive and approximately of $\mathcal{O}(N_\D^2)$, where $N_\D$ is the number of data points.
	In addition, the number of required samples in order to obtain a valuable approximation of 
	the posterior distribution depends on the mixing properties of the Markov chain.
	From our experience based on the performed experiments one needs $>10^3$ samples after a burn-in phase of $>10^3$ iterations. 
	Therefore, our proposed method in its current implementation is computationally expensive but still feasible for small to intermediate data sets with approximately $N_\D \lessapprox 10^4$ events;
	which seems sufficient for many situations where site specific seismic analysis takes place. 
	

	\newpage
	%------------------------------------------------------------------------------------------------------------------------
	\subsection{Case study: L'Aquila, Italy}
	%------------------------------------------------------------------------------------------------------------------------
	Now we apply GP-ETAS
	to real data and compare the performance with 
	the other models 
	ETAS--classical and ETAS--Silverman.
	
	\medskip
	
	The L'Aquila region in central Italy is seismically active and experiences from time 
	to time severe earthquakes.
	The most famous example is the $M_w=6.2$ earthquake on 6 April 2009, 
	which occurred directly below the City of L'Aquila, 
	and caused large damage and more than $300$ deaths \citep{Marzocchi2014}.
	This event was followed by a seismic sequence with a largest earthquake of $M_w=4.2$, 
	latter occurred almost one year later on 30 March 2010 \citep{Marzocchi2014}.
	
	\medskip
	
\begin{figure}
		\centering
		\small
		\includegraphics[width=0.43\linewidth]{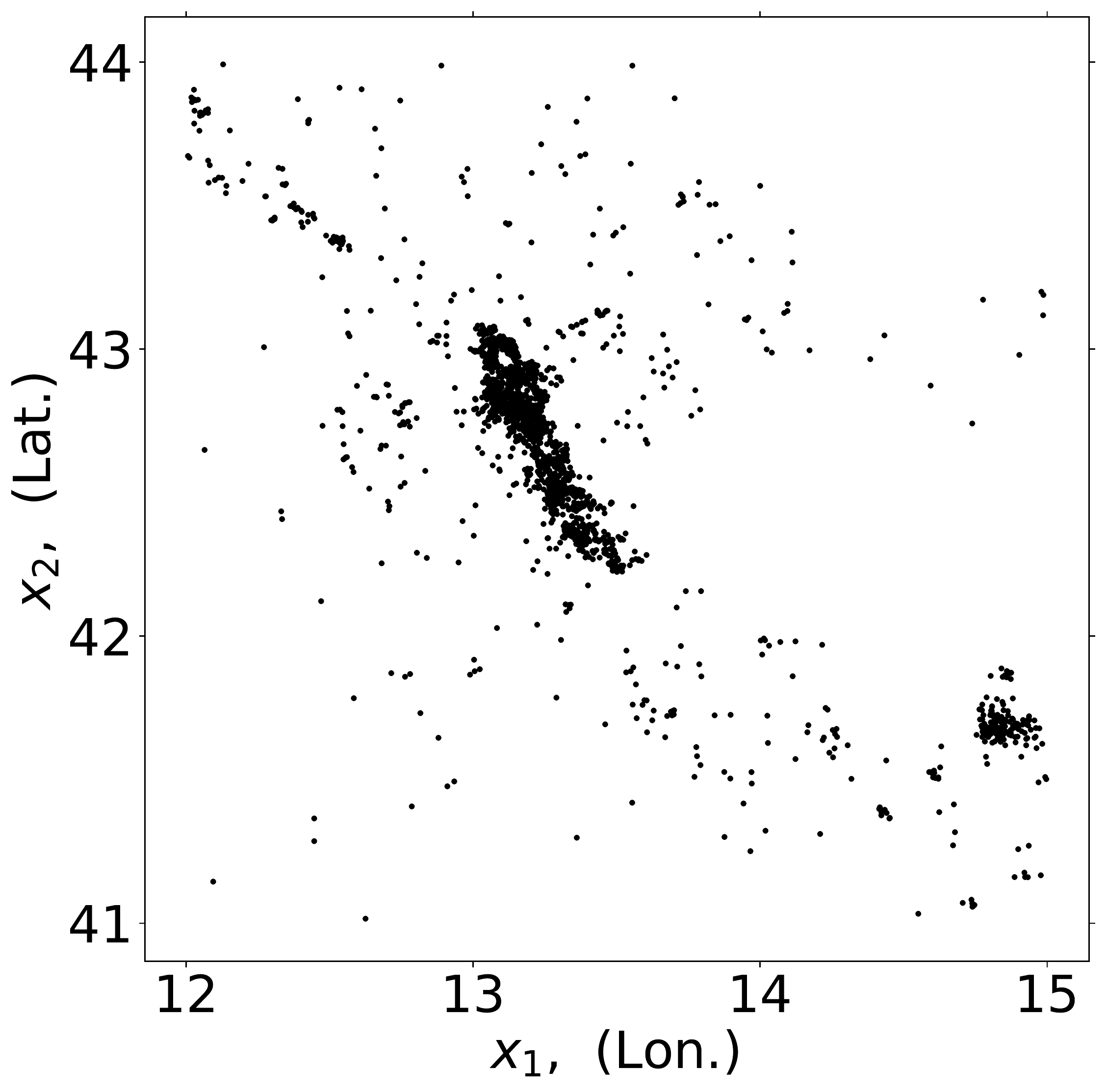}
		\includegraphics[width=0.463\linewidth]{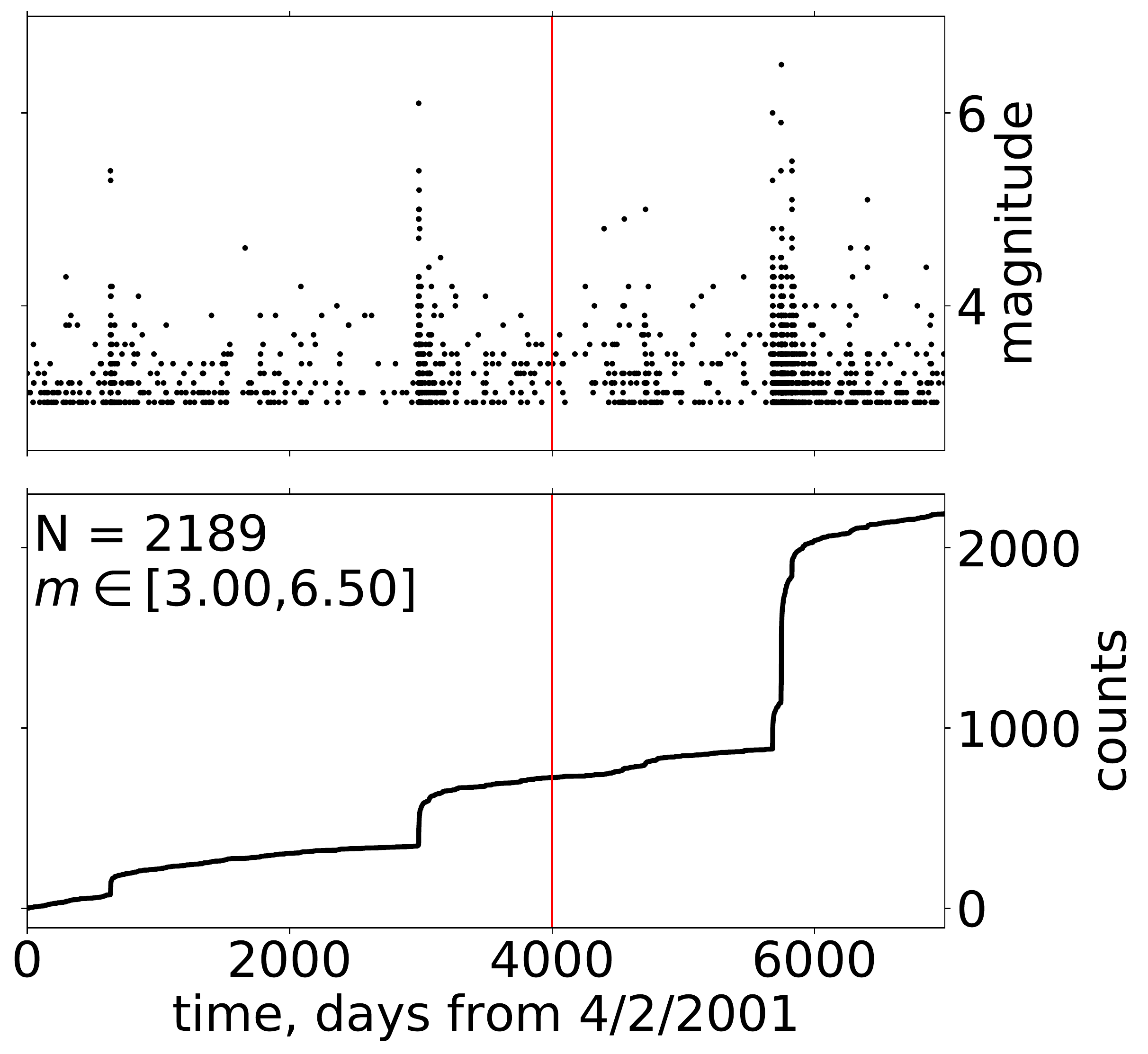}\\
		\caption{
			Earthquake data from central Italy: Epicentre plot (left) and 
			visualisation of the data as earthquake sequence over time (right).
		}
		\label{fig:setting_aquila}
	\end{figure}

	The L'Aquila data set comprises $N=2189$ events which occurred in a time period 
	from 04/02/2001 to the 28/3/2020, 
	on a spatial domain $\calX=[12° E,15° E]\times [41° N,44° N]$
	with earthquake magnitudes $3.0\le m \le 6.5$. The data was
	obtained from the website of the 
	National Institute of Geophysics and Vulcanology of Italy
	(http://terremoti.ingv.it/, Istituto Nazionale della Geofisica e Vulcanologia, INGV).
	We split the data set into training data, all events with event times $t_i\le4000$ days
	($N_{\rm training}=723$ events, $\calT_{\rm training}=[0,4000]$ days),
	and \textit{test data},  all events with $t_i>4000$ days 
	($N_{\rm test}=1466$, $\calT_{\rm test}=[4000,6992]$ days), 
	as shown in Figure \ref{fig:setting_aquila}. 
	The training data is used for the inference 
	and the test data is used to evaluate the 
	performance of the different models.
	
	\medskip
	
\begin{figure}
		\centering
		\small
		\includegraphics[width=0.45\linewidth]{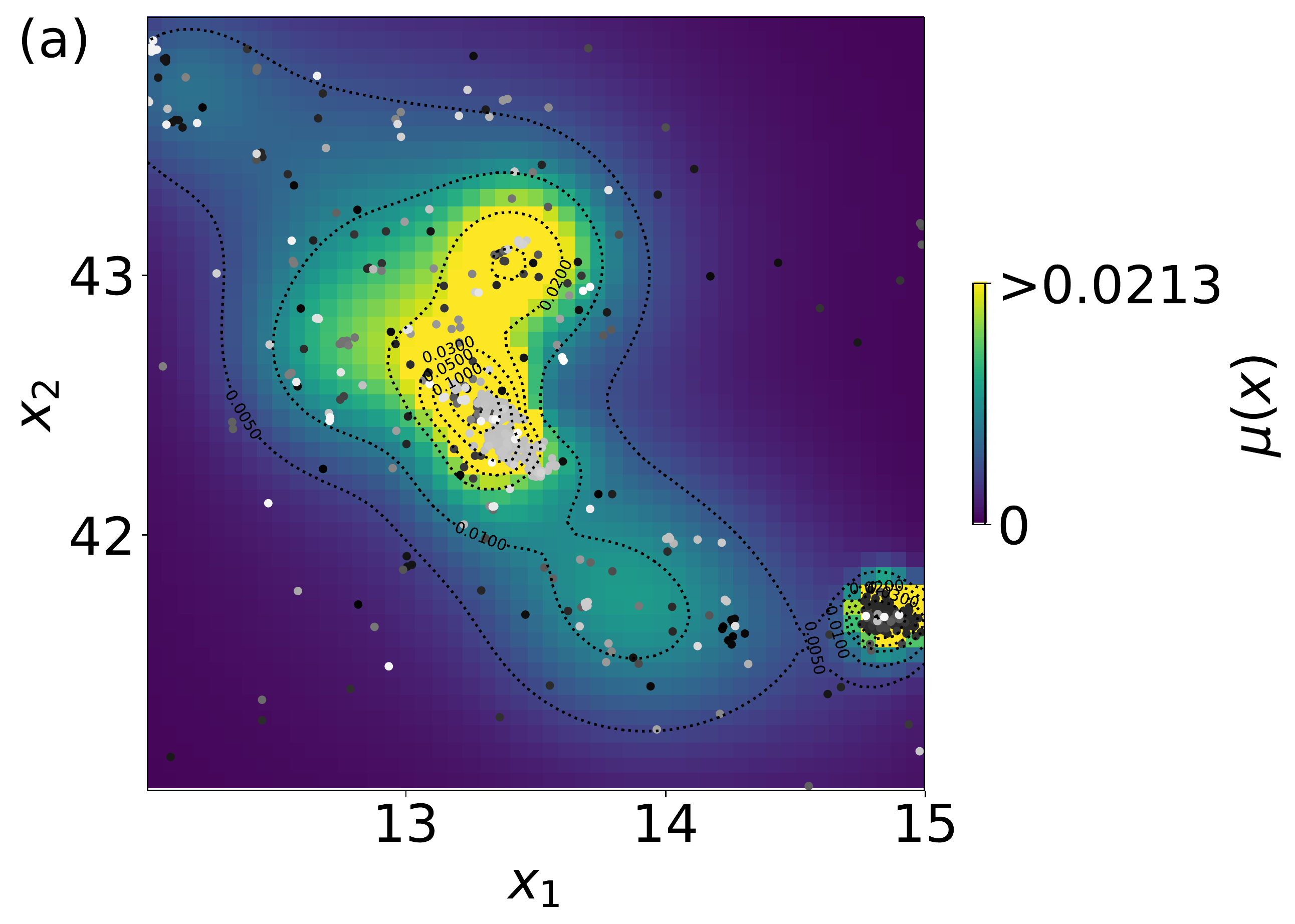}
		\includegraphics[width=0.45\linewidth]{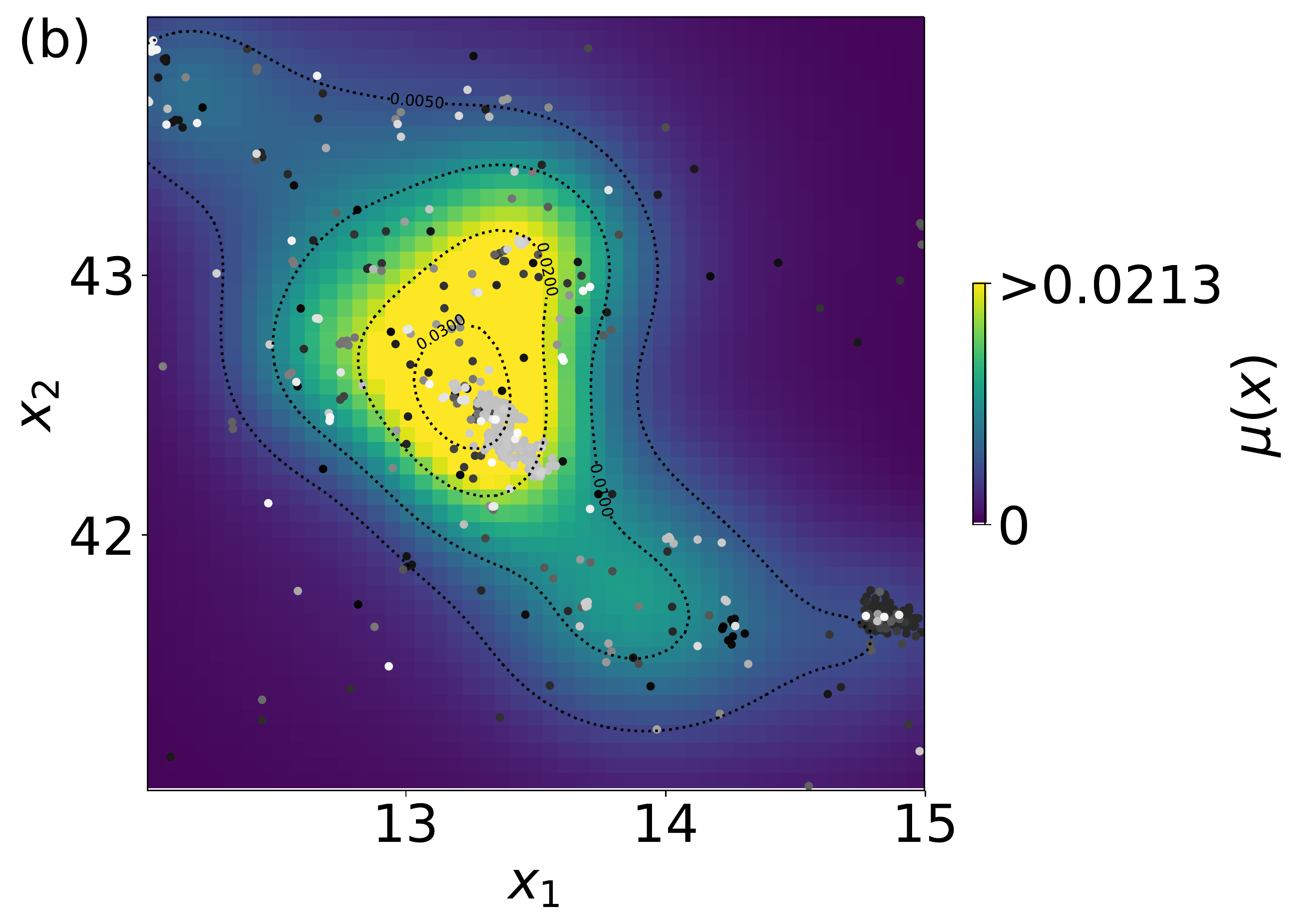}
		\includegraphics[width=0.45\linewidth]{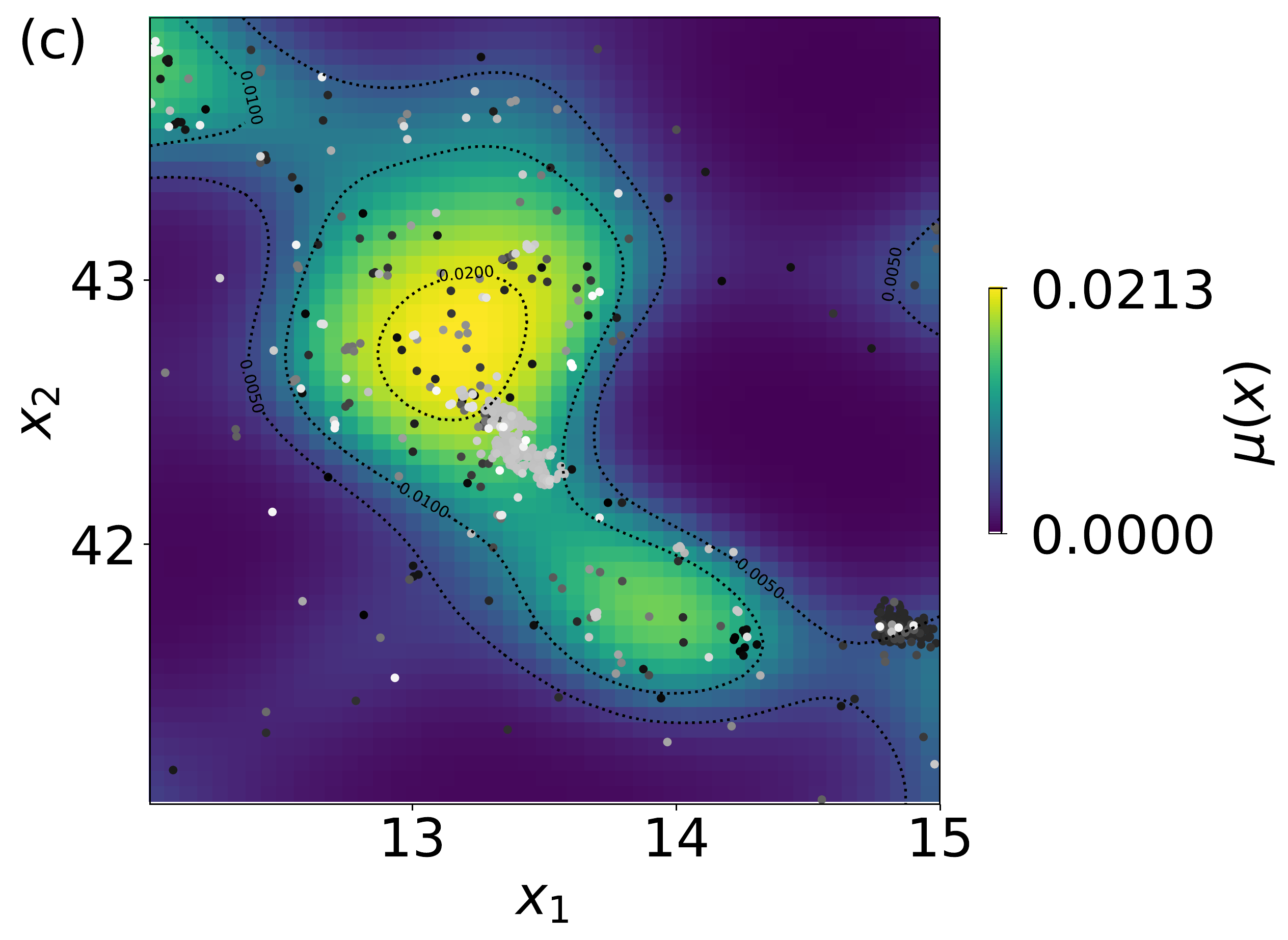}
		\includegraphics[width=0.44\linewidth]{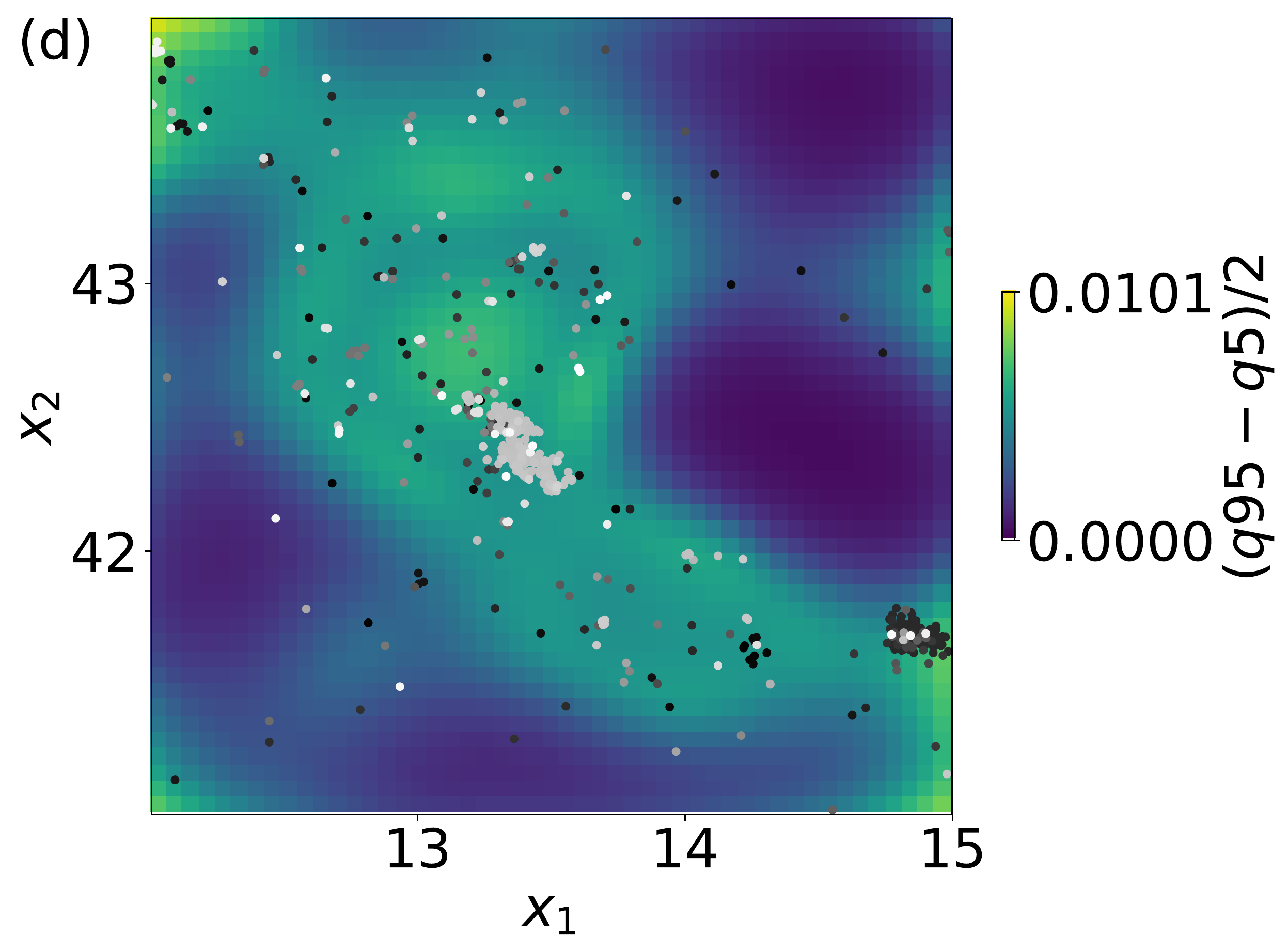}
		\caption{Results real data, L'Aquila data set: background intensity $\mu(\x)$ [number of shocks with $m \ge 3$ /day/degree$^2$]
			(a) ETAS--classical MLE, 
			(b) ETAS--Silverman MLE,
			(c) median GP-ETAS, 
			(d) uncertainty GP-ETAS: semi inter quantile 0.05, 0.95 distance, and
			dots are the events of the training data, where the grey scaling depicts the event times, 
			from black (older events) to white (current events).
			Note, (a--c)  have the same scale.}
		\label{fig:BG_aquila}
	\end{figure}

\begin{figure}
		\centering
		\includegraphics[width=0.41\linewidth]{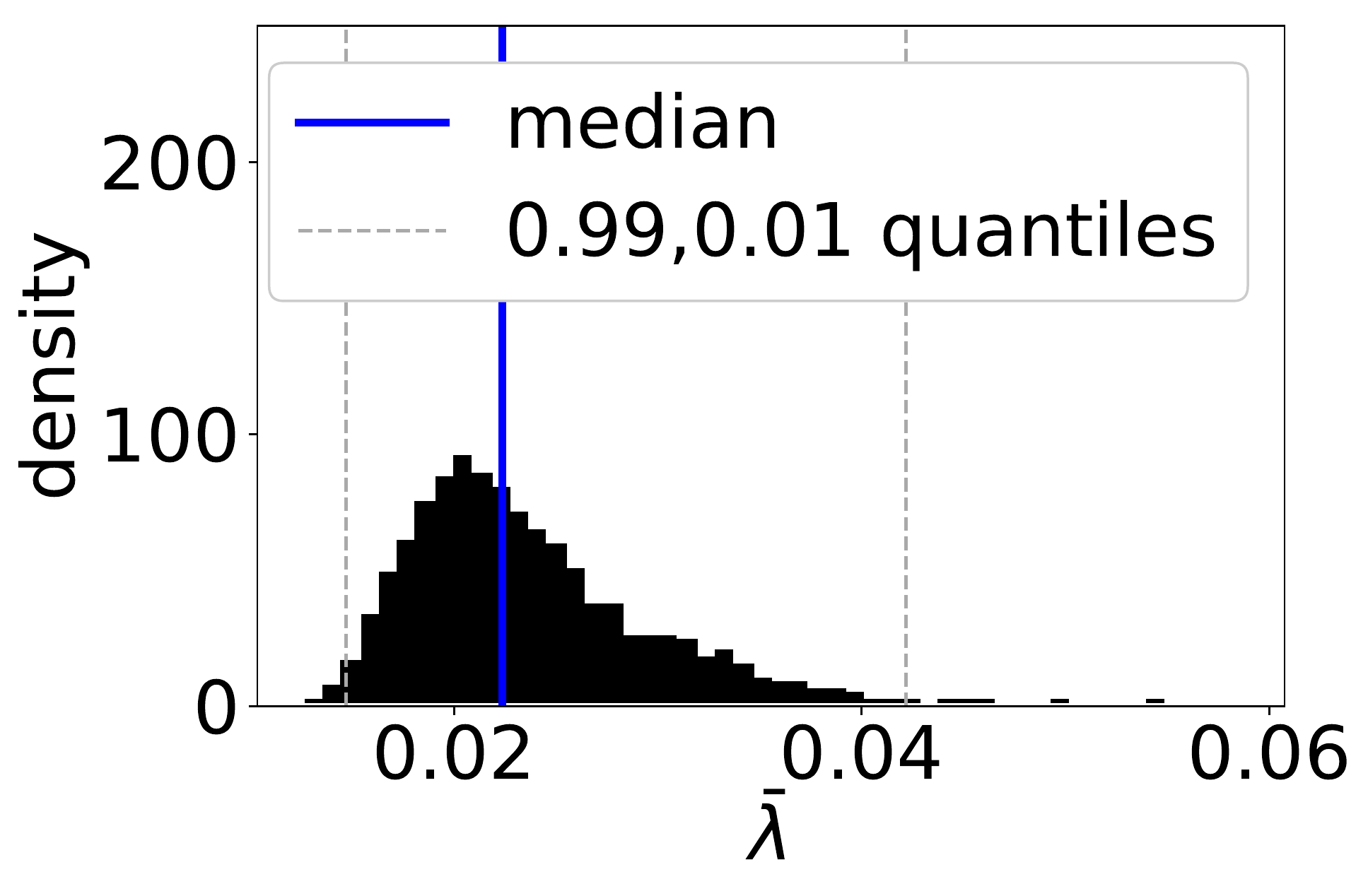}
		\caption{Normalised histogram of the sampled posterior of  the upper bound $\barlambda$}
		\label{fig:UB_aquila}
	\end{figure}

	The inference setup is the same as described for the synthetic data.
	We simulate 15000 posterior samples after a burn in of 2000.
	The priors are -- as for the synthetic data --  given in Table \ref{tab:GP-ETAS_setup}.
	The inference results for $\mu(\x)$ and $\thetas_\varphi$ are shown 
	in Figure \ref{fig:BG_aquila} and Table \ref{tab:theta_varphi_Aquila}.
	The performance metric $\ell_{\rm test}$ for different unseen 
	data sets (next 30 days, one year in future, five years in future, 
	and the total test data $\approx 8$ years) is shown 
	in Table \ref{tab:test_likelihood_Aquila}. The  posterior distribution 
	of the upper bound of the background intensity is shown in 
	Figure \ref{fig:UB_aquila}.
	
	\medskip
	
\begin{figure}
		\centering
		\includegraphics[width=0.325\linewidth]{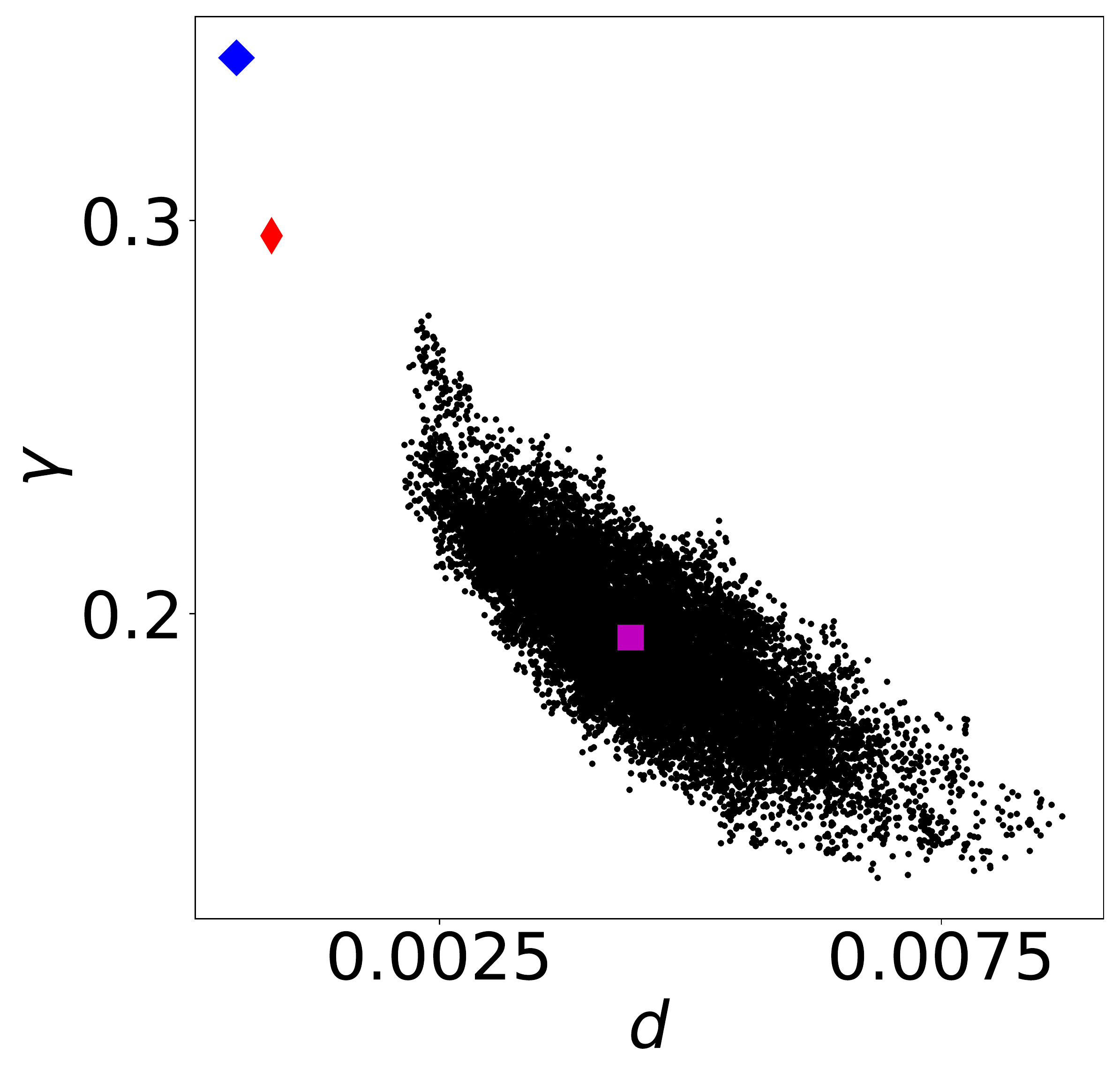}
		\includegraphics[width=0.325\linewidth]{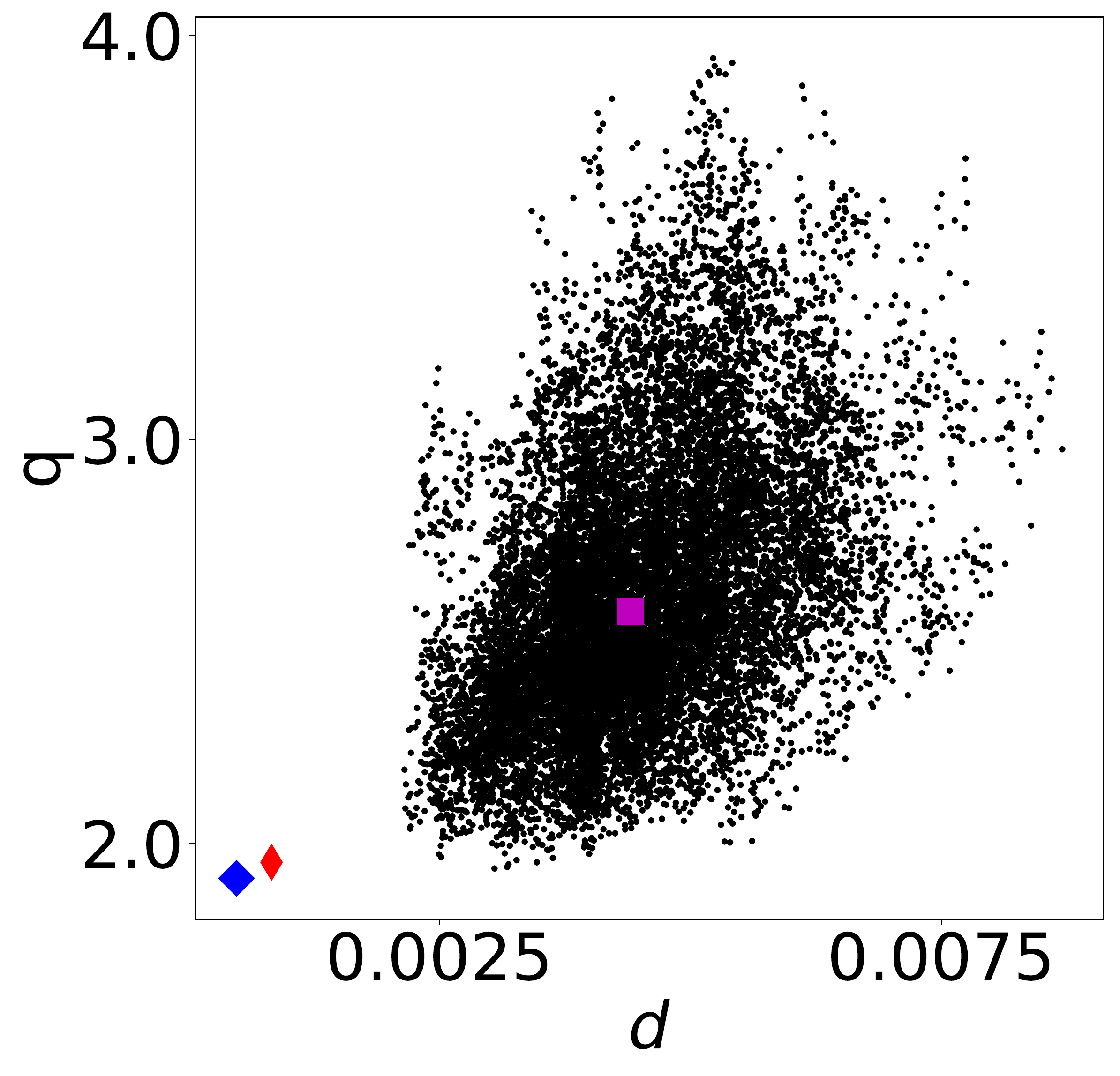}
		\includegraphics[width=0.325\linewidth]{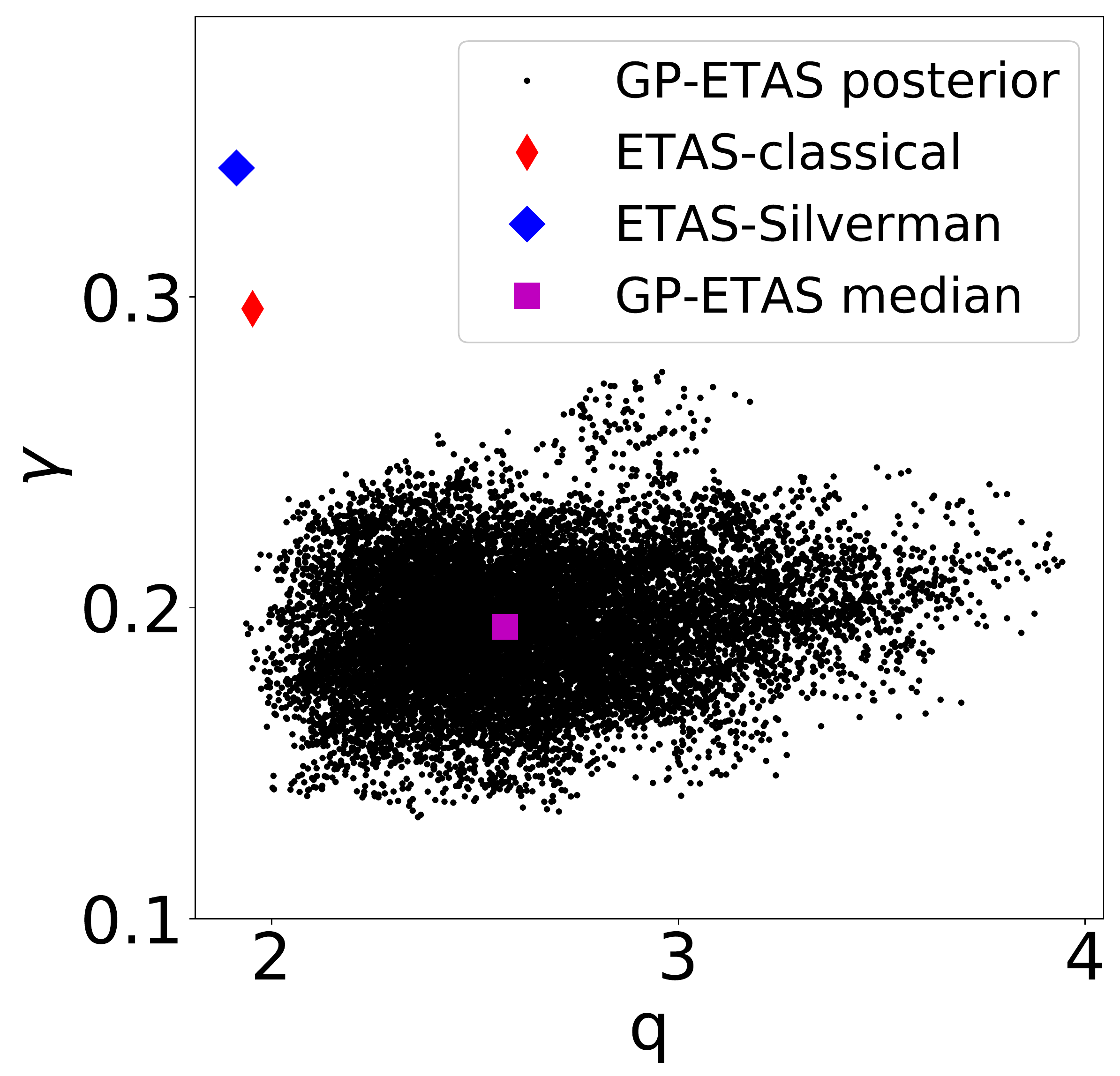}
		\caption{Scatter plot of the posterior samples of $d,\gamma,q$}
		\label{fig:case3fig013ascatter}
	\end{figure}

\begin{figure}
		\centering
		\includegraphics[width=0.49\linewidth]{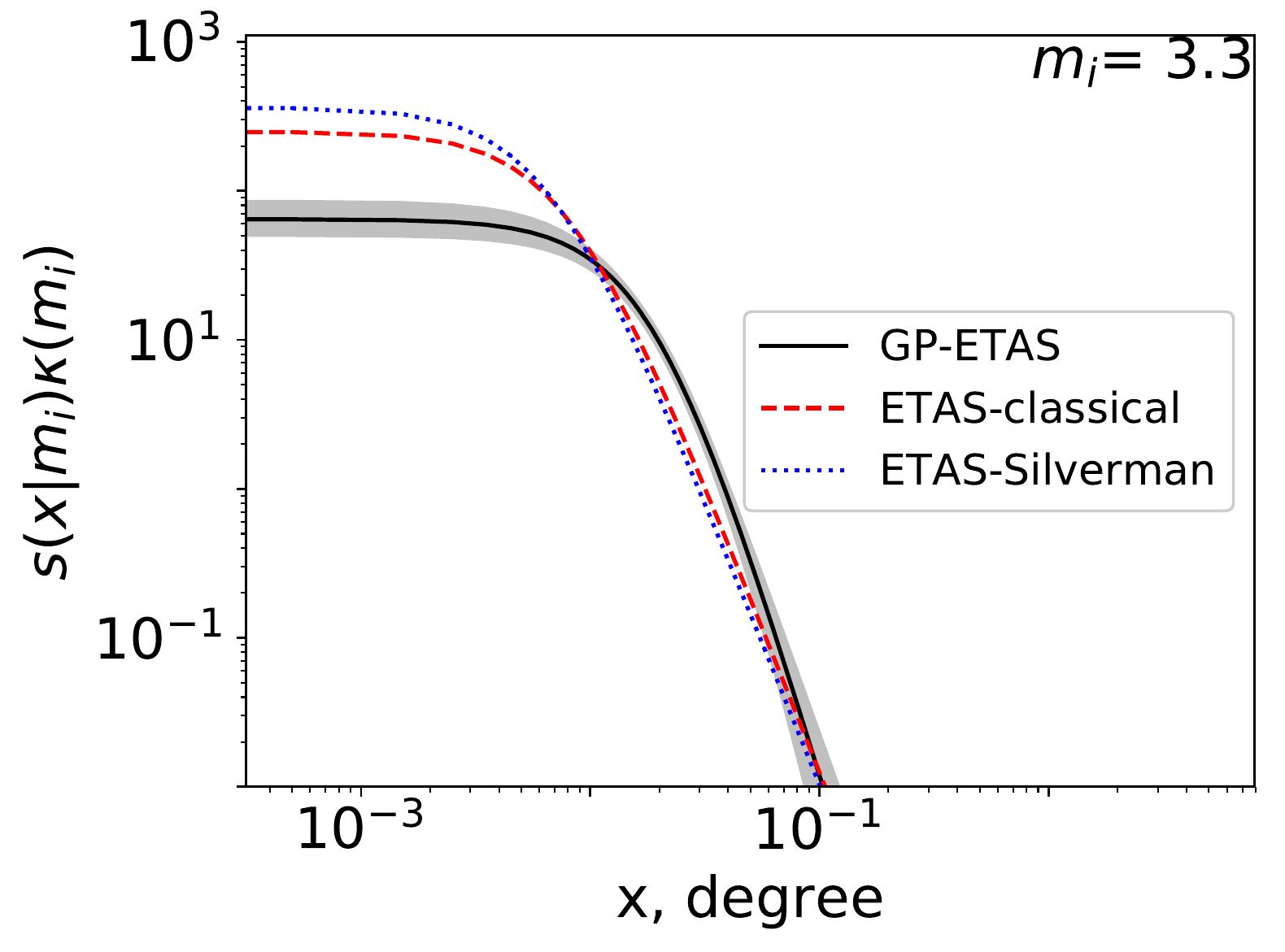}
		\includegraphics[width=0.49\linewidth]{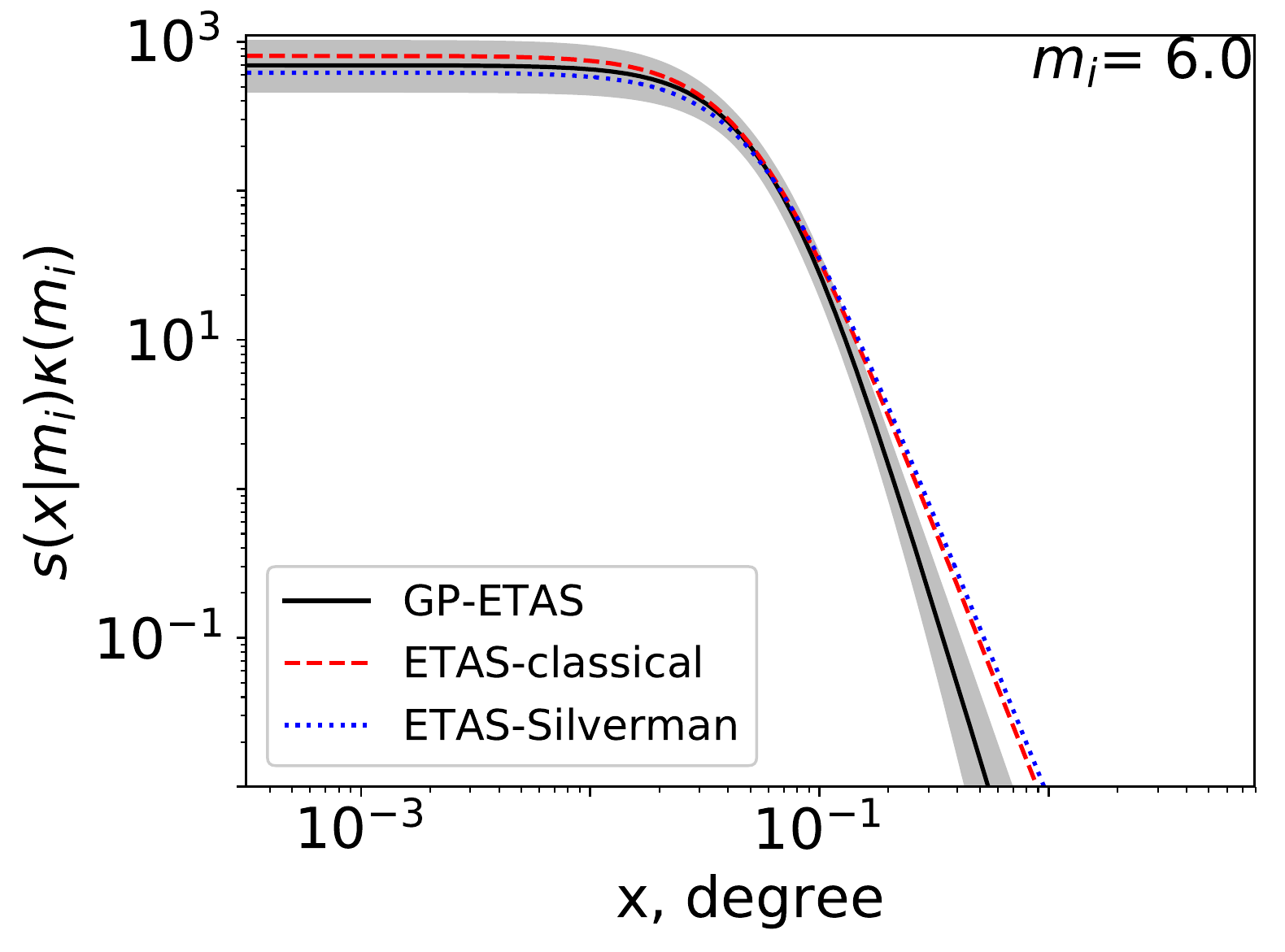}
		\caption{Estimated spatial kernel \eqref{eq:s_x_pwl} multiplied by the productivity \eqref{eq:kappa_m} for the mean 
			magnitude $m=3.3$ and for a large $m=6$ mainshock at position (0,0). Shaded area is 0.05, 0.95 percentile of GP-ETAS.}
		\label{fig:spatial_kernel_case3}
	\end{figure}

	For the l'Aquila data set GP-ETAS
	performs slightly better than the other models
	in terms of $\ell_{\rm test}$.
	Note, that ETAS--classical estimates fairly large values for $\mu(\x)$ in regions 
	with many aftershocks (Figure \ref{fig:BG_aquila}).
	This is similar to Case 1 for the synthetic experiments.
	Hence, as for the synthetics one may assume that
	ETAS--classical overshoots in these regions;
	the posterior of $\barlambda$ supports this hypothesis, see Figure \ref{fig:UB_aquila}.
	The estimated $\thetas_\varphi$ are similar for $K_0,c,p,\alpha$, differ for $d,\gamma,q$.
	Recall, that the latter describe the spatial distribution of the aftershocks.
	The discrepancies are to be seen in the context of (almost)
	linear trade-offs between the parameters $d,\gamma$; 
	for $d,\gamma$ this can be discerned from Figure \ref{fig:case3fig013ascatter}.  
	The spatial kernels of the three models are shown in Figure \ref{fig:spatial_kernel_case3}
	for the mean magnitude and a large magnitude.
	\begin{table}
		\centering
		\small
		\caption{L'Aquila data set: Parameter values $\thetas_\varphi$ of the triggering function.}
		\begin{tabular}{lcccccccc}
			\hline
			model & quantiles & $K_0$ & $c$ & $p$ & $\alpha$ & $d$ & $\gamma$ & $q$  \\ 
			\hline
			&  &  &  &  &  &  &  &   \\ 
			ETAS-classical &  & 0.0293 & 0.0400 & 1.21 & 1.801 & 0.0008  & 0.30   & 1.95 \\ 
			&  &  &  &  &  &  &  &  \\ \hline
			&  &  &  &  &  &  &  &   \\ 
			ETAS-Silverman &  & 0.0300 & 0.0352 & 1.18 & 1.773 & 0.0005  & 0.34   & 1.91 \\ 
			&  &  &  &  &  &  &  &  \\ \hline
			& median & 0.0269 & 0.0276 & 1.16 & 1.780 & 0.0044  & 0.19   & 2.57 \\ 
			GP-ETAS & 0.05 & 0.0224 & 0.0164 & 1.11 & 1.660 & 0.0029  & 0.16   & 2.17 \\
			& 0.95 & 0.0321 & 0.0451 & 1.20 & 1.887 & 0.0063  & 0.23   & 3.27 \\ 
			\hline 
		\end{tabular}
		\label{tab:theta_varphi_Aquila}
	\end{table}
	\begin{table}
		\centering
		\small
		\caption{Test likelihood $\ell_{\rm test}$ of unseen test data sets.}
		\begin{tabular}{lcccc}
			\hline
			testing period &  $N_{
				\rm test}$ & ETAS--classical & ETAS--Silverman & GP-ETAS \\ 
			\hline 
			30.0 days & 2 & -13.4 & -13.3 & \textbf{-12.4} \\ 
			1.0 years & 18 & -79.8 & -80.1 & \textbf{-77.4} \\ 
			5.0 years & 1116 & 5058.8 & 5050.5 & \textbf{5076.1} \\
			total test period ($\approx 8.2$ years) & 1466 &5748.3 & 5735.5 & \textbf{5749.0} \\ 
			\hline 
		\end{tabular} 
		\label{tab:test_likelihood_Aquila}
	\end{table}

	
	%###############################################################################
	\section{Discussion and Conclusions}
	%###############################################################################
	We have demonstrated that the  proposed GP-ETAS model in combination with 
	augmentation techniques~\citep{HawkesAndOakes1974,Adams2009,Donner2018} provides 
	the means of assessing the Bayesian posterior of a semi-parametric spatio-temporal ETAS model. 
	We have shown for three examples that the predictive performance improves over  classical methods.
	In addition, we can quantify parameter uncertainties via their empirical posterior density. 
	The developed framework is flexible and allows for
	several extensions that deserve consideration in future research, e.g.~a time depending background rate. 
	Another  obvious extension of our work would be a Bayesian non-parametric treatment of the triggering function $\varphi$. 
	See also \citep{Zhang2019}.
	
	\medskip
	Future research will deal with a more geology informed choice of the prior GP.
	Sometimes a given catalog comes with information about, e.g. fault locations, which are not straightforward to incorporate in traditional treatments of the spatio-temporal ETAS model. For the GP-ETAS model, however, incorporation of 
	informative priors is possible within our Bayesian setting. For example, spatial information about fault zones can be incorporated by an adequate choice of the mean of the GP, which was chosen to be $0$ throughout this work. 
	While we restricted ourselves to the squared exponential~\eqref{eq:cov_fun} as covariance function for the GP prior of $f$ the framework is not restricted to this either and other covariance function can be used to incorporate prior information, 
	e.g.~from the Mat\'ern class, or any other function that ensures that the covariance matrix is positive definite.
	\medskip
	
	Another important issue is the computational effort. Having to sample the GP at all observed events in $\D$ and at positions of the latent Poisson process $\Pi$, resulting in a cubic complexity of $\mathcal{O}\left((N_{\D\cup\Pi})^3\right)$, implies an undesirable computational complexity of the current 
	GP-ETAS Gibbs sampler. There are  however several possibilities to mitigate this complexity via model approximations 
	and/or alteration. For example, one could resort to approximations to the posterior distribution in order to be able to scale the GP-ETAS model to larger catalogues ($N_D\gg10^3$). While those always come with the sacrifice of asymptotic exactness, some approaches are likely to provide good estimates in the large data regime. One of such approximations is provided by variational inference, which was already proposed for the SGCP by~\citet{Donner2018} utilising sparse GPs~\citep{Titsias2009}. This approach makes use of the same model augmentations utilised in this work. The variational posterior of the triggering parameters %$\thetas_\varphi$ 
	could be inferred, e.g., via black-box variational inference~\citep{ranganath2014blackboxVBI}. Alternatively, one could restrict the calculations to finding the MAP estimate of the GP-ETAS model. For the background intensity this can be efficiently done by an expectation-maximisation algorithm based on the model augmentations presented here and sparse GPs~\citep{Donner2018}. This can be combined with a Laplace approximation to provide an approximate Gaussian posterior. The limiting factor under such approximations will most likely arise from the branching structure for which
	the required computations scale like ${\cal O}(N_\D (N_\D - 1) / 2)$. Finally, one could also investigate gradient-free affine invariant sampling methods as proposed by \cite{GIHLS19,GINR19,RW19}.
	
	\medskip

	We conclude by reemphasising the importance of semi-parametric Bayesian approaches to spatio-temporal 
	statistical earthquake modelling and the need for developing efficient tools for their computational inference. Within this work we
	have followed the Gibbs sampling approach in combination with data augmentation and have demonstrated its applicability to
	realistic earthquake catalogs. 
	
	\section*{Acknowledgments}
	
	This research has been partially funded by Deutsche Forschungsgemeinschaft (DFG, German Science Foundation) - SFB 1294/1 - 318763901.
	
	%\clearpage
	%\bibliographystyle{plain}
	\newpage
	\printnomenclature[0.9in]
	
	\bibliographystyle{abbrvnat}
	\bibliography{references}
	
	\pagebreak
	\appendix
	\appendixpage
	\addappheadtotoc
	%\newpage
	%####################################################################################
	\section{GP-ETAS generative model}
	\label{app:simGP-etas}
	%####################################################################################
	The generative model of GP-ETAS consists of two parts given in 
	Algorithm \ref{algo:simGP_part1_BG} -- \ref{algo:simGP_part2_offspring},
	and requires several inputs.
	Here, we chose a GP with zero mean and covariance function $k(\x,\x\prime \vert \nus)$  given in $\eqref{eq:cov_fun}$ with hyperparameters $\nus$, a triggering function $\varphi(\cdot|\thetas_\varphi)$ given in (\ref{eq:var_phi} -- \ref{eq:gt}) with spatial kernel \eqref{eq:s_x_pwl} and therefore
	$\thetas_\varphi=(K_0,c,p,\alpha,d,\gamma,q)$, and a mark distribution, 
	that is an exponential distribution
	$m_i-m_0 \sim {\rm Exponential}(\beta)$
	(Gutenberg--Richter relation) with parameters $\beta,m_0$.
	The simulation algorithm can be easily adjusted for other choices.
	
	\begin{algorithm}
		\algrenewcommand\algorithmicprocedure{\textbf{}}
		\caption{: GP-ETAS generative model Part 1: generating background events $\D_0$ }\label{algo:simGP_part1_BG}
		\hspace*{\algorithmicindent} \textbf{Input}: 
		Spatio-temporal domain $\calX\times\calT$; GP mean function and covariance function 
		\hspace*{\algorithmicindent} \hspace*{1.3cm} with hyperparameters $\nus$;
		upper bound $\bar{\lambda}$; parameters of the mark density $\beta$, $m_0$\\
		%	\\
		\hspace*{\algorithmicindent} \textbf{Output}:
		Background events $\D_0=\{t_i,x_i,m_i,z_i=0\}_{i=1}^{N_{\D_0}}$
		%	\\
		\begin{algorithmic}[1]
			\State $N\sim \mathrm{Poisson}(\bar{\lambda}|\calX||\calT|)$\Comment{Sample number of candidate events from Poisson distribution %on $\calX\times\calT$
			}
			\State $\{\x_i\}_{i=1}^N\sim \mathcal{U}(\calX)$\Comment{Distribute candidate events uniformly in $\calX$}
			\State $\{f(\x_i)\}_{i=1}^N\sim \mathcal{GP}(\boldsymbol{0},\Ks_{\nus})$
			\Comment{Draw function values from the GP with $\Ks_{\nus}=\{k(\x_i,\x_j|\nus)\}_{i,j}^N$}
			\State $\D_0 \gets \emptyset$ \Comment{Initialise the set of background events $\D_0$}
			\State $N_{\D_0} \gets 0$ \Comment{Initialise number of background events}
			\For{$i \gets 1,...,N$}\Comment{\textit{Thinning} procedure}
			\State $r_i \sim \mathcal{U}(0,\bar{\lambda})$ \Comment{Draw a uniform random variable on the interval $[0,\bar{\lambda}]$}
			\If{$r_i<\bar{\lambda}\sigma(f(\x_i))$} \Comment{Acceptance criteria}
			\State $\x_i$ is accepted
			\State $t_i\sim \mathcal{U}(\calT)$ \Comment{Distribute background event uniformly in $\calT$}
			\State $m_i-m_0 \sim \mathrm{Exponential}(\beta)$ \Comment{Sample a mark from a shifted exponential distribution}
			\State $z_i \gets 0$ \Comment{Assign branching variable $z_i$, index of parent event}
			\State $\D_0 \gets \D_0  \cup \{t_i,\x_i,m_i,z_i\}$ \Comment{Add event to accepted background events $\D_0$}
			\State $N_{\D_0} \gets N_{\D_0} +1$ \Comment{Count number of background events}
			\EndIf
			\EndFor
			\State Sort $\D_0$ by event times $t_i$
			%		\\
			\\
			\Return{$\D_{0}$} 
		\end{algorithmic}
	\end{algorithm}
	\begin{algorithm}
		\algrenewcommand\algorithmicprocedure{\textbf{}}
		\caption{: GP-ETAS generative model Part 2: generating and adding offspring events}\label{algo:simGP_part2_offspring}
		\hspace*{\algorithmicindent} \textbf{Input}: $\D_0$, $N_{\D_0}$ from Algorithm (\ref{algo:simGP_part1_BG});
		spatio-temporal domain $\calX\times\calT$; 
		triggering function 
		\hspace*{\algorithmicindent} \hspace*{1.3cm}$\varphi(\cdot)$ with defining 
		parameters 
		$\thetas_\varphi=(K_0,c,p,\alpha,d,\gamma,q)$; parameters of the mark \hspace*{\algorithmicindent} \hspace*{1.3cm}distribution $\beta$, $m_0$\\
		%	\\
		\hspace*{\algorithmicindent} \textbf{Output}:
		Background events $\D=\{t_i,x_i,m_i,z_i\}_{i=1}^{N_\D}$
		%	\\
		\begin{algorithmic}[1]
			\State $\D \gets \D_0$ \Comment{Initialise the set of simulated events $\D$ with the background events $\D_0$}
			\State $N_{\D} \gets N_{S_0}$  \Comment{Initialise number of simulated events with the number of background events}
			\State $j \gets 1$  \Comment{Initialise the index of potential parent event}
			\While {$j<N_{\D}$}  \Comment{Consider all events in $\D$ for producing potentially offspring}
			\State $\{t_j,\x_j,m_j,z_j\}\gets \D [j]$ \Comment{Obtain entries of the $j$th simulated event in $\D$}
			\State $\lambda_{\mathrm{max},j} \gets {\rm max}(\kappa(m_j)g(t-t_j))=Ke^{\alpha(m_j-m_0)}c^{-p}$ \Comment{Get upper bound $\lambda_{\mathrm{max},j}$ of $j$th PP}
			\State $|\calT_j| \gets |\calT|-t_j$\Comment{Compute the size of the time window $|\calT_j|$ of direct offspring}
			\State $N_j \gets \mathrm{Poisson}(\lambda_{\mathrm{max},j}|\calT_j|)$\Comment{Sample number of candidate offspring events}
			\If{$N_j>0$}\Comment{Check if there are candidate offspring events}
			\State $\{t_i\}_{i=1}^{N_j}\sim \mathcal{U}(0,|\calT_j|)+t_j$\Comment{Distribute candidate events uniformly in $[t_j,t_{max}]$}
			\For{$i \gets 1,...,N_j$}\Comment{\textit{Thinning} procedure}
			\State $r_i \sim \mathcal{U}(0,\lambda_{\mathrm{max},j} )$ \Comment{Draw a uniform random variable on the interval $[0,\lambda_{\mathrm{max},j}]$}
			\If{$r_i<\lambda_{\mathrm{max},j} (t_i-t_j+c)^{-p}$} \Comment{Acceptance criteria using (\ref{eq:kappa_m},\ref{eq:gt})}
			\State $t_i$ is accepted
			\State $\x_i \sim s(\x-\x_j|m_j,d,\gamma,q)$ \Comment{Sample position of the offspring event \eqref{eq:s_x_pwl}}
			\State $m_i-m_0 \sim \mathrm{Exponential}(\beta)$ \Comment{Sample marks, see above}
			\State $z_i \gets j$ \Comment{Assign branching variable $z_i$, index of parent event}
			\State $\D \gets \D  \cup \{t_i,\x_i,m_i,z_i\}$ \Comment{Add offspring event to $\D$}
			\State $N_{\D} \gets N_{\D}+1$ \Comment{Count number of simulated events $D$}
			\EndIf
			\EndFor
			\State $j \gets j+1$ \Comment{Advances to next possible parent event}
			\EndIf
			\EndWhile
			\State Sort $\D$ by event times $t_i$, while taking care of properly mapping the branching variables
			%		\\
			\\
			\Return{$\D$} 	
		\end{algorithmic}
	\end{algorithm}

	\newpage
	%####################################################################################
	\section{Definition of the P\'olya--gamma density}
	\label{app:PG_def}
	%####################################################################################
	Here, we briefly define the P\'olya-gamma density~\citep{Polson2013}. First we define the $\PG(\omega\vert b,0)$, which is completely defined through its Laplace transform
	\begin{equation}\label{eq:laplace pg}
	\int_0^\infty e^{-\omega t}\PG(\omega\vert b, 0)d\omega = \cosh^{-b}(\sqrt{t / 2}).
	\end{equation}
	With this definition it can be shown that
	\begin{equation}
	\omega \stackrel{d}{=} \frac{1}{2\pi^2}\sum_{k=1}^\infty \frac{g_k}{(k-1/2)^2},
	\end{equation}
	where $g_k\sim\mathrm{Gamma}(b,1)$ and the equality is in distribution. With this result one can then define a {\it tilted} P\'olya--gamma density given by
	\begin{equation}\label{eq:tilted_PG}
	\PG(\omega\vert b, c) \propto e^{-\frac{c^2}{2}\omega}\PG(\omega\vert b, 0),
	\end{equation}
	where $b\in \R^+$ and $c\in \R$, 
	and the normalisation can be straightforwardly obtained with~\eqref{eq:laplace pg}. Also for this the tilted density we can derive the Laplace transform
	\begin{equation}\label{eq:mgf pg}
	\int_0^\infty e^{-\omega t}\PG(\omega\vert b, c)d\omega = \frac{\cosh^b(c/2)}{\cosh^{b}\left(\frac{\sqrt{c^2/2 + t}}{2}\right)}.
	\end{equation}
	From \eqref{eq:mgf pg} all moments of the P\'olya--gamma density can be derived analytically and furthermore an acceptance-rejection algorithm with high acceptance rate was derived by~\cite{Polson2013}.

	%\newpage
	%####################################################################################
	\section{Conditional posteriors for the background intensity}
	\label{appendix:BG_gibbs_sampler_in_detail}
	%####################################################################################
	Here we derive the conditional posterior distributions given in~\eqref{eq:cond_post_Pi_latent}--\eqref{eq:cond_post_GP}. For each of those our starting point is the augmented likelihood of the background intensity in~\eqref{eq:aug2} with the prior of interest. Note, that for the augmented variables $\Pi,\omegas_{\D},\omegas_\Pi$ there are no additional priors, and hence their conditionally distribution will only be determined by~\eqref{eq:aug2}.
	%\medskip
	
	\paragraph{The latent  Poisson  process $\Pi$}
	To derive the conditional posterior for $\Pi$, we consider all terms 
	in~\eqref{eq:aug2} that depend on $\Pi=\{x_l\}_{l=1+N_\D}^{N_{\D\cup\Pi^{(k)}}}$ and marginalise over the $\omegas_{\Pi}$. This results in
	\begin{equation}\label{eq:cond_latent_poisson}
	\begin{split}
	p(\Pi \vert \barlambda,\fs)
	\propto 
	\prod_{l=N_\D+1}^{N_{\D\cup\Pi}}
	\barlambda\sigma(-f_l)
	\exp\left(-|\calT|\int_\calX \barlambda \sigma(-f(\x))\diff \bx \right),
	\end{split}
	\end{equation}
	which we identify as an unnormalised Poisson process density with rate $\barlambda\sigma(-f_l)$. Again note, that the process is defined over $\calT\times\calX$.
	To sample $\Pi^{(k)}$ in the $k^{\rm th}$ iteration, we can utilise the \textit{thinning} procedure~\citep{Lewis1976}, where we first sample a homogeneous point process with intensity $\bar{\lambda}^{(k-1)}$. At the resulting points we draw the the GP $f$ given the previous sample $\fs^{(k-1)}$ from the predictive distribution (see below). Then, we keep all events $l$ with probability $\sigma(-f_l)$ which yields  $\Pi^{(k)}$  according to \eqref{eq:cond_post_Pi_latent}.

	\paragraph{The P\'olya--gamma random variables $\omegas$:}
	Next we sample the P\'olya--gamma random variables at %the observations in 
	$\D_0$. From \eqref{eq:aug2} we see, that the components in $\omegas_{\D_0}$ factorise, meaning that the conditional posteriors are independent. Hence, we get for each $i:z_i=0$
	\begin{align}
	p(\omega_i\vert\fs,\D,Z) 
	\propto 
	e^{-\frac{\left(f_i\right)^2}{2}\omega_i}\PG(\omega_i\vert 1, 0) 
	\propto 
	\PG(\omega_i\vert 1, |f_i|),
	\end{align}
	and hence 
	$\omegas_{\D}^{(k)}$ 
	can be sampled independently from a tilted 
	P\'olya-gamma distribution where $i:z_i=0$, given branching structure $Z^{(k)}$ and GP $\fs^{(k-1)}$. For $i:z_i\neq0$ we set $\omega_i=0$.
	The last equivalence follows from a property of an \emph{tilted} P\'olya-gamma distribution, see \eqref{eq:tilted_PG}.
	In effect, we get \eqref{eq:cond_post_Omega_S0}.
	
	\medskip
	
	The conditional posterior $\omegas_{\Pi}$ at the positions of the latent events 
	$\Pi$ also factorises in all components and hence we get for each $l=N_\D + 1, \ldots,N_{\D\cup\Pi}$ 
	\begin{align}
	p(\omega_l\vert \fs,\Pi) 
	\propto 
	%\prod_{\bx_j\in \Pi}
	e^{-\frac{\left(f_l\right)^2}{2}\omega_l}\PG(\omega_l\vert 1, 0) 
	\propto 
	\PG(\omega_l\vert 1, |f_l|),
	\end{align}
	and also $\omegas_{\Pi}^{(k)}$ can be sampled independently from a tilted 
	P\'olya--gamma distribution \eqref{eq:cond_post_Omega_Pi} given the samples of $\Pi^{(k)}$ and $\fs^{(k-1)}$;
	we get \eqref{eq:cond_post_Omega_Pi}.
	
	\paragraph{The upper bound on the intensity $\barlambda$:}
	For $\bar{\lambda}$ we assume a Gamma prior $p(\bar{\lambda}\vert\alpha_0,\beta_0)$ 
	with shape parameter $\alpha_0$ and rate parameter $\beta_0$. 
	Together with~\eqref{eq:aug2}  we derive the conditional posterior being,
	\begin{align}
	p(\barlambda \vert \D_0,\Pi, Z) 
	\propto
	\barlambda^{N_{\D_0\cup\Pi}}
	e^{-\barlambda|\calX||\calT|}
	p(\barlambda\vert\alpha_0,\beta_0)\propto{\rm Gamma}(\bar{\lambda}\vert \alpha_1,\beta_1).
	\end{align}
	where $\alpha_1 = N_{\D_0\cup\Pi}+\alpha_0$ and $\beta_1 = |\calT||\calX|+\beta_0$. 
	Hence, given $\Pi^{(k)}$ and $D_0^{(k)}$ we can sample $\bar{\lambda}^{(k)}$, 
	and one gets \eqref{eq:cond_post_upper_bound}.
	
	\paragraph{The posterior Gaussian Process $f$:}
	For the conditional posterior of the Gaussian process $f$ we rewrite \eqref{eq:aug2} with the terms depending on $f$ as follows
	\begin{equation}\label{eq:23 in mat form}
	\begin{split}
	p(\D_0,\omegas_{\D},\Pi,\omegas_{\Pi} \vert \fs, \barlambda, Z)
	&\propto  
	\prod_{\begin{subarray}{l}
		i:z_i=0
		\end{subarray}}^{N_\D}
	e^{f_iu_i-\frac{f_i^2}{2}\omega_i} 
	\prod_{\begin{subarray}{l}
		i:z_i\neq 0
		\end{subarray}}^{N_\D}
	e^{f_iu_i-\frac{f_i^2}{2}\omega_i} 
	\prod_{l=N_\D+1}^{N_{\D\cup\Pi}}
	e^{f_lu_l-\frac{f_l^2}{2}\omega_l}\\
	&=  e^{-\frac{1}{2}\fs^\top\Omegas\fs + \bu^\top\fs},
	\end{split}
	\end{equation}
	where we define $u_i=\frac{1}{2}$ if $z_i=0$, $u_i=0$ if $z_i\neq0$, and $u_l=-\frac{1}{2}$ for $l=N_\D+1,\ldots,N_{\D\cup\Pi}$. It follows that $\Omegas={\rm diag}(\omegas_\D,\omegas_{\Pi})$. The GP prior $\fs$ at a finite set $\D,\Pi$ of points is given by
	\begin{equation}
	p(\fs) = \mathcal{N}(\fs\vert \mathbf{0}, \Ks_{\fs,\fs})
	\end{equation}
	where the entry of row $i$ and column $j$ of $\Ks_{\fs,\fs}$ is given by the covariance function~\eqref{eq:cov_fun} $k(\x_i, \x_j \vert \nus)$. Together with~\eqref{eq:23 in mat form} we identify the conditional posterior
	\begin{subequations}
		\begin{align}
		p(\fs \vert \D, \omegas_{\D}, \Pi, \omegas_{\Pi},Z) &\propto e^{-\frac{1}{2}\fs^\top\Omegas\fs + \bu^\top\fs}\mathcal{N}(\fs\vert \mathbf{0}, \Ks_{\fs,\fs})\\ &
		\propto \mathcal{N}\left(\fs\vert \left[\Omegas + \Ks_{\fs,\fs}^{-1}\right]^{-1}\bu,\left[\Omegas + \Ks_{\fs,\fs}^{-1}\right]^{-1}\right),
		\end{align}
	\end{subequations}
	which defines the conditional posterior at $\fs$, and it is easy to sample $\fs^{(k)}$ given instances of $\omegas_{\D}, \Pi, \omegas_{\Pi}$ from previous samples. However, the algorithm requires us to sample the posterior also at other points $\x^\ast\notin\D\cup\Pi$, e.g. for sampling the next instance of $\Pi$ or for visualisation of the background intensity $\mu(\x)$ on a grid. Here, we denote the GP at all additional points $\fs^\ast$. The GP prior defines a predictive distribution~\citep{Williams2006gaussian} of any set $\fs^\ast$ given $\fs$
	\begin{equation}\label{eq:predictive GP}
	p(\fs^*|\fs,\nus)
	=\mathcal{N}
	\left(
	\fs^*|
	\Ks_{\fs^*,\fs}\Ks_{\fs,\fs}^{-1}\fs,\quad
	\Ks_{\fs^*,\fs^*}-\Ks_{\fs^*,\fs}\Ks_{\fs,\fs}^{-1}\Ks_{\fs,\fs^*}
	\right),
	\end{equation}
	where $\Ks_{\fs^*,\fs^*}$ contains the covariances between the points $\bx^\ast$ of $\fs^\ast$ and $\Ks_{\fs^*,\fs}=\Ks_{\fs^*,\fs}^\top$ between the points $\bx^\ast$ of $\fs^*$ and $\fs$ at $\x$. Note, that the posterior of $\fs^\ast$ given $\fs$ is equal to the conditional prior, since~\eqref{eq:aug2} does not depend on $\fs^\ast$. 
	
\end{document}